\documentclass[11pt]{article}
\usepackage{amssymb,amsmath}
\usepackage{bm} 
\usepackage{booktabs} 
\usepackage{array}
\usepackage{latexsym}
\usepackage{graphicx}
\usepackage{color}
\usepackage{datetime}
\usepackage[nosort]{cite}
\usepackage{verbatim}
\usepackage{enumerate}
\usepackage{chngpage} 
\usepackage{mathrsfs}
\usepackage{euscript}
\usepackage{psfrag}
\usepackage{subfig}
\usepackage{anyfontsize}
\usepackage{braket}
\usepackage[normalem]{ulem}

\usepackage{blindtext}%
\usepackage[theorems,breakable]{tcolorbox}%

\usepackage{multirow,graphicx}
\usepackage{makecell}
\usepackage{float}
\usepackage{tikz}
\usetikzlibrary{decorations.text}
\usepackage{enumitem}
\setlist{itemsep=0pt}
\usetikzlibrary{math}
\usepackage{setspace}
\usepackage{comment}

\usepackage{mciteplus}

\usepackage[colorlinks=true,      linkcolor=blue,      urlcolor=blue,      
            filecolor=blue,      citecolor=blue,       pdfstartview=FitH,     
						pdfpagemode=UseNone,      bookmarksopen=true]{hyperref}  
\usepackage[all]{hypcap}     

\usepackage{tikz}
\usepackage{pgfplots}
\usetikzlibrary{decorations.text,intersections, pgfplots.fillbetween}
\usetikzlibrary{math}
 \usetikzlibrary{3d,shapes.geometric,shadows.blur}
\usepackage{tikz-3dplot}

\usepackage{blindtext}%
\usepackage[theorems,breakable]{tcolorbox}%

\newtcbtheorem{mytheo}{}%
{colback=gray!5,colframe=gray!35!black,fonttitle=\bfseries}{th}

\newtcbtheorem{lem}{}{%
        colback=green!5,%
        colframe=green!35!black,%
        fonttitle=\bfseries,title %
    }{lem}

\definecolor{amaranthred}{rgb}{0.83,0.13,0.18}
\definecolor{amazon}{rgb}{0.23,0.48,0.34}
\definecolor{bdazzledblue}{rgb}{0.18,0.35,0.58}
\definecolor{absolutezero}{rgb}{0.0,0.28,0.73}
\definecolor{bitterlemon}{rgb}{0.79,0.88,0.05}
\definecolor{byzantine}{rgb}{0.74,0.2,0.64}
\definecolor{turquoise}{rgb}{0.19, 0.84, 0.78}
\definecolor{burgundy}{rgb}{0.5, 0.0, 0.13}
\definecolor{airforceblue}{rgb}{0.36, 0.54, 0.66}
\definecolor{arsenic}{rgb}{0.23, 0.27, 0.29}

\newcommand{\comm}[1]{} 

\def\({\left(}
\def\){\right)}
\def\[{\left[}
\def\]{\right]}

\def\One{{\hbox{ 1\kern-.8mm l}}}

\def\barray{\begin{array}}
\def\earray{\end{array}}
\def\be{\begin{equation}}
\def\ee{\end{equation}}
\def\bea{\begin{eqnarray}}
\def\eea{\end{eqnarray}}
\def\bal{\begin{align}}
\def\eal{\end{align}}
\def\nn{\nonumber}

\def\-{\,-\,}
\def\={\,=\,}
\def\+{\,+\,}
\def\equi{\,\equiv\,}

\numberwithin{equation}{section} 

\definecolor{cardinal}{rgb}{0.6,0,0}
\definecolor{darkgreen}{rgb}{0,0.4,0}
\definecolor{golden}{rgb}{0.92, 0.7, 0}
\definecolor{midnight}{rgb}{0, 0, 0.5}
\definecolor{darkblue}{rgb}{0, 0, 0.7}
\definecolor{purple}{rgb}{0.5, 0, 0.5}

\definecolor{amaranthred}{rgb}{0.83,0.13,0.18}
\definecolor{amazon}{rgb}{0.23,0.48,0.34}
\definecolor{bdazzledblue}{rgb}{0.18,0.35,0.58}
\definecolor{absolutezero}{rgb}{0.0,0.28,0.73}
\definecolor{bitterlemon}{rgb}{0.79,0.88,0.05}
\definecolor{byzantine}{rgb}{0.74,0.2,0.64}
\definecolor{turquoise}{rgb}{0.19, 0.84, 0.78}
\definecolor{burgundy}{rgb}{0.5, 0.0, 0.13}

\def\IR{\mathbb{R}}

\def\cB{{\cal B}}

\def\cE{{\cal E}}

\def\cO{{\cal O}}

\usetikzlibrary{positioning}

\begin{document}

\phantom{AAA}
\vspace{-10mm}

\begin{flushright}
%
%
\end{flushright}

\vspace{2cm}

\begin{center}

{\fontsize{19}{23}\selectfont{\bf Branes and Antibranes in AdS$_3$:}}

\vspace{0.3cm}
{\fontsize{19}{23}\selectfont{\bf The Impossible States in the CFT Gap}
}

\vspace{0.2cm}

\vspace{1cm}

{\bf \normalsize Roberto Emparan$^{1,2}$ and  Pierre Heidmann$^{3}$}
\vspace{0.5cm}\\

\centerline{$^1$ Instituci\'o Catalana de Recerca i Estudis Avançats (ICREA),}
\centerline{Passeig Lluis Companys, 23, 08010 Barcelona, Spain}
\centerline{$^2$ Departament de F\'isica Qu\`antica i Astrof\'isica and Institut de Ci\`encies del Cosmos,}
\centerline{Universitat de Barcelona, Martí i Franquès 1, 08028 Barcelona, Spain}
\centerline{$^3$ Department of Physics and Center for Cosmology and AstroParticle Physics (CCAPP),}
\centerline{The Ohio State University, Columbus, OH 43210, USA}

\vspace{0.5 cm}

{\footnotesize\upshape\ttfamily emparan@ub.edu,~ heidmann.5@osu.edu } 

 \vspace{0.5 cm}

\end{center}

\begin{adjustwidth}{5mm}{5mm} 
 
\begin{abstract}
\noindent 
We construct a new family of type IIB supergravity solutions corresponding to states of the D1-D5-P-KKm system that carry the same charges and energy as the non-extremal four-charge black hole and are asymptotic to AdS$_3 \times ($S$^3/\mathbb{Z}_{N_k}) \times$ T$^4$. The solutions consist of static binaries of two extremal D1-D5-P black holes with S$^3$ horizons and charges of opposite signs, held in equilibrium by a topological bubble supporting $N_k$ units of KKm charge. Although dynamically unstable, the spacetimes remain smooth on and outside the horizons. The equilibrium condition discretizes the black hole separation, producing a quantized spectrum labeled by the number of antibranes and antimomenta at the anti-BPS center. Strikingly, the lowest-energy states lie within an energy window smaller than the dual CFT mass gap. We also show that these solutions admit regular finite-temperature deformations, which slightly lift the two black holes above extremality while remaining within the gap.

These results challenge the expectation that no states exist within the CFT gap, realizing \emph{impossible states}. We discuss two possible resolutions. First, Schwarzian-type quantum corrections could lift these solutions above the gap. Alternatively, though less likely, higher-genus corrections to the two-dimensional effective super-JT theory allow a sparse spectrum of exponentially suppressed states within the gap. In either case, our construction provides explicit realization of a dense set of intricate, highly non-perturbative, low-energy excited states of holographic CFTs.

\end{abstract}
\end{adjustwidth}

\vspace{8mm}
 

\thispagestyle{empty}

\newpage



\tableofcontents

\newpage

\section{Introduction}

The microscopic spectrum of a black hole is expected to be highly intricate and likely chaotic. Nevertheless, when a black hole admits a degenerate supersymmetric (BPS) ground state, a remarkably robust feature emerges: an energy gap separating the supersymmetric ground states from the first excited level.
This gap scales as an inverse power of the black hole charges, making it parametrically much larger than the typical level spacing, which is suppressed exponentially in the entropy.

A first instance of such a gap was identified in \cite{Maldacena:1996ds} for D1-D5-P black holes using the string theory model of Strominger and Vafa \cite{Strominger:1996sh}.
The lowest excitations arise in the maximally twisted sector of the dual CFT, corresponding to oscillations with wavelength $N_1 N_5$ times the length of the wrapped circle, yielding a gap
\begin{equation}\label{eq:cftgap}
    \Delta E_\text{CFT}=\frac{3}{c}
\end{equation}
in units of the circle radius, where $c=6N_1 N_5$ is the central charge of the CFT.
Although derived at weak coupling---describing a weakly gravitating D-brane configuration rather than a true black hole---the result matched an earlier thermodynamic estimate for near-extremal black holes~\cite{Preskill:1991tb}.
Subsequent work rederived the same gap in the gravitational regime, either by exploiting the AdS$_3$ structure of the near-horizon geometry and its dual two-dimensional CFT~\cite{Maldacena:1997ih}, or from linear perturbations on supersymmetric coherent black hole microstates known as superstrata \cite{Bena:2018bbd,Bena:2019azk}.
In the former picture, the gap is set by the rotational energy of the first fermionic excitation, which carries spin $1/2$.
More recently, its existence was confirmed by a direct computation of the spectrum of near-BPS black holes with AdS$_2$ throats from the gravitational path integral of $\mathcal{N}=4$ super-JT gravity in the disk, reproducing the same quantitative value when applied to the D1-D5-P system~\cite{Heydeman:2020hhw}.

In this article, we present an explicit construction of states that appear to challenge this lore. Building on recent progress in constructing static, non-supersymmetric configurations in supergravity \cite{Heidmann:2021cms,Bah:2022pdn,Bah:2022yji,Bah:2023ows,Heidmann:2023kry,Bena:2024gmp}, we derive exact solutions of type IIB supergravity with
AdS$_3\times($S$^3/\mathbb{Z}_{N_k})\times$T$^4$ asymptotics, which places them within the same dual CFT that describes four-dimensional black holes with D1-D5-P-KKm charges \cite{Johnson:1996ga}---analogous to near-extremal Reissner-Nordström solutions.\footnote{Holography in AdS$_3\times($S$^3/\mathbb{Z}_{N_k})$ with $N_k>1$ and its dual CFT are less well understood than in AdS$_3 \times S^3$  (e.g., \cite{Kutasov:1998zh,Larsen:1999dh}). It is unclear how this could affect our analysis, but we will bear this point in mind again in Section~\ref{sec:Interpretation}.}

For fixed total charge (and hence fixed CFT central charge and momentum), these non-BPS configurations consist of brane-antibrane bound states in AdS$_3$. Quantized by construction, they have zero spin (along the S$^3$) and positive energy, producing a discrete spectrum of nonperturbative, non-BPS states. Strikingly, we find that the lowest-energy members of this spectrum lie deep within the expected mass gap, with energies above the BPS bound $\Delta E_\text{min}=\mathcal{O}(c^{-\frac{5}{3}})$.

These solutions describe bound states of two extremal (Strominger-Vafa) black holes with opposite D1-D5-P charges of unequal magnitude, held apart by a Taub-bolt bubble supported by KKm flux.
As classical supergravity solutions, they are well-behaved: despite being dynamically unstable, they exhibit neither regions of large curvature nor curvature or conical singularities on or outside their horizons.
Because the properties of extremal horizons have recently been shown to be very subtle \cite{Turiaci:2023wrh,Horowitz:2024kcx}, we also construct smooth finite-temperature, non-extremal versions of these bound states that remain within the CFT gap, further reinforcing the challenge they pose as apparently forbidden states.

\subsection{Summary of the results}
\label{sec:Summary}

The solutions presented in this paper are mathematically intricate and complex in form, but the principles governing their construction and existence are intuitive. 

To motivate them, let us recall a line of work that aims at microscopically accounting for the entropy of non-extremal black holes using both branes and antibranes in string theory \cite{Callan:1996dv,Horowitz:1996fn,Horowitz:1996ay,Horowitz:1996ac}. While successful for near-extremal black holes with momenta and antimomenta, \cite{Callan:1996dv,Horowitz:1996fn}, this approach has not yielded a satisfactory microscopic picture once all required brane and antibrane species (four for a four-dimensional black hole) are included, as needed far from extremality.

A proposal to make these degrees of freedom manifest, first introduced in \cite{Heidmann:2023kry}, is to consider configurations in which branes and antibranes are localized in distinct regions of spacetime, allowing the system to remain in an (unstable) equilibrium. This corresponds to replacing the non-extremal black hole with a bound state of two extremal holes---one sourced by branes and another by antibranes---held apart by a smooth topological structure. This bound state has the same total energy and charges as the non-extremal black hole, but, unlike the latter, the microscopic constituents that lift the system above the BPS extremal limit are explicit.

This construction was successfully realized for configurations that are neutral and non-rotating in four-dimensional asymptotically flat spacetime---the setting of the Schwarzschild black hole---
providing a microscopic account of half of its entropy in terms of brane/antibrane bound states \cite{Heidmann:2023kry}. Although these states span only a subspace of dimension $e^{S_\text{BH}/2}$ and are therefore atypical within the full ensemble of black hole microstates---a feature shared by the constructions in this paper---it is nevertheless striking that such a framework appears to capture so large a fraction of the entropy.\footnote{An important caveat to these entropy counts arises from potential quantum effects on the extremal black hole/anti-black hole bound states, which we discuss in Section \ref{sec:Interpretation}.\label{foot:caveat}}

In this work, we extend this construction to the four-charge black hole---which includes the Reissner-Nordström and Schwarzschild solutions as special cases---focusing on the near-extremal regime where the net charges nearly saturate the BPS bound. Instead of the standard four-dimensional Minkowski asymptotics, we consider the AdS$_3$ decoupling limit in type IIB, which allows a more controlled holographic description. With these steps, we aim to provide new insight into the structure of near-extremal states in the strong-coupling regime.

\paragraph*{Framework and solutions.} We work in type IIB supergravity on T$^4$ with D1, D5, P, and KKm charges. In this setting, the decoupling limit of the four-charge black hole yields a non-extremal BTZ geometry \cite{Banados:1992wn} with internal space S$^3/\mathbb{Z}_{N_k}\times$T$^4$, where $N_k$ is the KKm charge. This BTZ black hole carries $N_1$ units of D1 brane charge, $N_5$ units of D5 brane charge 
and $N_p$ units of momentum along the AdS$_3$ circle, with total energy $E \geq N_p$. 

\begin{figure}[t]
    \centering
    \includegraphics[width=1\textwidth]{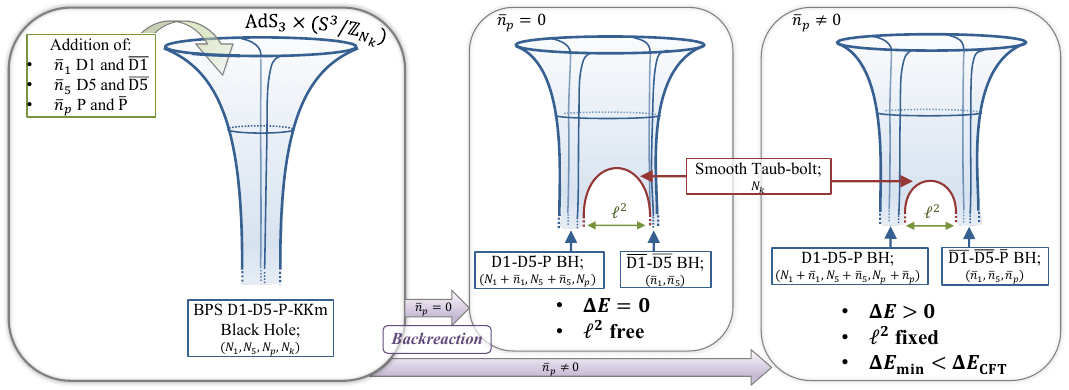}
    \caption{The bound states formed by a BPS and an anti-BPS Strominger-Vafa black hole in AdS$_3\times$S$^3/\mathbb{Z}_{N_k}\times$T$^4$. These solutions are constructed by adding pairs of branes/antibranes and momenta/antimomenta to the four-charge BPS black hole (left panel). We distinguish between zero-energy bound states, which do not include momentum/antimomentum pairs (middle panel), and positive-energy bound states, which do include antimomenta (right panel).}
    \label{fig:Intro1}
\end{figure}

We construct brane-antibrane bound states with the same energy and charges as the near-extremal D1-D5-P-KKm black hole, and therefore the same mass and momentum parameters as the associated non-extremal BTZ black hole. This is achieved using the generalized static Ernst formalism pioneered in \cite{Heidmann:2021cms,Bah:2022pdn,Bah:2022yji,Bah:2023ows}, and can be broken down into several steps, illustrated in Fig.\ref{fig:Intro1}:
\begin{itemize}
    \item We start with the single-center supersymmetric black hole, sourced by $N_1$ D1 branes, $N_5$ D5 branes, $N_p$ momentum modes, and $N_k$ units of KKm charge \cite{Maldacena:1996gb} (Fig.~\ref{fig:Intro1}, left panel).
    \item We then add, ``one by one,'' $\overline{n}_1$, $\overline{n}_5$, and $\overline{n}_p$ pairs of D1 branes and $\overline{\text{D1}}$ antibranes, D5 branes and $\overline{\text{D5}}$ antibranes, and P momenta and $\overline{\text{P}}$ antimomenta, respectively. This breaks supersymmetry while leaving the net charges unchanged (Fig.~\ref{fig:Intro1}, left panel).
    \item Rather than collapsing into a non-extremal black hole, the system is arranged so that the branes/momenta and antibranes/antimomenta polarize into two distinct loci (Fig.~\ref{fig:Intro1}, right panels). This is achieved by blowing up the KKm NUT center into a smooth Taub-bolt geometry supported by $N_k$ units of KKm charge. The constituents are an extremal D1-D5-P black hole at the South pole of the bolt, and another extremal black hole with $\overline{\text{D1}}$-$\overline{\text{D5}}$-$\overline{\text{P}}$ negative-sign charges at the North pole.
    \item Regularity at the bolt imposes a constraint between the quantized charges and the parameter $\ell^2$ controlling the center separation.\footnote{The parameter $\ell^2$ does not represent the physical distance in ten dimensions (which is infinite, since the horizons are extremal) but rather the distance in the two-dimensional $(r,\theta)$ base, where $r$ and $\theta$ parametrize the AdS$_3$ and S$^3$ positions. Nevertheless, it is appropriately regarded as a measure of the distance and interaction between the two throats.} We find two distinct classes of states:
    \begin{itemize}
        \item \underline{Zero-energy bound states} (middle panel). When no antimomenta are present, the regularity condition does not constrain $\ell^2$, which remains a flat direction that can be large or small compared to other length scales in the geometry---the AdS radius, the string length, or the asymptotic S$^1$ radius in AdS$_3$. However, the number of antibranes is strongly constrained, leading to a small, discrete set of regular solutions (roughly scaling with the number of divisors of $N_k$). These states are non-BPS due to the presence of antibranes, yet have the same energy as the BPS four-charge black hole. They are also not perfectly regular, since the $\overline{\text{D1}}$-$\overline{\text{D5}}$ solution without antimomenta is a ``small black hole."
        \item \underline{Positive-energy bound states} (right panel). When antimomenta are present, the number of antibranes is subject only to an upper bound $\overline{n}_{1,5} <N_k N_{1,5}/4$, while regularity fixes the separation $\ell^2$ in terms of the quantized charges. These states exhibit a positive energy gap above the BPS four-charge black hole, determined by the number of branes/antibranes and momenta/antimomenta, thereby forming a rich spectrum of fully backreacted excited states above the BPS ground state.
    \end{itemize}
\end{itemize}

The quantized spectrum of these states can be studied and compared directly with CFT expectations. Their entropy is equal to that of two non-interacting D1-D5-P black holes,
\begin{equation}\label{eq:2SVent}
    S \= 2\pi \left(\sqrt{(N_1+\overline{n}_1)(N_5+\overline{n}_5)(N_p+\overline{n}_p)} + \sqrt{\overline{n}_1\overline{n}_5\overline{n}_p}\right).
\end{equation}
To the extent that the entropy of each of the two black holes admits a microscopic counting, this provides a natural microscopic origin for the bound-state entropy.
We then show that when the expression \eqref{eq:2SVent} is recast in terms of the total energy and charges of the system, it is of the same order as the entropy of the non-extremal four-charge black hole in AdS$_3$, 
\begin{equation}\label{eq:btzent}
    S_\text{BTZ}=2\pi\sqrt{N_1 N_5 N_k}\left(\sqrt{\frac{E+N_p}{2}}+\sqrt{\frac{E-N_p}{2}}\right)\,,
\end{equation}
and typically approximates half of it for the positive-energy states, $S\lesssim \tfrac{1}{2} S_\text{BTZ}$, and slightly more for the zero-energy states, $S\lesssim \tfrac{1}{\sqrt{2}} S_\text{ext}$, where $S_\text{ext}$ is the entropy of the extremal hole.

The entropy of the nonextremal BTZ black hole was microscopically accounted for in~\cite{Horowitz:1996fn,Strominger:1997eq} by considering left- and right-moving open strings along the D1-D5 intersection. Our result provides a fundamentally different picture of microstates in this system: first, it involves antibranes, and second, the momentum and antimomentum excitations are of a different nature than the left- and right-moving modes of~\cite{Horowitz:1996fn,Strominger:1997eq}. Here, each is carried by a different set of D1-D5 branes.

\paragraph*{Low-energy spectrum.} When we analyze the low-energy spectrum of the positive-energy bound states at fixed net charges, we find that it is entirely determined by the numbers of antibranes and antimomenta in the configuration, with two key features emerging:
\begin{itemize}
    \item The energy of our states above the BPS bound, $\Delta E=E-N_p$, 
    has a minimum significantly smaller than the expected CFT mass gap. For equal net quantized charges, we find
    \begin{equation}
        \Delta E_\text{min} \= \cO \left(\frac{1}{c^{5/3}} \right) \,\ll\, \Delta E_\text{CFT}=\frac{3}{c}=\frac1{2N_k N_1 N_5}\,.
    \end{equation}
    \item The lowest-energy bound states are \emph{not} those involving the smallest numbers of antibranes and antimomenta, $(\overline{n}_1,\overline{n}_5,\overline{n}_p)=\cO(1)$ (those have a large energy above extremality of order $N_k N_p$). Instead, they result from a fine-tuning of the anti-element numbers: the antibrane numbers scale with the net charges $(N_1,N_5)$, while the antimomenta can even exceed the net momentum charge, scaling as $N_k N_p$. Therefore, these are highly non-perturbative excitations above the BPS four-charge black hole. 
    \item We show that their unexpectedly low energy, despite the large numbers of anti-elements, is due to the emergence of a long AdS$_2$ throat between the boundary and the cap containing the two extremal holes. The energy localized at the cap is then strongly redshifted and becomes extremely small as seen from the boundary. We find that the length $L$ of this AdS$_2$ throat matches that of the near-extremal black hole with the same energy and charges,
    \begin{equation}
    L \,\approx\, \frac{R_{\text{AdS}_2}}{2} \,\log \frac{N_p}{\Delta E}\,. \end{equation}
    This length is much larger than the maximal value $L_\text{max} = R_{\text{AdS}_2} \log S_\text{ext}$ that \cite{Lin:2022rzw} found in $\mathcal{N}=2$ JT gravity.\footnote{Bear in mind that the throats of near-BPS asymptotically flat black holes in four and five dimensions are instead described by $\mathcal{N}=4$ JT gravity.} In that case, quantum effects halt the growth of the throat at temperatures $\sim \Delta E_\text{CFT}$. The large length of the throats in our states, $L\gg L_\text{max}$, is closely related to their below-gap energies; the relevance for our solutions of the quantum effects that limit their values will be discussed later.
\end{itemize}

\begin{figure}[t]
    \centering
    \includegraphics[width=1\textwidth]{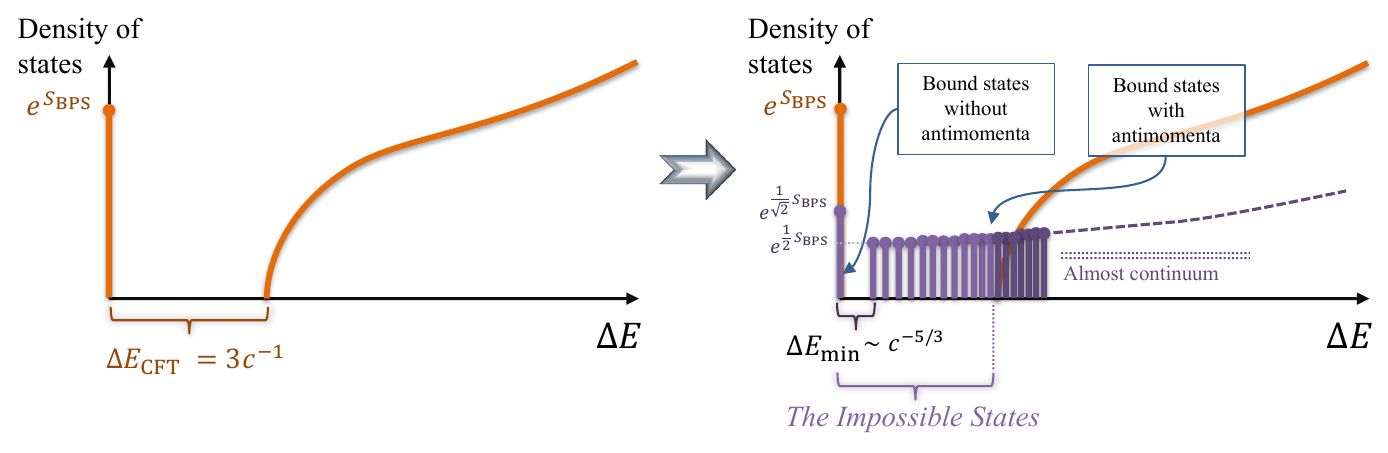}
    \caption{Schematic representation of the density of states as a function of the energy above the BPS bound. Left: CFT expectations, featuring a mass gap above the supersymmetric states. Right: inclusion of the quantized spectrum of brane-antibrane bound states, including the ``impossible states'' populating the gap.}
    \label{fig:Intro2}
\end{figure}

\paragraph*{The impossible states.} These bound states in AdS$_3\times$ S$^3/\mathbb{Z}_{N_k}\times$T$^4$ appear to challenge standard CFT expectations. Fig.~\ref{fig:Intro2} depicts the density of states as a function of the energy above the BPS bound. The left panel displays the spectrum predicted from the CFT and the supergravity analysis in \cite{Heydeman:2020hhw}: a large degeneracy of BPS states at $\Delta E=0$, corresponding to the entropy of the supersymmetric four-charge black hole, separated by a gap of order $c^{-1}$ from the nearly continuous spectrum of non-BPS black hole states.

Our quantized spectrum of non-BPS states reveals the presence of near-extremal states within the CFT mass gap, which we refer to as \emph{the impossible states} (Fig.~\ref{fig:Intro2}, right). Although they have a large degeneracy \eqref{eq:2SVent}, their density is exponentially suppressed by at least $e^{S_\text{BPS}/2}$ relative to the BPS sector. Crucially, the corresponding geometries feature no regions of large curvature or compact cycles at the string scale.

There are two distinct classes of impossible states, corresponding to the two kinds of bound states described above: the zero-energy bound states at $\Delta E=0$, which do not involve antimomenta, and the positive-energy bound states in the range 
\begin{equation}
    c^{-5/3} \lesssim \Delta E \lesssim c^{-1}\,,
\end{equation}
which do. 

\paragraph*{Stability and finite-temperature.} We assess the sensitivity of these bound states to perturbations. First, following \cite{Elvang:2006dd}, we show that the positive-energy bound states are ``top-of-the-hill'' configurations, unstable under axisymmetric perturbations. Nevertheless, the instability potential is surprisingly shallow, implying a slow decay rate. In contrast, the zero-energy bound states, owing to their flat direction, are insensitive to such instabilities. 

Second, we investigate whether the bound states admit finite-temperature deformations. To this end, we construct solutions in type IIB supergravity where the extremal black holes at the poles of the Taub-bolt are replaced by near-extremal, finite-temperature versions with the same charges. We find that the zero-energy bound states do not admit a $T\neq 0$ extension without modifying the local charges, whereas the positive-energy states do. This suggests that positive-energy bound states can support thermal fluctuations, while the zero-energy bound states are isolated at $T=0$, with any infinitesimal increase in temperature likely triggering a non-perturbative transition.

\subsection{Quantum fluctuations and the fate of the impossible states}
\label{sec:Interpretation}

Figure~\ref{fig:Intro1} highlights the central role of long throats in our construction. The positive-energy impossible states lie inside an AdS$_2$ throat (the \emph{UV throat}) which, deep in the IR, splits into two AdS$_2$ throats (the \emph{IR throats}) for the extremal (or near-extremal) black hole and anti-black hole constituents. Recent work shows that such throats---whether near-BPS or not---undergo large quantum fluctuations in their length, effectively described by a Schwarzian theory \cite{Iliesiu:2020qvm,Heydeman:2020hhw}.

\subsubsection*{Quantum effects in the throat} 
A full derivation of the Schwarzian dynamics in our solutions is beyond the scope of this paper, but these effects are unlikely to significantly alter their local geometries, since the Schwarzian mode is global, primarily controlling throat length. Instead, the most relevant consequences expected in our context are:
\begin{enumerate}
\item[(1)] The equilibrium maintaining our configurations may be perturbed, though the unstable maximum of the potential is likely to persist, albeit shifted.
\item[(2)] The degeneracies (entropies) of states near extremality will be modified.
\item[(3)] The energy levels of all configurations are expected to shift.
\end{enumerate}

Effects (1) and, more importantly (2), affect mostly the IR throats, while (3) is of more consequence in the UV throat.

A key distinction between the UV and IR throats is that only the former is governed by near-BPS dynamics.
Although the bound-state constituents can be extremal D1-D5-P black holes, supersymmetry is almost certainly broken in their throats since the presence of the anti-partner introduces an angular distortion of $S^3$ which persists arbitrarily down the throat. Although this distortion leaves the black hole area unchanged, it should lift light fermion modes, yielding the non-supersymmetric Schwarzian theory of near-extremal, non-BPS throats \cite{Iliesiu:2020qvm}. Since supersymmetry is only broken deep in the IR, the larger, system-encompassing UV throat is expected to remain near-BPS, governed by the supersymmetric Schwarzian of \cite{Heydeman:2020hhw}. We emphasize that, in the strict extremal limit, quantum fluctuations diverge, making supergravity saddles unreliable---though locally the geometry may remain sensible---which motivates the study of finite-temperature regularizations.

The energy scale of Schwarzian fluctuations in AdS$_2\times S^2$ is set by the AdS$_2$ radius, with $\Delta E_q \sim 1/R_{\text{AdS}_2}^3$ in four-dimensional Planck units. In our configurations, the larger UV throat sets the energy gap, while the constituent IR throats, being smaller, fluctuate over length scales that are parametrically similar or possibly higher.\footnote{We find that the AdS$_2$ radii of the IR (constituent) throats cannot be larger than $\sim N_k^{-1/4}$ times the radius of the UV AdS$_2$ throat.} 

These effects enter in point (2): the entropies derived from the areas of black hole/anti-black hole constituents, such as \eqref{eq:2SVent}, will receive significant corrections. Since supersymmetry might be broken in the IR throats, their entropy could be much smaller than the Bekenstein-Hawking value, with a density of states approaching $O(1)$ at zero temperature \cite{Iliesiu:2020qvm}. At temperatures comparable to the gap in each throat, the degeneracies remain similar to the semiclassical ones, up to order-one factors. This provides the proper context for interpreting the entropy counts in \eqref{eq:2SVent} and in \cite{Heidmann:2023kry}---with finite-temperature regularizations of the constituents.

Point (3) has important implications for interpreting the impossible states. Even if the local geometries remain reliable, global quantities such as the energy and throat length can shift appreciably---by amounts given by the size of the gap, or possibly larger, since the constituent throats have higher energy scales. As already noted, the throats in our geometries appear to be much longer than what quantum fluctuations in JT gravity would seem to allow, suggesting that sizable quantum corrections to the energies of these states may arise. Such effects could help mitigate the tension posed by the existence of the impossible states, as we will discuss later. 

\subsubsection*{Unitarity and the gap}

Recall that in \cite{Maldacena:1996ds}, the gap was derived for weakly coupled string states and does not directly apply to the black hole spectrum at strong coupling. Similarly, the gaps in \cite{Bena:2018bbd,Bena:2019azk} apply to particular states that are not typical black hole microstates. The argument in \cite{Maldacena:1997ih}, derived at strong coupling, relies on the Sugawara construction of the stress tensor in a theory with an $SU(2)$ Kac-Moody current. For a primary $|h,j\rangle$, this implies that
\begin{equation}\label{eq:suga}
    h\geq \frac{j(j+1)}{k+2}\,.
\end{equation}
To relate this to our setup, we use that the theory has left- and right-moving sectors with $E=h+\bar{h}$ and $N_p=\bar{h}-h$, and we are interested in the energy above the left-moving state with $E=N_p$, i.e., with $h=0$. This energy is
\begin{equation}
    \Delta E=E-N_p=2 h\,. 
\end{equation}
Now assume that the right-moving sector is supersymmetric, so that the state $E=N_p$ is BPS.\footnote{If the theory is not the $(4,4)$ theory of the D1-D5 black hole, but rather a chiral $(4,0)$ theory with supersymmetry only in the right sector---as in our setup--- the state $E=N_p$ is BPS but $E=-N_p$ is not. The constructions in this article could be readily applied to the latter case, but since there is no gap, we will not pursue it further.}
Using that the superconformal algebra fixes the Kac-Moody level to be $k=c/6$, and we are in a regime of $c\gg 1$, \eqref{eq:suga} gives
\begin{equation}
    \Delta E \geq \frac{12 j(j+1)}{c}\,.
\end{equation}
Evaluating this bound for a supermultiplet whose highest-spin component has $j = 1/2$ yields a bound that is weaker than \eqref{eq:cftgap}. In addition, the  Virasoro and $SU(2)$ couplings of the fermions in the $\mathcal{N}=4$ theory may further complicate the analysis at strong coupling.

However, there is a purely kinematic argument---independent of the coupling---that derives the gap as a consequence of unitarity\footnote{This argument was communicated to us by Joaquín Turiaci, whom we thank for permission to reproduce it here.}. Given a primary $|h,j\rangle$, consider linear combinations of the descendants
\begin{equation}
    L_{-1} |h,j\rangle\,,\qquad J_{-1}^3|h,j\rangle\,.
\end{equation}
Unitarity requires the $2\times2$ matrix of their inner products to be positive definite. This matrix, which does not receive fermionic contributions, takes the form
\begin{equation}
    \begin{pmatrix}
2 h & j \\[4pt]
j & \tfrac{k}{2}
\end{pmatrix}\,.
\end{equation}
Positivity of its determinant implies
\begin{equation}
     h\geq \frac{j^2}{k}=\frac{6j^2}{c}\,,
\end{equation}
and hence
\begin{equation}\label{eq:unitarygap}
     \Delta E \geq \frac{12j^2}{c}\,.
\end{equation}
Considering a supermultiplet with $j = 1/2$, we recover the expected gap \eqref{eq:cftgap}. 

Therefore, states lying within the gap of the CFT would be incompatible with unitarity, independent of the coupling strength. The only conceivable loophole in our setting is that the CFT dual to AdS$_3\times($S$^3/\mathbb{Z}_{N_k})$ with $N_k>1$ remains insufficiently understood. While we regard this as an unlikely way out, we cannot entirely exclude it.

\subsubsection*{The persistence of the impossible}

We are now led to conjecture two possible fates or interpretations of the impossible states, which merit detailed exploration in future work:
\begin{itemize}
\item \textbf{Quantum Lifting of Impossible States:}
As discussed above, it is possible that incorporating quantum fluctuations and backreaction within the multiple AdS$_2$ throats lifts the energies of these states above the CFT mass gap. On general grounds, one would expect that when the string coupling is $\sim 1/c$, the corrections to the energy should be of that order too. At the very least, the zero-energy bound states, in the absence of any protection, should be lifted to non-zero energy. Quantum corrections could also modify the balance that allows such states to exist classically, possibly restoring the expected CFT gap. 

\item \textbf{Stable States Consistent with JT Expectations:}  
Alternatively, the main features of the states may survive quantum effects while remaining impossible, implying that the CFT gap is not entirely depopulated. While this would require that the unitarity argument leading to \eqref{eq:unitarygap} is not applicable in our setting, it could be consistent with the result in \cite{Heydeman:2020hhw}, which does not account for exponentially suppressed contributions from saddles with other topologies. Indeed, the number of impossible states is at least a factor $\sim e^{-S_\text{BPS}/2}$ smaller than the black hole density. In this view, they would constitute a legitimate, albeit highly suppressed, sector of the theory, enriching the near-BPS spectrum beyond conventional expectations.

\end{itemize}

Before concluding this section, we add a cautionary note, related to the first possibility above, which is worth spelling out separately. In the classical supergravity limit, the gap is strictly zero---so how can classical solutions be used to describe states within this gap? The implicit assumption is that the properties of our supergravity solutions, where the charges of branes and momenta vary continuously (except the KKm charge $N_k$, which is discrete already in supergravity), remain valid to a good approximation when these charges are quantized according to the usual rules (see eq.~\eqref{eq:QuantizedAsymCharges}\footnote{In the supergravity limit, $g_s\to 0$, with the charges $Q_i$ finite.}). This assumption---also underlying the derivation in \cite{Maldacena:1997ih} of the gap from supergravity solutions---becomes increasingly delicate for states deeper inside the gap. Nevertheless, since we find states spanning nearly the entire gap for which the assumption appears reasonable, we are left to confront one of the two alternatives discussed above.

\paragraph*{Structure of the paper.} In Section~\ref{sec:Section1}, we establish the type IIB framework, review known results about the four-charge, non-extremal black hole in AdS$_3$, and present the type IIB generalization of the electrostatic Ernst formalism developed in \cite{Heidmann:2021cms,Bah:2022pdn,Bena:2024gmp}. Readers primarily interested in the explicit solutions may proceed directly to Section~\ref{sec:3}, where we construct the bound states and analyze their main properties and regularity conditions. Section~\ref{sec:Sec4} discusses the discrete family of states, with particular attention to the low-energy spectrum. In Section~\ref{sec:BSNonExt}, we construct the nonzero-temperature bound states. We conclude in Section~\ref{sec:Conclusion} with a summary of results and future outlook. Several appendices complement the main text: Appendices~\ref{App:Ernst} and~\ref{App:ErnstTypeIIB} detail the solution-generating techniques, while Appendix~\ref{App:SUSYLim} discusses the supersymmetric limit of our bound states in the absence of antibranes and antimomenta.

\section{D1-D5-P-KKm system and asymptotically AdS$_3$ geometries}
\label{sec:Section1}

Here, we introduce our setup, review basic solutions, and describe the formalism for the construction of the non-BPS bound states that will be presented in the next section.

\subsection{D1-D5-P-KKm system}

We consider static configurations of the D1-D5-P-KKm system in type IIB supergravity on T$^4$. The ten-dimensional spacetime splits into the time direction, a rigid internal T$^4$, a circle along $y$ carrying momentum, a circle along $\psi$ supporting Kaluza-Klein monopole (KKm) flux, and a three-dimensional base $\cB$ transverse to the fluxes. All fields depend solely on the coordinates of $\cB$---actually, only on two of them, $(r,\theta)$, since we also assume that the base is axisymmetric with polar angle $\phi$. The brane and flux configuration is summarized in Table~\ref{tab1}.

\begin{table}[h]
 \centering
\begin{tabular}{|c|c|c|c|c|c|c|c|}
\hline
 & $t$ & $\cB$ & $\psi$ & $y$ & T$^4$\\
\hline
D1& $\leftrightarrow$ & $\bullet$ & $|$ & $\leftrightarrow$ &  $|$  \\
\hline
D5 & $\leftrightarrow$ & $\bullet$ & $|$ & $\leftrightarrow$ &  $\leftrightarrow$ \\
\hline
P &  & $\bullet$ & & $\leftrightsquigarrow$ &   \\
\hline
KKm &  & $\bullet$  & $\leftrightsquigarrow$ &  &   \\
\hline
\end{tabular}
\caption{\textit{Brane and spacetime configuration in type IIB supergravity. Arrows indicate brane extension directions; vertical bars denote smearing; curly arrows indicate momentum or KKm flux; bullets mark localized sources on the external base.} \label{tab1}}
\end{table}

For all our solutions, the type IIB supergravity fields (in the string frame\footnote{The string-frame metric relates to the Einstein-frame metric such as: $g = e^{\Phi/2} g_E$.}) take the form
\begin{align}
ds_{10}^2 &=  \frac{1}{\sqrt{Z_1 Z_5}} \left[ - \frac{f dt^2}{Z_p}+ Z_p(dy-A_p dt)^2\right]  + \sqrt{Z_1 Z_5} \left[\frac{1}{Z_0} \left(d\psi+H_0 \,d\phi\right)^2 +Z_0 \,ds(\cB)^2 \right] +\sqrt{\frac{Z_1}{Z_5}}ds(T^4)^2,\nn\\
C^{(2)} &= H_5 \,d\phi \wedge d\psi -A_1 \,dt\wedge dy \,,\qquad  e^\Phi \= \sqrt{\frac{Z_1}{Z_5}}\,,\qquad C^{(0)}=C^{(4)}=B_2=0\,,\label{eq:TypeIIBAnsatz}
\end{align}
The fields organize into four sectors: $(Z_1,A_1)$,  $(Z_5,H_5)$,  $(Z_p,A_p)$,  and $(Z_0,H_0)$, corresponding respectively to the D1, D5, P, and KKm gravitational and gauge potentials. The $A$ and $H$ denote electric and magnetic gauge potentials, respectively.

We define $R_{y}$ as the radius of the $y$ circle, $g_s$ as the string coupling, $l_s$ as the string length, and $(2\pi)^4 V_4$ as the volume of T$^4$. The ten-dimensional Newton constant is $G_{10} = 8\pi^6 g_s^2 l_s^8$.

\subsection{The four-charge black hole}

Let us first review how the static, non-extremal four-charge black hole fits within the type IIB ansatz \eqref{eq:TypeIIBAnsatz}. It is given by the following base metric and fields, labeled by $\Lambda = 0,1,5,p$,\footnote{A factor 2 difference in normalization has been used between the P and the D1-D5-KKm sectors, chosen to match later conventions.}
\begin{align}
    &Z_\Lambda = 1+ \frac{M_\Lambda}{4 \bar{r}},\quad f= 1-\frac{a^2}{4 \bar{r}},\quad A_p = - \frac{Q_p}{4 \bar{r} Z_p}, \quad A_1 = -\frac{Q_1}{2 \bar{r} Z_1},\quad H_5 = \frac{Q_5}{2} \cos \bar{\theta},\quad H_0 = \frac{Q_0}{2} \cos \bar{\theta},\nn\\
   & ds(\cB)^2 = \frac{d\bar{r}^2}{f} + \bar{r}^2 \left(d\bar{\theta}^2 +\sin^2 \bar{\theta} \, d\phi^2 \right),
\end{align}
where the mass parameters for each constituent are
\begin{equation}
     \quad M_p \=\frac{1}{2}\left( \sqrt{a^4+4 Q_p^2}-a^2\right) , \quad M_I \=\frac{1}{2}\left( \sqrt{a^4+16 Q_I^2}-a^2\right), \quad I=0,1,5.
\end{equation}
The solution is asymptotically $\IR^{1,3}\times$T$^6$, characterized by five parameters: the four supergravity charges $(Q_1,Q_5,Q_p,Q_0)$ and the non-extremality parameter $a^2$. When all charges are equal, $Q = Q_0 = Q_1 = Q_5 = Q_p/2$, the solution reduces to the well-known embedding of the Reissner-Nordström black hole as a BPS solution in type IIB supergravity.

Each charge satisfies an extremality bound,
\begin{equation}
    M_p \,\geq\, Q_p\,,\qquad M_I \,\geq\, 2 Q_I\,,\qquad I=0,1,5.
\end{equation}
The extremal limit $M_p = Q_p$ and $M_I = 2 Q_I$ corresponds to $a^2 = 0$, where the solution becomes the static supersymmetric D1-D5-P-KKm black hole \cite{Johnson:1996ga}.

The horizon is located at $\bar{r} = a^2/4$, and the temperature, mass, and entropy are\footnote{The temperature is defined from the $\widetilde{r} \to 0$ limit of the $(\widetilde{r},t)$ plane: $d\widetilde{r}^2 - (2\pi T)^2 \widetilde{r}^2 \frac{dt^2}{R_y^2}$, with $\widetilde{r}^2=4\bar{r}-a^2$.}
\begin{equation}
    M \= \frac{2 a^2+M_0+M_1+M_5+M_p}{16},\quad T \= \frac{a^2 R_y}{\pi \sqrt{\prod (a^2+M_\Lambda)}},\quad S \= \frac{\pi \sqrt{\prod (a^2+M_\Lambda)}}{16 G_4}, 
\end{equation}
where $G_4$ is the four-dimensional Newton constant.

\subsection{AdS$_3$ decoupling limit and the BTZ black hole}
\label{sec:BTZ}

Being interested in asymptotically AdS$_3$ geometries, we take a decoupling limit of the non-extremal black hole via the coordinate rescaling
\begin{equation}
    \bar{r} \,\to\, \epsilon^2 \,\frac{r^2+a^2}{4},\qquad (t,y) \,\to\, \frac{1}{\epsilon}(t,y), \qquad \bar{\theta} \,\to\, 2\theta,
    \label{eq:NearHorizonLim}
\end{equation}
together with the near-extremal limit
\begin{equation}
    a^2 \,\to\, \epsilon^2 a^2\,,\qquad Q_p \,\to\, \epsilon^2 Q_p\,,
    \label{eq:NearExtremalLim}
\end{equation}
with $\epsilon \to 0$, which generates a near-horizon AdS$_3$ region,

Thus, the D1-D5-KKm sectors become extremal ($M_I \to 2 Q_I$), while the P sector remains non-extremal ($M_p \geq Q_p$). The non-extremality of the resulting geometry is therefore entirely sourced by the P flux.

In this limit, the four-charge black hole becomes
\begin{align}
ds_{10}^2 \= &\frac{1}{2\sqrt{Q_1 Q_5}} \left[-\frac{r^2}{1+\frac{M_p}{r^2+a^2}}\,dt^2 +\left(r^2+a^2+M_p\right) \,\left(dy+ \frac{Q_p}{r^2+a^2+M_p} dt\right)^2 \right]\nn \\
&+ N_k \sqrt{Q_1 Q_5} \left[\frac{dr^2}{r^2+a^2} + d\Omega_3^2 \right]+ \sqrt{\frac{Q_1}{Q_5}}\,ds(T^4)^2  \,,\nn\\
C^{(2)} \= & N_k Q_5 \cos^2 \theta \,d\varphi_1 \wedge d\varphi_2 +\frac{r^2+a^2}{2Q_1} \,dt\wedge dy \,,\qquad  e^\Phi \= \sqrt{\frac{Q_1}{Q_5}}\,,\label{eq:metBTZ}
\end{align}
where $d\Omega_3^2$, ($\varphi_1$, $\varphi_2$) denote the metric and the Hopf angular coordinates of a round three-sphere,
\begin{align}
d\Omega_3^2& =d\theta^2+\cos^2 \theta\,d\varphi_1^2+\sin^2 \theta\,d\varphi_2^2 \,,\label{eq:DefHyperspher}\\
\varphi_1 &\equi \frac{1}{2}\left(\phi -\frac{2\psi}{N_k}\right)\,,\qquad \varphi_2 \equi  \frac{1}{2}\left(\phi +\frac{2\psi}{N_k}\right)\quad \Leftrightarrow \quad \phi \= \varphi_1 +\varphi_2 \,,\qquad \psi= \frac{N_k}{2} \,(\varphi_2 -\varphi_1)\,. \nn
\end{align}
We also redefined the charge $Q_0 = N_k \in \mathbb{Z}$ such that the sphere is regular with the lattice of periodicities for the compact directions given by
\begin{equation}
\begin{split}
(\psi,\phi) &\= (\psi,\phi) \+ (2\pi,0)\,,\qquad (\psi,\phi) \= (\psi,\phi) \+ (N_k\pi,2\pi) \,,\qquad y \= y \+2\pi  R_y \,,
\end{split}
\label{eq:psi&phiPerio}
\end{equation}
The geometries are asymptotic to AdS$_3\times($S$^3/\mathbb{Z}_{N_k})\times$T$^4$ for which the AdS$_3$ and S$^3$ radii at the boundary are equal to $(N_k^2 Q_1 Q_5)^\frac{1}{4}$ \cite{Kutasov:1998zh,Bena:2005ay}. \medskip

The decoupling limit \eqref{eq:metBTZ} is the product of a nonextremal BTZ black hole with a S$^3/\mathbb{Z}_{N_k}\times$T$^4$ internal space. The outer and inner horizons are at $r=0$ and $r^2=-a^2$, but the latter will not play any role in our discussion.

Applying the generic procedure reviewed in the Appendix \ref{app:EnergyAdS}, the net quantized D1-D5-P charges, the central charge, and the energy are given by
\begin{align}\label{eq:QuantizedAsymCharges}
&N_1 \= \frac{V_4\,Q_1}{g_s l_s^6}\,,\quad N_5 \= \frac{Q_5}{g_s l_s^2}\,,\quad N_p \= \frac{V_4 R_y^2\,Q_p}{2g_s^2 l_s^8}, \\ 
&c  \= 6N_ k N_1 N_5, \quad E \= \frac{V_4 R_y^2\,\sqrt{a^2+4 Q_p^2}}{4g_s^2 l_s^8}.
\end{align}
From the expression for the energy, it is clear that only the P flux contributes to the surplus energy above extremality, and the extremal bound reads
\begin{equation}
    E\,\geq\, N_P.
    \label{eq:EnergyRangeBTZ}
\end{equation}

The temperature and entropy of the black hole are
\begin{equation}
T\= \frac{R_y a^2}{2\pi \sqrt{2N_k Q_1 Q_5(a^2+M_p)}}\,,\qquad S_\text{BTZ}\= \frac{\pi c}{3} \frac{R_y \sqrt{a^2+M_p}}{\sqrt{2N_k Q_1 Q_5}}\,.
\label{eq:TempEntropEnBTZ}
\end{equation}
This entropy can also be expressed in terms of energy and momentum as
\begin{equation}
S_\text{BTZ}=2\pi\sqrt{\frac{c}{12}}\, \left(\sqrt{E+N_p}+ \sqrt{E-N_p} \,.\right)
\label{eq:EntropyBTZ}
\end{equation}
In this form, it admits a microscopic counting both in the supersymmetric extremal limit $E = N_p$, where \cite{Maldacena:1996gb}
\begin{equation}
T \= 0\,,\qquad S_\text{ext}\= 2\pi \sqrt{N_k N_1 N_5 N_p}\,,\qquad E\= N_p\,,
\end{equation}
and when  $E > N_p$, with $E+N_p$ and $E-N_p$ corresponding to left and right movers in the dual CFT$_2$ \cite{Horowitz:1996fn,Strominger:1997eq}.\footnote{The counting in \cite{Horowitz:1996fn,Strominger:1997eq} is made for the D1-D5 CFT. It readily extends to $N_k=1$ \cite{Johnson:1996ga}, while the extension to $N_k>1$ can be justified through U-duality from other CFTs for the four-charge black hole.} 

\subsection{Ernst formalism of the D1-D5-P-KKm system}
\label{sec:ErnstTypeIIB}

Ref.~\cite{Heidmann:2021cms} uncovered a new integrable structure for brane intersection solutions across several supergravity frameworks. Under suitable assumptions---specifically for static, cohomogeneity-two configurations—the Einstein equations decompose into decoupled sectors of electrostatic Ernst equations \cite{Ernst:1967by,Ernst:1967wx}. This has enabled the adaptation of the many known solutions of the Ernst equations from four-dimensional Einstein–Maxwell theory to string theory, leading to new constructions such as black hole and black brane bound states \cite{Bah:2020pdz,Bah:2022pdn,Dulac:2024cso,Bena:2024gmp,Chakraborty:2025ger,Dima:2025tjz}, non-supersymmetric smooth horizonless geometries \cite{Bah:2021owp,Bah:2021rki,Bah:2022yji,Bah:2022pdn,Heidmann:2022zyd,Bah:2023ows}, and brane–antibrane bound states \cite{Heidmann:2023kry,Heidmann:2023thn}.

When applied to the D1-D5-P-KKm system in type IIB supergravity \cite{Heidmann:2021cms,Bah:2022pdn,Bena:2024gmp}, the ansatz is given by \eqref{eq:TypeIIBAnsatz} with
\begin{equation}
ds(\cB)^2 \=e^{2(\nu_{0}+\nu_{1}+\nu_{5}+\nu_{p})} \left( d\rho^2 +dz^2\right) + \rho^2d\phi^2 ,\qquad f=1.
\label{eq:BaseMetGen}
\end{equation}
The scalars $e^{2\nu_\Lambda}$ are conformal factors that are sourced by the D1-D5-P-KKm sectors, and the coordinates $(\rho, z, \phi)$ are Weyl-Papapetrou coordinates such that all fields depend only on $(\rho, z)$.

The fields organize into four decoupled sectors: $(Z_1, A_1, \nu_1)$, $(Z_5, H_5, \nu_5)$, $(Z_p, A_p, \nu_p)$, and $(Z_0, H_0, \nu_0)$, corresponding respectively to the D1, D5, P, and KKm fluxes. Each sector satisfies the electrostatic Ernst equations detailed in \cite{Bah:2022pdn},
\begin{align}
& \Delta \log Z + Z^2 \,  \nabla A . \nabla A \,=\, 0\,, \qquad  \nabla .  \left( \rho Z^2 \nabla A\right) \,=\, 0 \,,\nn\\
&\frac{2\partial_z \nu}{\rho} = \partial_\rho \log Z \,\partial_z \log Z - Z^2 \partial_\rho  A\partial_z  A,\label{eq:EOMErnst} \\ 
&\frac{4\partial_\rho \nu}{\rho} = \left( \partial_\rho \log Z\right)^2 - \left(\partial_z \log Z\right)^2 - Z^2  \left((\partial_\rho A)^2-(\partial_z A)^2 \right), \nonumber
\end{align}
where we have dropped the indices for clarity and defined the gradient and Laplacian,  $\Delta \equi \frac{1}{\rho} \,\partial_\rho \left( \rho \,\partial_\rho\right) + \partial_z^2$ and $\nabla \equi (\partial_\rho,\partial_z)$.  Moreover,  for the KKm and D5 sectors involving a magnetic gauge field $H$ instead of an electric gauge field $A$,  those are connected by the duality relations
\begin{equation}
dA_I \= -\frac{1}{\rho Z_I^2} \,\star_2 dH_I\,, \qquad I=0,5,
\label{eq:EMDualEq}
\end{equation}
where $\star_2$ is the Hodge star operator in the flat two-dimensional $(\rho,z)$ space.

Solutions with asymptotic behavior
\begin{align}
&e^{2\nu},Z_p \,\to \,1\,,\qquad Z_1 \,\to\, \frac{2Q_1}{r^2} \,, \qquad Z_5 \,\to\, \frac{2Q_5}{r^2} \,, \qquad Z_0 \,\to\, \frac{2N_k}{r^2} \,, \nn \\
&  H_0   \,\to \,\frac{N_k}{2}\cos 2\theta\,,  \qquad H_5  \,\to \, \frac{Q_5}{2} \cos 2\theta\,,  \qquad A_1 \,\to \, -\frac{r^2}{2Q_1}\,,\qquad A_p \,\to\, -\frac{Q_p}{r^2}\,,
\label{eq:AsympBehav}
\end{align}
at large $r$, where 
\begin{equation}
\rho \equi \frac{r^2}{4}\, \sin 2\theta \,,\qquad z \equi \frac{r^2}{4}\, \cos 2\theta\,,
\label{eq:AsympCoord}
\end{equation}
are asymptotic to the same AdS$_3 \times (S^3/\mathbb{Z}_{N_k}) \times T^4$ geometry as the decoupling limit of the nonextremal four-charge black hole,
\begin{equation}
\begin{split}
ds_{10}^2& \,\to \,  N_k\sqrt{Q_1 Q_5} \left[\frac{r^2}{2N_k Q_1 Q_5}\,(-dt^2+dy^2)+ \frac{dr^2}{r^2} \+ d\Omega_3^2\right]  +\sqrt{\frac{Q_1}{Q_5}}ds(T^4)^2 \,, \\ \label{eq:AdS3Asymp}
C^{(2)} &\,\to \, \frac{Q_5N_k}{2} \cos 2\theta \, d\varphi_1 \wedge d\varphi_2 + \frac{r^2}{2Q_1} dt\wedge dy\,, \qquad e^{\Phi } \to \sqrt{\frac{Q_1}{Q_5}}\,,
\end{split}
\end{equation}
where $d\Omega_3^2$, $\varphi_1$, and $\varphi_2$ are defined in \eqref{eq:DefHyperspher} with the periodicity lattice given in \eqref{eq:psi&phiPerio}.

Similarly, the supergravity charges $Q_1$, $Q_5$, and $Q_p$ relate to the quantized net D1-D5-P charges as in \eqref{eq:QuantizedAsymCharges}.

Therefore, this Ernst-type formalism can be used to construct new geometries with the same boundary conditions, energy, and conserved charges as the decoupling limit of the four-charge non-extremal black hole, but with a richer internal structure, as in \cite{Bah:2022pdn,Bena:2024gmp,Sakamoto:2025jtn}.

\section{Bound states of D1-D5-P and $\overline{\text{D1}}$-$\overline{\text{D5}}$-$\overline{\text{P}}$ black holes}
\label{sec:3}

The cohomogeneity-two ansatz \eqref{eq:TypeIIBAnsatz} enables the construction of new type IIB supergravity solutions from known Ernst equation solutions in each sector $(Z_I, A_I, \nu_I)$. We will consider a regular bound state of two extremal D1-D5-P black holes: one with positive (brane) charges and momenta, the other with negative (antibrane) charges and antimomenta. The two centers, separated by a distance $\ell^2/4$ in the base, are held apart by a smooth Taub-bolt geometry---a non-contractible sphere---within AdS with a KKm charge $N_k$ \cite{Page:1978hdy}.

Similar neutral, asymptotically flat configurations were found in \cite{Heidmann:2023kry}. Instead, here we construct bound states that (1) carry nonzero net charge and (2) are asymptotic to AdS$_3$. The first follows from choosing unequal constituent charges, and the second from modifying the field asymptotics according to \eqref{eq:AsympBehav}, which requires a nontrivial decoupling limit and a Harrison transformation of the Ernst solution \cite{harrison_new_1968,Harrison:1980fr}. The technical details appear in Appendices \ref{App:Ernst} and \ref{App:ErnstTypeIIB}; below we present the resulting geometry and analyze its properties.

\subsection{The solution in bolt-centered coordinates}
\label{sec:BoltCoor}

The solutions, written in the spherical coordinates centered around the Taub-bolt separating the two extremal centers, take the form
\begin{align}
ds_{10}^2 \= & \frac{1}{\sqrt{Z_1 Z_5}} \,\left[ - \frac{dt^2}{Z_p}+Z_p\,(dy-A_p dt)^2\right]  \+ \sqrt{Z_1 Z_5} \,ds_4^2\+ \sqrt{\frac{Z_1}{Z_5}}\,ds(T^4)^2, \nn \\
C^{(2)} \= &H_5 \,d\phi \wedge d\psi -A_1 \,dt\wedge dy \,,\qquad  e^\Phi \= \sqrt{\frac{Z_1}{Z_5}}\,.\label{eq:MetBS}
\end{align}
where the D1-D5-P fields are given by
\begin{align}
Z_I &\= \frac{2(q_I+\overline{q}_I) \left(r^2+\frac{\ell^2}{2} \left(1- \frac{q_I+\overline{q}_I}{q_I-\overline{q}_I}\cos 2\theta \right) \right)}{r^2(r^2+\ell^2)+\frac{\ell^4 (q_I+\overline{q}_I)^2}{4(q_I-\overline{q}_I)^2} \sin^2 2\theta}\,, \qquad I=1,5,\nn\\
 Z_p &\= 1+2\frac{M_p(2r^2+\ell^2+M_p)-Q_p\left(Q_p+\sqrt{\ell^4-M_p^2+Q_p^2}\cos2\theta \right)}{4r^2(r^2+\ell^2)+(\ell^4-M_p^2+Q_p^2)\sin^22\theta},\label{eq:FieldsBS} \\
 A_1 &\= -\frac{r^2}{2(q_1+\overline{q}_1)}-\frac{\ell^2}{4(q_1-\overline{q}_1)} \left( \cos 2\theta - \frac{2\ell^2 q_1 \overline{q}_1}{(q_1^2 -\overline{q}_1^2)\left( r^2+\frac{\ell^2}{2} \left(1- \frac{q_1+\overline{q}_1}{q_1-\overline{q}_1}\cos 2\theta \right)\right)} \right), \nn \\
  H_5 &=  \frac{q_5+\overline{q}_5}{2}\left( \cos 2\theta  + \frac{\ell^2 \,Z_5}{4(q_5-\overline{q}_5)}\,\sin^2 2 \theta \right)\,, \nn \\
 A_p &\= -\frac{2Q_p(2r^2+\ell^2)-2M_p \sqrt{\ell^4-M_p^2+Q_p^2}\,\cos 2\theta}{(2r^2+\ell^2+M_p)^2-\left( Q_p+\sqrt{\ell^4-M_p^2+Q_p^2}\cos2\theta \right)^2},\nn
\end{align}
and the four-dimensional base is
\begin{align}
    ds_4^2 &= \frac{N_k\left(r^2+\frac{\ell^2}{2}\right)}{2} \left[G \left( \frac{dr^2}{r^2+\ell^2}+d\theta^2\right)+\frac{\sin^2 2\theta}{4}\,d\phi^2+ \frac{r^2(r^2+\ell^2)}{4\left(r^2+\frac{\ell^2}{2}\right)^2}\left(\frac{2d\psi}{N_k}+ \cos 2\theta \,d\phi \right)^2\right] \label{eq:BaseMetric}
\end{align}
with
\begin{align}
    G& \=  \left(  1- \frac{(M_p^2-Q_p)^2 \,\sin^2 2\theta}{4(r^2+\ell^2 \cos^2\theta)(r^2+\ell^2 \sin^2\theta)}\right)\prod_{I=1,5} \left(  1+ \frac{\ell^4q_I \overline{q}_I \,\sin^2 2\theta}{(q_I-\overline{q}_I)^2(r^2+\ell^2 \cos^2\theta)(r^2+\ell^2 \sin^2\theta)}\right).
\end{align}

\begin{figure}[t]
\centering
    \begin{tikzpicture}
\def\deb{-10} 
\def\inter{0.7} 
\def\ha{2.8} 
\def\zaxisline{4.5} 
\def\rodsize{1.7} 
\def\numrod{1.7} 
\def\fin{\deb+1+2*\rodsize+\numrod*\rodsize} 


\draw (\deb+0.5+\rodsize+0.5*\numrod*\rodsize,\ha+2-\inter) node{{{\it Two Strominger-Vafa Black Holes on a Taub-Bolt}}}; 

\draw[draw=black] (\deb+0.5+\rodsize-0.3+0.3,\ha+0.5*\inter) rectangle (\deb+0.5+\rodsize-0.3-0.3,\ha-\zaxisline*\inter-0.3) ;
\draw[gray] (\deb+0.5+\rodsize-0.3,\ha+0.5*\inter+0.2) node{{\tiny BPS SV BH $(n_1,n_5,n_p)$}};

\draw[draw=black] (\deb+0.5+3*\rodsize-0.3+0.3,\ha+0.5*\inter) rectangle (\deb+0.5+3*\rodsize-0.3-0.3,\ha-\zaxisline*\inter-0.3) ;
\draw[gray] (\deb+0.5+3*\rodsize-0.3,\ha+0.5*\inter+0.2) node{{\tiny anti-BPS SV BH $(\overline{n}_1,\overline{n}_5,\overline{n}_p)$}};

\draw [decorate, 
    decoration = {brace,
        raise=5pt,
        amplitude=5pt},line width=0.2mm,gray] (\deb-1,\ha-1.5*\inter+0.05) --  (\deb-1,\ha+0.5*\inter-0.05);
        \draw [decorate, 
    decoration = {brace,
        raise=5pt,
        amplitude=5pt},line width=0.2mm,gray] (\deb-1,\ha-2.5*\inter+0.05) --  (\deb-1,\ha-1.5*\inter-0.05);
                \draw [decorate, 
    decoration = {brace,
        raise=5pt,
        amplitude=5pt},line width=0.2mm,gray] (\deb-1,\ha-3.5*\inter+0.05) --  (\deb-1,\ha-2.5*\inter-0.05);
        
\draw[gray] (\deb-2,\ha-0.5*\inter) node{S$^3$};
\draw[gray] (\deb-2,\ha-2*\inter) node{{\scriptsize time}};
\draw[gray] (\deb-2.2,\ha-3*\inter) node{{\scriptsize S$^1$ in AdS$_3$}};


\draw[black,thin] (\deb+1,\ha) -- (\fin-1,\ha);
\draw[black,thin] (\deb,\ha-\inter) -- (\fin,\ha-\inter);
\draw[black,thin] (\deb,\ha-2*\inter) -- (\fin,\ha-2*\inter);
\draw[black,thin] (\deb,\ha-3*\inter) -- (\fin,\ha-3*\inter);

\draw[black,->,line width=0.3mm] (\deb-0.4,\ha-\zaxisline*\inter) -- (\fin+0.2,\ha-\zaxisline*\inter);

\draw (\deb-0.8,\ha) node{$\phi$};
\draw (\deb-0.8,\ha-\inter) node{$\psi$};
\draw (\deb-0.8,\ha-2*\inter) node{$t$};
\draw (\deb-0.8,\ha-3*\inter) node{$y$};

\draw (\fin+0.2,\ha-\zaxisline*\inter-0.3) node{$z$};


\draw[black, dotted, line width=1mm] (\deb,\ha) -- (\deb+0.5,\ha);
\draw[black,line width=1mm] (\deb+0.5,\ha) -- (\deb+0.5+\rodsize-0.36,\ha);
\draw[black,line width=1mm] (\fin-0.44-\rodsize+0.2,\ha) -- (\fin-0.58,\ha);
\draw[black, dotted,line width=1mm] (\fin-0.5,\ha) -- (\fin,\ha);


\draw[amazon,line width=1mm] (\deb+0.5+\rodsize-0.24,\ha-\inter) -- (\deb+0.44+3*\rodsize-0.3,\ha-\inter);

\draw[black,line width=1mm] (\deb+0.5+\rodsize-0.3,\ha-2*\inter) circle[radius=2pt];
\draw[black,line width=1mm] (\deb+0.5+3*\rodsize-0.3,\ha-2*\inter) circle[radius=2pt];


\draw[amazon,line width=1mm,opacity=0.25] (\deb+0.5+\rodsize-0.27,\ha-\zaxisline*\inter) -- (\deb+0.5+3*\rodsize-0.32,\ha-\zaxisline*\inter);
\draw[black,line width=1mm,opacity=0.5] (\deb+0.5+\rodsize-0.3,\ha-\zaxisline*\inter) circle[radius=2pt];
\draw[black,line width=1mm,opacity=0.5] (\deb+0.5+3*\rodsize-0.3,\ha-\zaxisline*\inter) circle[radius=2pt];


\draw[gray,dotted,line width=0.2mm] (\deb+0.5+\rodsize-0.3,\ha) -- (\deb+0.5+\rodsize-0.3,\ha-\zaxisline*\inter);
\draw[gray,dotted,line width=0.2mm] (\deb+0.5+3*\rodsize-0.3,\ha) -- (\deb+0.5+3*\rodsize-0.3,\ha-\zaxisline*\inter);

\draw[line width=0.3mm] (\deb+0.5+\rodsize-0.3,\ha-\zaxisline*\inter+0.1) -- (\deb+0.5+\rodsize-0.3,\ha-\zaxisline*\inter-0.1);
\draw[line width=0.3mm] (\deb+0.5+3*\rodsize-0.3,\ha-\zaxisline*\inter+0.1) -- (\deb+0.5+3*\rodsize-0.3,\ha-\zaxisline*\inter-0.1);

\draw (\deb+0.5+1*\rodsize-0.3,\ha-\zaxisline*\inter-0.6) node{{\small $-\frac{\ell^2}{8}$}};
\draw (\deb+0.5+3*\rodsize-0.3,\ha-\zaxisline*\inter-0.6) node{{\small $\frac{\ell^2}{8}$}};

\draw[gray] (\deb+0.5+2*\rodsize-0.3,\ha-\zaxisline*\inter-0.3) node{{\tiny Taub-bolt along $\psi$}};
\draw[gray] (\deb+0.5+2*\rodsize-0.3,\ha-\zaxisline*\inter-0.6) node{{\tiny ($N_k$ KKm charge)}};

\node[anchor=south,inner sep=-1cm] at (\fin+2.6,0) {\includegraphics[width=.2\textwidth]{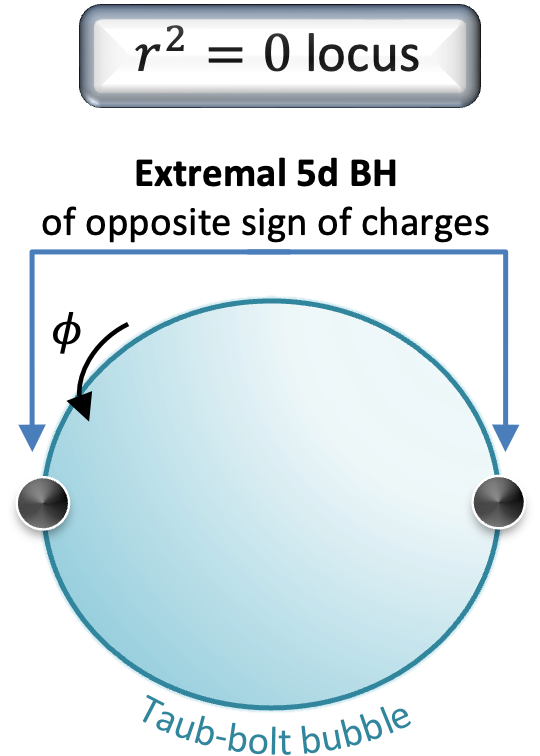}};

\draw (\fin+0.5,\ha-2.5*\inter) node{{\LARGE $\Rightarrow$}};

\end{tikzpicture}
\caption{Spacetime structure of the bound state of two Strominger-Vafa black holes in AdS$_3\times$S$^3/\mathbb{Z}_{N_k}\times$T$^4$, one BPS and one anti-BPS. The black holes are separated by a Taub-bolt of KKm charge $N_k$. The thick lines indicate which direction degenerates on the symmetry axis, while the dots are the black hole loci. The $r^2=0$ locus corresponds to the segment between $-\ell^2/8$ to $\ell^2/8$ on the symmetry axis.}
\label{fig:D1D5PBS}
\end{figure}
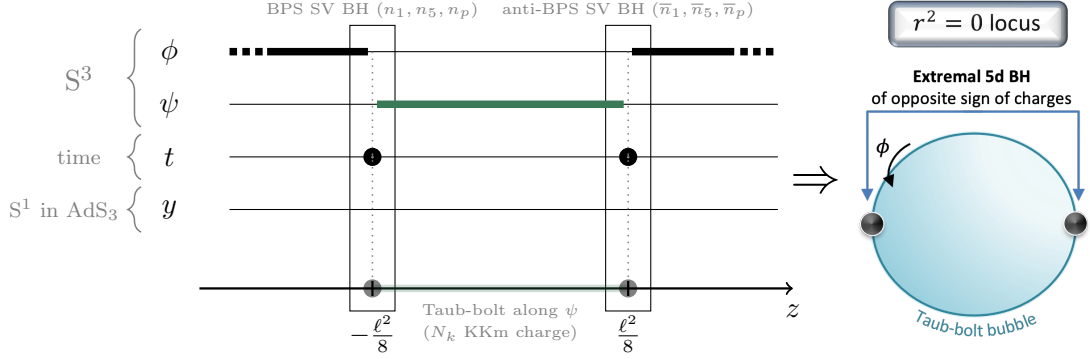 

The spacetime structure is depicted in Fig.~\ref{fig:D1D5PBS}, and the solutions have 8 parameters: 
\begin{itemize}
\item[•] \underline{Black hole brane charges:} $(q_1,q_5)$ and $(\overline{q}_1,\overline{q}_5)$ are the brane charges at the first center and the antibrane charges at the second, respectively,
\begin{equation}
q_I \,>\, 0\,,\qquad \overline{q}_I \,\leq \, 0\,,\qquad q_I + \overline{q}_I \,>\, 0\,.
\end{equation}
The last condition ensures positive net charges $Q_I$.
\item[•] \underline{Separation between the black holes and bubble size:} $\ell^2$ is associated with the distance between the two black holes and is also related to the size of the Taub-bolt bubble. It is not the proper distance in the full ten-dimensional spacetime, but rather the coordinate distance measured in the two-dimensional $(r,\theta)$ base. 
\item[•] \underline{KKm charge at the bubble:} $N_k$ is the KKm charge leading to the quotient S$^3/\mathbb{Z}_{N_k}$.
\item[•] \underline{Momentum charge and energy:} $Q_p$ and $M_p$ are the net momentum charge and the energy associated with the momenta-antimomenta, respectively.  The solution is well-defined when
\begin{equation}
M_p \geq Q_p \,,\qquad M_p^2-Q_p^2\leq \ell^4\,. \label{eq:PmomentaParam}
\end{equation}
The first inequality corresponds to the extremality bound for the bound state: its saturation indicates the absence of antimomenta.  The second inequality bounds the energy according to the separation between the black holes: for higher energy,  the black holes are within the ``Schwarzschild radius'' of the whole bound state and the solution is invalid.
\end{itemize}

The solutions are manifestly regular for $r > 0$ and $0 < \theta < \pi/2$, where all fields remain finite and nonzero. There are five coordinate degeneracies, which we will analyze in detail in an upcoming section. In brief: at $r = 0$ and $0 < \theta < \pi/2$, the $\psi$ circle degenerates, as at the origin of a Taub-bolt space. At $\theta = 0$ or $\theta = \pi/2$ for $r > 0$, a linear combination of $\phi$ and $\psi$ shrinks, corresponding to the polar degeneracy of the $\text{S}^3$ along its symmetry axes. Finally, at the centers $r = 0$ and $\theta = 0$ or $\pi/2$, the functions $Z_1$, $Z_5$, and $Z_p$ diverge, indicating the presence of two extremal D1-D5-P black holes in type IIB supergravity. Figure~\ref{fig:D1D5PBS} illustrates these various loci and the topology along the symmetry axis.

Note that we retrieve the extremal BTZ black hole ($a^2=0$) reviewed in Section \ref{sec:BTZ} by taking $\bar{q}_I=0$ (no antibranes),  $\ell=0$ (no bolt) and $M_p=Q_p$ (no antimomenta).

\subsection{The solution in pole-centered coordinates}
\label{sec:PoleCoor}

The coordinate system used in the previous section is centered on the bolt rather than its poles where the black holes lie. We can also express the solution in pole-centered coordinate systems that make the two-center structure of the solution manifest. In the $(r,\theta)$ coordinates, the distances to these centers are
\begin{equation}
    r_S \equi \frac{r^2 + \ell^2 \cos^2 \theta}{4}\,,\qquad r_N \equi \frac{r^2 + \ell^2 \sin^2 \theta}{4}\,,
\end{equation}
where ``$S$'' and ``$N$'' denote the South and North poles, respectively. For each pole, we define a polar angle satisfying
\begin{equation}
    r_S \cos^2 \frac{\theta_S}{2} \equi \frac{r^2}{4} \sin^2 \theta \,,  \qquad r_N \sin^2 \frac{\theta_N}{2} \equi \frac{r^2}{4} \cos^2 \theta \,.
\end{equation}
Thus, the two coordinate patches $(r_S,\theta_S)$ and $(r_N,\theta_N)$ are related by
\begin{equation}
    r_N \sin \theta_N \= r_S \sin \theta_S \,,\qquad r_N \cos \theta_N \= r_S \cos \theta_S + \frac{\ell^2}{4},
\end{equation}
which corresponds to the standard transformation between two spherical coordinate systems with a shift of $\ell^2/4$ along the $z$-axis.

In these coordinates, the metric and fields are given by \eqref{eq:MetBS} with
\begin{align}
    Z_I &= \frac{q_I+\overline{q}_I}{2(q_I-\overline{q}_I) K_I} \left( \frac{q_I}{r_S}-\frac{\overline{q}_I}{r_N}\right),\quad Z_p = \frac{\left(r_S+r_N+\frac{M_p}{4} \right)^2-\left(\frac{r_S-r_N}{\ell^2}\sqrt{\ell^4-M_p^2+Q_p^2}+\frac{Q_p}{4} \right)^2}{4 r_S r_N \,K_p},\nn\\
    A_1&= \frac{2}{q_1^2-\overline{q}_1^2} \left( \overline{q}_1 r_N -q_1 r_S+ \frac{\ell^4 q_1 \overline{q}_1}{16 (q_1 r_N- \overline{q}_1 r_S)} \right),\quad H_5 = \frac{q_5+\overline{q}_5}{2(q_5-\overline{q}_5) K_5} \left(\overline{q}_5 \cos \theta_N -q_5 \cos \theta_S\right),\nn\\
    A_p &= \frac{1}{Z_p}-1+ \frac{M_p-Q_p}{2\left(r_S+r_N-\frac{r_S-r_N}{\ell^2}\sqrt{\ell^4-M_p^2+Q_p^2}+\frac{M_p-Q_p}{4} \right)}, \label{eq:FieldsPoleCoor}
\end{align}
where we have defined 
\begin{equation}
    K_I \equi 1- \frac{q_I \overline{q}_I}{(q_I-\overline{q}_I)^2} \,\frac{(r_S-r_N)^2-\frac{\ell^4}{16}}{r_S r_N}\,,\qquad K_p \equi 1+ \frac{M_p^2-Q_p^2}{4\ell^4} \,\frac{(r_S-r_N)^2-\frac{\ell^4}{16}}{r_S r_N}.
    \label{eq:Kdef}
\end{equation}
Additionally, the four-dimensional base metric takes a particularly simple form when expressed as a $\phi$-fiber over a three-dimensional space, rather than a $\psi$-fiber,
\begin{equation}
    ds_4^2 \= \frac{1}{V} \left( d\phi + H \,d\psi \right)^2 + V \left[\frac{N_k^2 K_1 K_5 K_p}{4} \left(dr_P^2+{r_P}^2 d\theta_P^2 \right) + {r_P}^2 \sin^2 \theta_P\,d\psi^2\right]\,,\quad P=S\text{ or }N,
    \label{eq:BasePoleCoor}
\end{equation}
where $V$ and $H$ take the standard BPS two-center form,
\begin{equation}
    V \= \frac{1}{N_k}\left(\frac{1}{r_S}+\frac{1}{r_N}\right)\,,\qquad H \= -\frac{1}{N_k} \left(\cos \theta_S + \cos \theta_N \right)\,.
    \label{eq:HarmonicFunc}
\end{equation}
This is somewhat surprising since the base space is not hyper-Kähler in general.
Only when $N_k^2 K_1 K_5 K_p=4$ --- requiring $\overline{q}_I=0$ (no antibranes), $M_p=Q_p$ (no antimomentum), and $N_k=2$ (KKm charge equal to two) --- does the base space reduce to a hyper-Kähler Gibbons-Hawking metric, which is typical of BPS solutions in type IIB supergravity on T$^4$ with two isometries \cite{Gutowski:2003rg}. For more general configurations, the term $K_1 K_5 K_p$ accounts for the brane/antibrane and momentum/antimomentum interactions that break the hyper-Kähler structure and supersymmetry.

In the pole-centered coordinates, the bolt locus is less explicit. It lies at $\theta_P=0$ or $\pi$, where $H=0$, corresponding to the segment connecting the two centers, as expected. The advantage of this coordinate choice is that it highlights the two-center structure of the solution, where the only sources are point-like sources at $r_S=0$ and $r_N=0$. A key feature is that the roles of the $\psi$- and $\phi$-fibers is exchanged in the two forms of the solution: in bolt-centered coordinates, the $\psi$-circle is fibered over an intricate $(r,\theta,\phi)$ base with a bolt source, while in pole-centered coordinates, the $\phi$-circle is fibered over an almost-flat $(r_P,\theta_P,\psi)$ base with two NUT centers.

From the expression of the fields in \eqref{eq:HarmonicFunc}, it might seem that the KKm charges at the centers are $N_k^{-1}$. However, this would only be the case if there were no additional factor in front of the $d\theta_P^2$ component. As we will see shortly, regularity at the bolt (specifically, the condition $N_k^2 K_1 K_5 K_p\big|_{\text{bolt}} = 4$) ensures that the $r_P = 0$ slice of the four-dimensional base is topologically a smooth $\IR^4$, without any orbifold singularity, for arbitrary $N_k$. The absence of an orbifold structure implies that the KKm charge at each pole is 1. Consequently, all regular solutions, for arbitrary $N_k$ are sourced by two unit-charge NUT centers with fiber along the $\phi$ direction.

\subsection{Energy and conserved charges}

At large $r$,  the solution asymptotes to AdS$_3\times$S$^3/\mathbb{Z}_{N_k}\times$T$^4$ as given in \eqref{eq:AdS3Asymp}.  Following the generic procedure reviewed in the Appendix \ref{app:EnergyAdS}, we find that the energy and net quantized charges are given by
\begin{equation}
E=\frac{R_y^2 V_4}{2 g_s^2 l_s^8}\,M_p\,,\qquad N_1 = \frac{V_4}{g_s l_s^6}\,(q_1+\bar{q}_1)\,,\qquad N_5 = \frac{1}{g_s l_s^2}(q_5+\bar{q}_5)\,,\qquad N_p = \frac{R_y^2 V_4}{2 g_s^2 l_s^8}\,Q_p\,.
\label{eq:NetCharges}
\end{equation}
Thus, the momentum components of the geometry have generated energy $M_p$ and momentum charge $Q_p$ measured at the boundary. Since $M_p \geq Q_p$, the solution lies within the same energy range $E \geq N_p$ as the BTZ black hole in the decoupling limit of the nonextremal four-charge black hole.

Like in the black hole, only the momenta and antimomenta---and not the branes and antibranes--- contribute to the energy above the BPS bound. However, the presence of antibranes necessarily breaks supersymmetry---these bound states without antimomenta are extremal, meaning $E=N_p$, but not BPS.
The nature of these states will be explored in detail in section~\ref{sec:ZeroEnBS}.

\subsection{Regularity and internal structure}

In this section, we analyze the coordinate degeneracies and derive the conditions required for regular geometries. In bolt-centered coordinates, there are five such degeneracies: along the $z$-axis above and below the bolt, on the bolt itself, and at its poles, where the black holes are located. In pole-centered coordinates, these translate into four coordinate singularities: along the $z$-axis, at $\theta_P = 0$ and $\theta_P = \pi$, and at the extremal centers where $r_P = 0$.

\subsubsection{On the symmetry axis outside the bolt}
\label{sec:RegAxisOutside}

The symmetry axis above and below the bolt corresponds to the $\theta = 0$ and $\theta = \pi$ slices at fixed $r > 0$ in the bolt-centered coordinates. From the base metric \eqref{eq:BaseMetric}, the $\phi$ coordinate degenerates along these loci at fixed $\psi \pm \frac{N_k}{2} \phi$. Introducing the angular coordinates $\varphi_1$ and $\varphi_2$ as defined in \eqref{eq:DefHyperspher}, one finds that at $\theta = 0$, the $\varphi_1$ circle degenerates at fixed $\varphi_2$, such as $$d\theta^2 + \sin^2 \theta\, d\varphi_1^2 + \# \,d\varphi_2^2,$$ where $\#$ denotes an irrelevant finite coefficient. At $\theta = \pi$, the roles of $\varphi_1$ and $\varphi_2$ are interchanged, with $\varphi_2$ shrinking instead.

The periodicities of $(\varphi_1,\varphi_2)$ are given by those of $(\phi,\psi)$ imposed at the boundary \eqref{eq:psi&phiPerio},
\begin{equation}
    (\varphi_1,\varphi_2) = (\varphi_1,\varphi_2) + \frac{1}{N_k} (-2\pi,2\pi),\qquad  (\varphi_1,\varphi_2) = (\varphi_1,\varphi_2) + (0,2\pi).
\end{equation}
Thus, the periodicity of $\varphi_1$ at fixed $\varphi_2$, and reciprocally, is $2\pi$, despite a nontrivial $2\pi/N_k$ identification in the lattice. Therefore, we have an angular direction smoothly degenerating on the symmetry axis outside the bolt without orbifold action as shown in Fig.\ref{fig:D1D5PBS}.

We can perform the same analysis using the pole-centered coordinates. We consider the coordinates centered around the South pole, $P=S$, so that the axis above and below the bolt corresponds to $\theta_S=0$ with $r_S>0$ and $\theta_S=\pi$ with $r_S>\ell^2/4$. For those values, one has $K_1=K_5=K_p=1$ \eqref{eq:Kdef}, and the $(\theta_S,\phi,\psi)$ component of the base \eqref{eq:BasePoleCoor} gives 
\begin{equation}
    d\theta_S^2 + \sin^2 \theta_S\, \frac{4}{N_k^2} \,d\psi^2 + \# \,\left(d\phi \pm \frac{2}{N_k} d\psi\right)^2\,,
\end{equation}
which corresponds to the same spacetime as above when expressed in terms of $\varphi_1$ and $\varphi_2$.

\subsubsection{At the bolt}

In the bolt-centered coordinates, the locus $r=0$ corresponds to the degeneracy of the $\psi$-fiber captured by the line element,  
\begin{equation}
    ds_{r\psi}^2 = dr^2 + \frac{4(q_1 -\bar{q}_1)^2 (q_5 - \bar{q}_5)^2}{N_k^2 (q_1 + \bar{q}_1)^2 (q_5 + \bar{q}_5)^2 \left( 1-\frac{M_p^2-Q_p^2}{\ell^4}\right)}\,r^2 \left(d\psi+\frac{N_k}{2}\cos 2\theta \,d\phi\right)^2.
\end{equation}
Since $\psi$ is $2\pi$ periodic at fixed $\phi$, the locus corresponds to a smooth $\IR^2$ once we impose the regularity condition\footnote{In the pole-centered coordinate, the regularity condition could have been obtained by imposing $\frac{N_k^2 K_1 K_5 K_p}{4}|_{bolt}=1$ in \eqref{eq:BasePoleCoor}, where ``bolt'' means the segment in between both centers (a two-dimensional sphere in the complete space): either $\theta_S=\pi$ and $r_S<\ell^2/4$ or $\theta_N=0$ and $r_N<\ell^2/4$.}
\begin{equation}
N_k (q_1 + \bar{q}_1) (q_5 + \bar{q}_5) \,\sqrt{1-\frac{M_p^2-Q_p^2}{\ell^4}}\=2(q_1 -\bar{q}_1) (q_5 - \bar{q}_5)\,.
\label{eq:RegCond}
\end{equation}
This equation determines the bolt length $\ell^2$ in terms of the charges of the extremal black holes, except in the extremal limit $M_p=Q_p$, which will be treated separately. Since the black holes have charges of opposite signs, they attract and compress the bubble at the bolt to a size where its pressure balances the black hole attraction. A careful analysis of this interaction will be provided in section~\ref{sec:Stability}.

Note that the only non-zero charge at the Taub-bolt is the KKm charge, $N_k$, as indicated by the two-dimensional metric above. Moreover, the regularity condition \eqref{eq:RegCond} requires $N_k > 2$, or else the left-hand side would be smaller than the right-hand side. Consequently, geometries without an orbifolded S$^3$ cannot feature a regular spacetime between the two extremal black holes. 

The KKm charge is therefore the key feature that enables regular bound states of branes and antibranes in AdS$_3$: its magnetic flux supports the bubble that balances the gravitational attraction between the two black holes in the bound state.

\subsubsection{At the black holes}

We now analyze the solution near each pole of the bolt, i.e., at $r_S=0$ and $r_N=0$ in the pole-centered coordinates. At the South pole, the metric and fields become, after imposing the regularity condition \eqref{eq:RegCond},\footnote{We have also gauged away irrelevant constants in the gauge fields and defined the shifted $y$ coordinates $\widetilde{y}=y+\frac{Q_p+\sqrt{\ell^4-M_p^2+Q_p^2}}{\ell^2+M_p}\,t$.}
\begin{align}
    ds^2 \to &\frac{K_p\sqrt{2 N_k K_1 K_5 q_1 q_5}}{8\sqrt{1-\frac{M_p^2-Q_p^2}{\ell^4}}}\Biggl[ \frac{dr_S^2}{r_S^2}-\frac{32\, r_S^2}{q_1 q_5 q_p} dt^2+\frac{2(\ell^4-M_p^2+Q_p^2)q_p}{\ell^4 q_1 q_5 K_p^2} \left(d\widetilde{y}-\frac{4r_S}{q_p} dt \right)^2 + 4d\Omega_{3S}^2\Biggr] \nn\\
    & + \sqrt{\frac{q_1 (q_1+\overline{q}_1)(q_5-\overline{q}_5) K_5}{q_5 (q_5+\overline{q}_5)(q_1-\overline{q}_1) K_1}}\,ds(T^4)^2,\qquad e^\Phi \to \sqrt{\frac{q_1 (q_1+\overline{q}_1)(q_5-\overline{q}_5) K_5}{q_5 (q_5+\overline{q}_5)(q_1-\overline{q}_1) K_1}}\,, \label{eq:BHmetric}\\
    C^{(2)} \to & \frac{2r_S}{q_1} \,dt \wedge dy + \frac{(q_5+\overline{q}_5)(\overline{q}_5- q_5 \cos \theta_S)}{2(q_5-\overline{q}_5) \,K_5} \,d\phi\wedge d\psi, \nn
\end{align}
where $d\Omega^2_{3S}$ is the line element of a S$^3$, $q_p$ is the local momentum charge, and $K_1$, $K_5$ and $K_p$ converge towards
\begin{align}
    d\Omega_{3S}^2 &\= d\left(\frac{\theta_S}{2} \right)^2 + \frac{1}{K_1 K_5 K_p} \left[ \sin^2 \frac{\theta_S}{2}\,d\phi^2 + \cos^2 \frac{\theta_S}{2} \left( d\phi- \frac{2}{N_k}\,d\psi\right)^2 \right]\,,\\
    K_I &\to 1+\frac{4 q_I \overline{q}_I \sin^2 \frac{\theta_S}{2}}{(q_I -\overline{q}_I )^2},\quad K_p \to 1-\frac{(M_p^2-Q_p^2) \sin^2 \frac{\theta_S}{2}}{\ell^4}\,,\quad q_p \equi \frac{1}{2} \left(Q_p+\frac{M_p(\ell^2+M_p)-Q_p^2}{\sqrt{\ell^4-M_p^2+Q_p^2}}  \right). \nn
\end{align}
One can check that the S$^3$, despite being stretched, has no orbifold singularity and leads to a regular S$^3$ when the regularity condition \eqref{eq:RegCond} is satisfied. Thus, the $r_S=0$ locus corresponds to the AdS$_2\times$S$^3\times$S$^1\times$T$^4$ near-horizon geometries of an extremal BPS D1-D5-P black hole, distorted by the $\theta$ dependence in $K_I$ and $K_p$. We will return to the latter feature below.\medskip

The black hole at the North pole has an identical near-horizon geometry but sourced by negative charges $(\overline{q}_1,\overline{q}_5,\overline{q}_p)$. The quantized charges at both black holes are obtained by integrating the flux in the AdS$_2$ throats (see App.\ref{app:EnergyAdS}),
\begin{equation}
\begin{split}
n_1 &\= \frac{V_4 q_1}{g_s l_s^6}\,, \quad  n_5 \= \frac{q_5}{g_s l_s^2}\,,\quad n_p \=  \frac{R_y^2 V_4}{4g_s^2 l_s^8}\,\left(Q_p+\frac{M_p(\ell^2+M_p)-Q_p^2}{\sqrt{\ell^4-M_p^2+Q_p^2}} \right)\,, \\
\overline{n}_1 &\= -\frac{V_4 \overline{q}_1}{g_s l_s^6}\,,  \quad \overline{n}_5 \= -\frac{ \overline{q}_5}{g_s l_s^2}\,,\quad  \overline{n}_p \=  \frac{R_y^2 V_4}{4g_s^2 l_s^8}\,\left(-Q_p+\frac{M_p(\ell^2+M_p)-Q_p^2}{\sqrt{\ell^4-M_p^2+Q_p^2}} \right)\,.
\label{eq:QuantizedChargesLocal}
\end{split}
\end{equation}
Here $\overline{n}$ denotes the number of antibranes and antimomenta, while $n$ corresponds to the number of branes and momenta. The net quantized charges \eqref{eq:NetCharges} are given by the difference between the number of branes and antibranes or momenta and antimomenta,
\begin{equation}
N_1 \= n_1 - \overline{n}_1\,,\qquad N_5 \= n_5 - \overline{n}_5\,,\qquad N_p \= n_p - \overline{n}_p\,.\label{eq:NetChargeQuantized}
\end{equation}
However, in crucial contrast to the BTZ black hole, the energy is not $E=n_p+\overline{n}_p$.

Moreover, we confirm that in the extremal limit $M_p=Q_p$ ($E=N_p$), the quantity $\overline{n}_p$ vanishes. In this case, the black hole at the North pole becomes a two-charge "small black hole" sourced by $\overline{\text{D1}}$-$\overline{\text{D5}}$ antibranes. Its near-horizon geometry is an S$^3\times$T$^4$ fibration over Poincaré-AdS$_3$. \medskip

The near-horizon geometries are AdS$_2$, which might suggest that the local geometry preserves some supersymmetry. However, the S$^3$ is distorted due to the warp factors $K_I$. These deformations are expected in a binary configuration, where each horizon is tidally stretched by the presence of its partner. As a consequence of this warping---even though we have not proven the absence of Killing spinors---supersymmetry is most likely broken in these AdS$_2$ throats.

Nonetheless, despite the intricate fiber structure and likely breaking of local supersymmetry, the area density---and hence the Bekenstein-Hawking entropy---of the local S$^3\times$S$^1\times$T$^4$ is independent of the warp factors $K_I$ and identical to that of isolated extremal black holes. That is, the entropy of each of the two black holes is
\begin{equation}
S_N \= 2\pi \sqrt{\overline{n}_1\overline{n}_5 \overline{n}_p},\qquad S_S \= 2\pi \sqrt{n_1 n_5 n_p}\,,
\end{equation}
and therefore the total entropy of the bound state is
\begin{equation}
S \= 2\pi \left( \sqrt{\overline{n}_1\overline{n}_5 \overline{n}_p} + \sqrt{ n_1 n_5 n_p}\right)\,. \label{eq:EntropyBS}
\end{equation}
When we eliminate the $n_I$ using \eqref{eq:NetChargeQuantized}, this yields equation \eqref{eq:2SVent} in the introduction.

It is notable that the black holes are entirely insensitive to the KKm charge $N_k$: their near-horizon geometries feature a non-orbifolded S$^3$, and their entropies are independent of $N_k$. The KKm flux serves solely to support the bolt that balances the attraction between the two black holes.

\subsection{Spectrum of regular bound states}
\label{sec:Spectrum}

To solve the regularity condition \eqref{eq:RegCond} analytically it is useful to introduce the quantity 
\begin{equation}
X\equi \frac{2}{N_k} \left(1+\frac{2\overline{n}_1}{N_1} \right)\left(1+\frac{2\overline{n}_5}{N_5} \right)\,, \label{eq:XDef}
\end{equation}
which depends solely on the integer numbers of branes, antibranes, and the KKm charge.
We will solve \eqref{eq:RegCond} separately for two distinct types of bound states: the positive-energy bound states, with $E>N_p$, and the zero-energy bound states with $E=N_p$.

\begin{itemize}
\item[•] \underline{Positive-energy states:}
\end{itemize}

We consider configurations with $\overline{n}_p\neq 0$, i.e., $M_p > Q_p$. The regularity condition fixes $\ell^2$ and the bound-state energy \eqref{eq:NetCharges} in terms of the quantized charges as
\begin{equation}
\begin{split}
E \= N_p+ \frac{\left(\sqrt{n_p(1-X)}-\sqrt{\overline{n}_p(1+X)} \right)^2}{X},\qquad \ell^2 \= \sqrt{\frac{E^2-N_p^2}{1-X^2}}\,\,\frac{2g_s^2 l_s^8}{R_y^2 V_4}.
\label{eq:EnergySpectrum}
\end{split}
\end{equation}
These equations show that the bound states are quantized, with discrete energy levels that, for fixed total charges $(N_1,N_5,N_p)$, are parametrized by the numbers of antibranes and antimomenta $(\overline{n}_1,\overline{n}_5,\overline{n}_p)$.

Requiring that $\ell^2$ is real and finite and satisfies \eqref{eq:PmomentaParam} imposes
\begin{equation}
\frac{n_p - \overline{n}_p}{n_p + \overline{n}_p} < X < 1\,,
\end{equation}
which implies that 
\begin{equation}
N_p = n_p - \overline{n}_p \,<\, E \,<\, n_p + \bar{n}_p\,.
\end{equation}
The first inequality here ensures that the bound state exists in the same regime $E-N_p> 0$ as the nonextremal BTZ black hole. The second inequality indicates that the energy remains lower than the sum of the energies of the momenta and antimomenta. This reflects the bound-state nature of our solution, where part of the energy of the individual constituents is converted into binding energy.

Moreover, the condition $X< 1$ constrains the number of antibranes that the bound state can support, reflecting the necessity of a nontrivial KKm charge $N_k$. Specifically, ensuring $X<1$ requires $N_k> 2$, with the number of antibranes bounded above by 
\begin{align}
    \overline{n}_I \leq \frac{N_I(N_k-2)}{4}\,,\quad I=1,5\,.
\end{align}

\begin{figure}[t]
    \centering
    \includegraphics[width=0.9\textwidth]{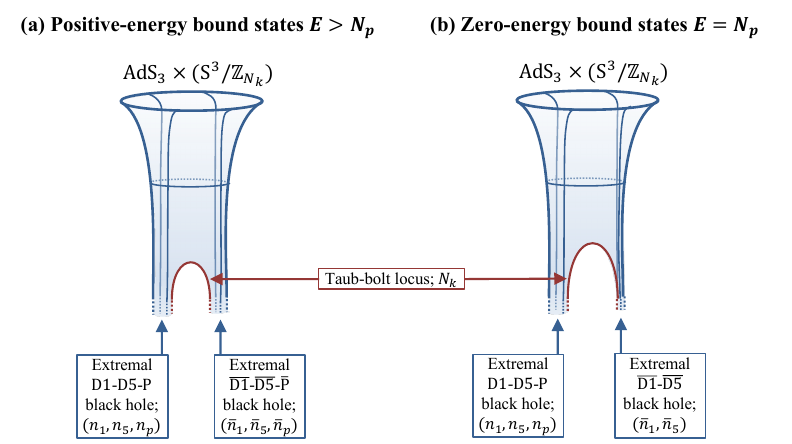}
    \caption{Spacetime of the bound states of two Strominger-Vafa black holes.}
    \label{fig:SpacetimeBS}
\end{figure}

We would like to compare the entropy of the bound state \eqref{eq:EntropyBS} to that of the BTZ black hole \eqref{eq:EntropyBTZ} at fixed energy $E$ and net momentum charge $N_p$. Unfortunately, the exact expression is quite lengthy and does not simplify easily. Using \eqref{eq:EnergySpectrum} inside \eqref{eq:EntropyBS} to replace $\overline{n}_p$ by $E$, and comparing with \eqref{eq:EntropyBTZ} we derive the bound
\begin{equation}
    S<\frac{S_\text{BTZ}}{2}\,.
\end{equation}
However, and more significantly, a numerical scan shows that for large central charge,  $S$ is actually always close to saturating this bound,
\begin{equation}
\frac{S}{S_\text{BTZ}} \,\lesssim\, \frac{1}{2}\,.
\end{equation}
for all allowed values of $E$ and $N_p$.
Thus, the bound state accounts for approximately half the entropy of the four-charge non-extremal black hole in AdS$_3$. As in \cite{Heidmann:2023kry} for the asymptotically flat case, insofar as the entropy of each constituent black hole is microscopically understood, the bound state furnishes a microscopic description of atypical states of the non-extremal four-charge black hole in terms of branes/antibranes and momenta/antimomenta.

\begin{itemize}
\item[•] \underline{Zero-energy bound states:}
\end{itemize}

We now consider the special case where $\overline{n}_p=0$, i.e., $M_p=Q_p$, where the black hole at the North pole degenerates into a ``small black hole" sourced by $\overline{\text{D1}}$-$\overline{\text{D5}}$ antibranes without any antimomentum.

Notably, the regularity condition \eqref{eq:RegCond} does not constrain the bolt size $\ell^2$, implying that the separation between the two black holes remains a flat direction and can be tuned arbitrarily. This suggests that, despite being sourced by branes and antibranes, the two black holes do not exert an attraction strong enough to constrain the Taub-bolt bubble to a specific size\footnote{Recall, however, that the "small black hole" is not fully regular and will undergo corrections that may modify this.}. However, they still interact significantly. If they were entirely noninteracting, the number of antibranes could be arbitrary. Instead, the regularity condition \eqref{eq:RegCond} imposes a strong constraint on the antibrane charges, enforcing $X=1$, which leads to the relation
\begin{equation}
N_k N_1 N_5 \= 2(N_1 +2\overline{n}_1)(N_5 +2\overline{n}_5)\,. \label{eq:RegCond2}
\end{equation}
This arithmetic constraint significantly limits the number of antibranes that the second black hole can carry, yielding only a handful of regular solutions. We will analyze this point in more detail in Section \ref{sec:ZeroEnBS} and show that the number of solutions of this equation is at most the number of divisors of $N_k$.

The most remarkable feature of these states is that their energy saturates the BPS bound,
\begin{equation}
E\= N_p\,.
\end{equation}
It might then seem that the BPS black hole could tunnel into this bound state.  However, this process is suppressed, since the regularity condition \eqref{eq:RegCond2} implies that 
\begin{equation}
    \frac{S}{S_\text{ext}} \, \lesssim\,  \frac{1}{\sqrt{2}}\,,
\end{equation}
and therefore these bound states are always sub-dominant in entropy. 

Although these two-center solutions saturate the BPS bound, they are not supersymmetric and lie outside the class of BPS multi-center solutions discussed in \cite{Bates:2003vx}. This will be made more precise in Section \ref{sec:ZeroEnBS}. As genuinely non-BPS states, they are unprotected and can be lifted under variations of continuous parameters (moduli). Accordingly, they do not contribute to the supersymmetric index of the IIB theory in AdS$_3\times($S$^3/\mathbb{Z}_{N_k})\times$T$^4$.

\subsection{Stability and bound-state interaction}
\label{sec:Stability}

In this section, we analyze the stability properties of the bound states and the characteristics of the interactions between the branes/antibranes and momenta/antimomenta.

For the stability, we follow the method introduced in \cite{Elvang:2006dd} by considering axial perturbations when the distance between the extremal centers is varied out of equilibrium. This involves studying the behavior of the conical singularity between the two centers as their separation changes. The conical singularity appears at $r=0$ in the geometry along the $r,\psi$ directions,
\begin{equation}
    ds^2_{r\psi} \= dr^2 + r^2 \delta_b^2 \left( d\psi- \frac{N_k}{2} \cos 2\theta \,d\phi \right)^2 \,,\qquad \delta_b \equi X \,\left(1-\frac{M_p^2-Q_p^2}{\ell^4}\right)^{-\frac{1}{2}}\,.
\end{equation}
Here $\delta_b$ is the conical defect at the bolt. It represents an excess for $\delta_b>1$, a deficit for $\delta_b<1$, and equilibrium at $\delta_b=1$. This locus corresponds to a cosmic string with tension $\tau$, defined as 
\begin{equation}
    \tau \equi 1-\delta_b\,,
\end{equation}
where we omit an irrelevant factor depending on the Newton constant and compactification circles. In \cite{Elvang:2006dd}, it was argued that the system is unstable under perturbations along its axis, $\ell^2 \to \ell^2 + \delta_b\ell^2$, if the conical singularity is such that
\begin{equation}
   - \frac{d\tau}{d\ell^2} \Bigl|_\text{equi} \,<\, 0. 
\end{equation}
When this condition is satisfied, we can regard the configuration as sitting at the top of a potential.

For the bound states, we aim to vary the black hole separation without varying the local quantized charges \eqref{eq:QuantizedChargesLocal}. This requires considering $M_p$ and $Q_p$ as functions of $\ell^2$ such that $\frac{dn_p}{d\ell^2}|_\text{equi}=\frac{d\overline{n}_p}{d\ell^2}|_\text{equi}=0$. We find
\begin{equation}
    \frac{dQ_p}{d\ell^2}\Bigl|_\text{equi}  = 0,\qquad \frac{dM_p}{d\ell^2}\Bigl|_\text{equi} = \frac{(1-X^2)\,\sqrt{M_p^2-Q_p^2}}{\sqrt{1-X^2} \,M_p+X^2\,\sqrt{M_p^2-Q_p^2}}. 
\end{equation}
Then, the variation of string tension under $\ell^2 \to \ell^2 + d\ell^2$ and at fixed charges leads to:
\begin{equation}
    - \frac{d\tau}{d\ell^2} \bigl|_\text{equi} \= - \frac{1-X^2}{ M_p +\frac{X^2}{\sqrt{1-X^2}}\, \sqrt{M_p^2-Q_p^2}}\,<\, 0.
\end{equation}
Thus, the bound states are generically unstable unless $M_p=Q_p$ and $X=1$, which corresponds to the zero-energy configurations with no antimomentum, resulting in a flat direction where the tension vanishes and is independent of the separation between the centers.
 
 However, for equilibrium configurations, one has $0<X\leq 1$ such that
 \begin{equation}
     \left|\frac{d\tau}{d\ell^2} \bigl|_\text{equi}  \right| \,<\, \frac{1}{M_p} \,<\, \frac{1}{Q_p}
 \end{equation}
This quantity is a measure of the curvature of the potential for perturbations along the axis. Since it is extremely small in the supergravity regime where charges are large compared to the string scale, the maximum of the potential is very broad, and hence the associated instability rate is small.

Furthermore, the cosmic string tension provides significant insight into the dynamics of the branes/antibranes and momenta/antimomenta in the bound states. Specifically, $\tau$ represents the force density acting along the string of length $\ell^2$. Consequently, the total force exerted by the string is given by
\begin{equation}
    F \propto \ell^2 \tau \= \ell^2 \left( 1- X \,\left(1-\frac{M_p^2-Q_p^2}{\ell^4}\right)^{-\frac{1}{2}} \right)\,.
\end{equation}
For large separations, $\ell^2\gg (M_p,Q_p)$, this simplifies to
\begin{equation}
    F \sim \ell^2(1-X) - \frac{2X n_p \overline{n}_p}{\ell^2}\, \frac{g_s^2 l_s^8}{R_y^2 V_4}\,,
\end{equation}
where we have replaced $M_p^2-Q_p^2$ with $n_p \overline{n}_p$, which holds at large $\ell^2$ \eqref{eq:QuantizedChargesLocal}. The force generated by the string precisely compensates for the attraction between the two centers, carrying essential information about their interaction.

The second term, $\sim -1/\ell^2$, represents an attractive \textit{Coulomb} interaction between the momenta and antimomenta, as one could expect. The momenta have energy and charge $(n_p,n_p)$, while the antimomenta carry $(\overline{n}_p,-\overline{n}_p)$, meaning that both the gravitational and electric interactions contribute attractively, scaling as $n_p \overline{n}_p / \ell^2$. The only difference from a standard Coulomb interaction is the modulation by the factor $X$ \eqref{eq:XDef}, which depends on the D1-D5-KKm fluxes. Generally, $X$ can take any value larger than $\frac{2}{N_k}$, is largely independent of $N_1$ and $N_5$, and increases with the number of antibranes $(\overline{n}_1,\overline{n}_5)$. Consequently, the attraction between momenta and antimomenta can be significantly reduced for large KKm charge, and it can be magnified by increasing the number of antibranes in the bound state.

The first term corresponds to a \textit{nonlinear Hookean interaction} (in the sense of growing with distance) between the D1-D5 branes and the $\overline{\text{D1}}$-$\overline{\text{D5}}$ antibranes, combined with the pressure from the KKm bubble. 
This interaction is attractive when $X>1$ and repulsive when $X<1$. Its generic nonvanishing character signals a non-supersymmetric interaction.

Let us consider $N_k\geq 3$ and $X$ as a function of the number of antibranes $(\overline{n}_1,\overline{n}_5)$. When $\overline{n}_I \ll N_I$, we find $X<1$, meaning that the interaction is repulsive despite the presence of branes and antibranes. Moreover, the repulsion increases with distance like a nonlinear Hooke's law with a negative spring constant and a quadratic dependence on the distance. This repulsion arises due to the KKm charge supporting the Taub-bolt bubble, as it disappears when $N_k\leq 2$ ($X>1$). Furthermore, the interaction between branes and antibranes is attractive and follows the same quadratic Hookean dependence with distance. Indeed, when we increase $\overline{n}_I$ by one unit, the force exerted by the string must become more repulsive by an amount $\delta F \sim -\frac{4\ell^2}{N_k N_I}$. 

Both the KKm and brane–antibrane fluxes produce a Hookean interaction that strengthens with bubble size or brane separation, while the KKm flux adds a contraction-resistant pressure that drives them apart and enables bound states.

As the number of antibranes increases, with $\overline{n}_I \gtrsim N_I$, the brane-antibrane attraction begins to dominate over the topological pressure of the Taub-bolt bubble. Critical configurations arise when $X=1$, at which point the “spring constant” vanishes---that is, the brane-antibrane attraction exactly balances the bubble pressure, producing an equilibrium that is independent of separation. These are the zero-energy bound states in equilibrium without antimomenta. When $X>1$, all interactions in the bound state become attractive, precluding the existence of stable equilibrium configurations. \medskip

This analysis not only reinforces previous observations but also provides valuable detail: the Taub-bolt bubble, along with its associated KKm flux, is the essential element that enables equilibrium bound states, in which the attraction between oppositely-charged constituents is counteracted by the bubble’s topological pressure. This equilibrium is unstable except for the zero-energy bound states, though the instability is mild, being suppressed by the system’s large momentum charge. While the fate of the instability is clear when the system is pushed to shrink the bubble---namely, it will collapse into a non-extremal D1-D5-P-KKm black hole---the evolution is more uncertain when the bubble expands; possibly, it will trigger a catastrophic decay of the system.

\section{The impossible states in the CFT gap}
\label{sec:Sec4}

The bound states of branes and antibranes introduced in the previous section define a discrete family of nonperturbative and regular excitations of the supersymmetric four-charge black hole in AdS$_3$.  Fixing the total charges $(N_1,N_5,N_p,N_k)$,  the solutions are parametrized uniquely by the numbers of antibranes and antimomenta $(\overline{n}_1,\overline{n}_5,\overline{n}_p)$ at the second center.  In this section,  we study in more detail the spectrum of states, focusing on the states with the closest energy to the BPS limit $E=N_p$.  We first discuss the positive-energy bound states that have $\overline{n}_p \neq 0$, and afterwards turn to the peculiar zero-energy bound states with $\overline{n}_p = 0$.

\subsection{The positive-energy states}
\label{sec:LowEnMomCharges}

The bound state consists of two extremal black holes, with respective brane/momentum and antibrane/antimomentum numbers $(n_1,n_5,n_p)$ and $(\overline{n}_1,\overline{n}_5,\overline{n}_p)$, sitting at the poles of a bubble of KKm charge $N_k$. 

We will analyze the discrete spectrum of states at fixed total charges, so we replace $n_a = N_a+\overline{n}_a$, focusing on those with the lowest energy. We will show that the spectrum populates the energy range between the BPS bound and the CFT mass gap, 
\begin{equation}
    N_p\,<\, E \,<\, N_p + \Delta E_\text{CFT},\qquad \Delta E_\text{CFT} \equi \frac{3}{c} \= \frac{1}{2N_k N_1 N_5}\,,
\end{equation}
and that these bound states have an entropy of order half the entropy of the BPS four-charge black hole.

Second, we will study the geometry of these specific low-energy states, showing that they correspond to non-perturbative excitations on top of the BPS black hole, corresponding to a strikingly large number of antibranes and antimomenta localizing in the near-horizon region of the BPS hole and that are smoothly prevented from collapsing by the Taub-bolt bubble.

\subsubsection{Energy}
\label{sec:MassGap}

The energy above the BPS bound is
\begin{equation}
\Delta E \= E-N_p \= \frac{\left(\sqrt{(N_p+\overline{n}_p)(1-X)}-\sqrt{\overline{n}_p(1+X)} \right)^2}{X}\,.
\end{equation}

If we consider $X$ fixed,  that is, we fix the number of antibranes $(\overline{n}_1,\overline{n}_5)$, then $\Delta E$ initially decreases as a function of $\overline{n}_p$, reaches a minimum, and then increases again. This behavior is highly counterintuitive, as it suggests that the lowest-energy states do not correspond to configurations with the fewest anti-brane and anti-momenta. More surprisingly, we will show that the lowest-energy states arise when the number of antimomenta is much larger than the total momentum charge, specifically when $\overline{n}_p \propto N_k N_p$.

At first glance, it appears that we could achieve $\Delta E = 0$ by imposing
\begin{equation}
\bar{n}_p \=\frac{N_p}{2} \left(\frac{1}{X}-1 \right)\,.
\label{eq:NantimomMassGap}
\end{equation}
This is a nontrivial arithmetic condition, but solutions may exist. However, this choice also enforces $\ell^2 = 0$ via \eqref{eq:EnergySpectrum}, indicating that such a solution is a degenerate limit where branes and antibranes coincide. It therefore corresponds to a single-center BPS solution and is not a genuine bound state.

To find genuine bound states with minimal energy gaps, we require $\overline{n}_p$ to be slightly above the value in \eqref{eq:NantimomMassGap},
\begin{equation}
\bar{n}_p \gtrsim \frac{N_p}{2} \left(\frac{1}{X}-1 \right)\,,
\end{equation}
so that $\ell^2 > 0$.
In this regime, the energy and separation scale as
\begin{equation}
\Delta E \sim \frac{2X^2\left( \bar{n}_p - \frac{N_p}{2} \left(\frac{1}{X}-1 \right)\right)^2}{(1-X^2) N_p} \,,\qquad \ell^2 \sim \frac{X\left( \bar{n}_p - \frac{N_p}{2} \left(\frac{1}{X}-1 \right)\right)}{1-X^2}\,\frac{4g_s^2 l_s^8}{R_y^2 V_4}\,.
\label{eq:lowestEnEstimate}
\end{equation}
This expansion shows that $\Delta E$ attains its minimum when $X$ is small and $\frac{N_p}{2} \left( \frac{1}{X} - 1 \right)$ lies just below an integer, so that $\bar{n}_p  \= \left\lfloor \frac{N_p}{2} \left(\frac{1}{X}-1 \right)  \right\rfloor +1$ corresponds to the lowest-energy configuration.

\begin{itemize}
\item[•] \underline{Equal net charges:}
\end{itemize}

We begin by considering the illustrative case where all net charges are equal: $N_k = N_1 = N_5 = N_p = N \gg 1$. In this setup, requiring $X$  to be small imposes $\overline{n}_1, \overline{n}_5 \ll N^{3/2}$.

Unfortunately, determining analytically when $\frac{N_p}{2} \left( \frac{1}{X} - 1 \right)$ is closest to an integer is difficult in general. We are only able to obtain an analytic expansion in the regime $\overline{n}_1, \overline{n}_5 \ll N^{1/3}$. In that case, expanding at large $N$ yields
\begin{align}
X \sim \frac{2}{N} \,,
\end{align}
and
\begin{align}
\frac{N_p}{2} \left(\frac{1}{X}-1 \right) &\sim \frac{N^2}{4} - \frac{N(1+\overline{n}_1+\overline{n}_5)}{2} + \overline{n}_1^2+\overline{n}_5^2+\overline{n}_1 \overline{n}_5 - \frac{2(\overline{n}_1+\overline{n}_5)(\overline{n}_1^2+\overline{n}_5^2)}{N}. \label{eq:ExpansionMinimum}
\end{align}
If $N$ is even, the leading three terms in this expansion are all integers. Setting the antimomentum accordingly gives
\begin{equation}
\overline{n}_p \=  \frac{N^2}{4} - \frac{N(1+\overline{n}_1+\overline{n}_5)}{2} + \overline{n}_1^2+\overline{n}_5^2+\overline{n}_1 \overline{n}_5\quad \Rightarrow \quad \bar{n}_p - \frac{N_p}{2} \left(\frac{1}{X}-1 \right) \sim  \frac{2(\overline{n}_1+\overline{n}_5)(\overline{n}_1^2+\overline{n}_5^2)}{N}\,.
\end{equation}
In this case, the energy gap is
\begin{equation}
\Delta E \sim \frac{32(\overline{n}_1+\overline{n}_5)^2(\overline{n}_1^2+\overline{n}_5^2)^2}{N^5} \,.\label{eq:MinEnergyGap}
\end{equation}
Since $\overline{n}_1, \overline{n}_5 \ll N^{1/3}$, this is much smaller than the CFT mass gap, $\Delta E_\text{CFT} = \frac{3}{c} = \frac{1}{2N^3}$. Thus, we find a large number of regular bound states with energy below the CFT gap.

However, if $N$ is odd, the analytic argument fails: the expansion \eqref{eq:ExpansionMinimum} results in an integer plus $\frac{1}{4}$, leading to $\overline{n}_p - \frac{N_p}{2} \left( \frac{1}{X} - 1 \right) \sim \frac{1}{4}$ and an energy gap of $\Delta E \sim \frac{1}{2N^3}$. 

However, these analytic estimates are limited to the regime $\overline{n}_1, \overline{n}_5 \ll N^{1/3}$ and do not capture the true minimum energy. We therefore performed a numerical search across the broader range $N^{1/3} < \overline{n}_1, \overline{n}_5 < N$, with $X$ in the interval $\frac{2}{N} < X < \frac{18}{N}$. We identified many bound states with energy scaling as $\Delta E = \mathcal{O}(N^{-5})$, independent of whether $N$ is even or odd. All such states involve a large number of anti-elements, with
\begin{equation}
    (\overline{n}_1,\overline{n}_5) = \mathcal{O}\left(N\right),\qquad \overline{n}_p = \mathcal{O}(N^2)\,,\qquad \Delta E \= \mathcal{O}\left(\frac{1}{N^5}\right) \= \mathcal{O}\left({\Delta E_\text{CFT}}^\frac{5}{3}\right) .\label{eq:NumSolLowEn}
\end{equation}
These conditions are necessary but not sufficient: generic bound states with $(\overline{n}_1, \overline{n}_5) = \mathcal{O}(N)$ and $\overline{n}_p = \mathcal{O}(N^2)$ can have widely varying energies. In fact, changing $\overline{n}_1$, $\overline{n}_5$, or $\overline{n}_p$ by just one unit can shift the energy by an entire order of magnitude in $N$.

These low-energy bound states defy CFT expectations, exhibiting a large number of excited configurations with energies below the CFT gap. Moreover, they have entropy scaling as $N^2$ \eqref{eq:EntropyBS}, comparable to that of the four-charge BPS black hole. More precisely, we find\footnote{For the solutions giving \eqref{eq:MinEnergyGap}, that is $\overline{n}_1, \overline{n}_5 \ll N^{1/3}$ and $\overline{n}_p\sim N^2/4$, we directly get $S\sim \frac{1}{2} S_\text{ext}$, while for the numerical solutions \eqref{eq:NumSolLowEn}, we checked it numerically.}
\begin{equation}
S \,\sim\, \frac{1}{2} S_\text{ext},
\end{equation}
where $S_\text{ext} = 2\pi N^2$ \eqref{eq:EntropyBTZ}. This indicates that these are not individual states but a large ensemble of bound states below the CFT gap. The factor of $\frac{1}{2}$ implies that their density is still exponentially suppressed compared to the typical BPS states.

A particularly striking feature (explored further in the next section) is that the lowest-energy states are \emph{not} small deformations involving just a few anti-BPS elements. Instead, they contain as many antibranes as net branes, and antimomentum scaling as the square of the net charge. This is exactly the scaling required for the entropy of the $\overline{\text{D1}}$-$\overline{\text{D5}}$-$\overline{\text{P}}$ black hole, $2\pi \sqrt{\overline{n}_1 \overline{n}_5 \overline{n}_p}$, to match the $N^2$ scaling of the four-charge BPS black hole. Despite their low energy, these states are large nonperturbative excitations. The only small parameter in these solutions is the spatial separation between the two centers in the $(r, \theta)$ base: 
\begin{equation}
    \ell^2 \sim \frac{1}{N^2} \frac{g_s^2 l_s^8}{R_y^2 V_4}\,.
\end{equation}
The condition $\ell^2 \ll l_s^2$ is unproblematic, since $\ell^2$ is not a physical ten-dimensional length scale such as a cycle size or a horizon radius. In the next subsection, we show that $\ell^2/l_s^2$ is simply related to $e^{-L/l_s}$, where $L$ denotes the length of the AdS$_2$ throat connecting the boundary to the bound state. We have also verified that the Kretschmann scalar is largely insensitive to $\ell^2$, so that $\ell^2 \ll l_s^2$ does not lead to regions of large string-scale curvature.

\begin{itemize}
\item[•] \underline{Generic charges:}
\end{itemize}

This analysis can be extended to arbitrarily large net charges $(N_k, N_1, N_5, N_p) \gg 1$. The goal is to find integers $(\overline{n}_1, \overline{n}_5)$ such that $ \frac{N_p}{2} \left(\frac{1}{X} - 1 \right)$ is close to an integer, and then take the number of antimomenta to be $\lfloor \frac{N_p}{2} \left(\frac{1}{X} - 1 \right) \rfloor + 1$ to minimize the energy gap. This can only be determined precisely through case-by-case numerical analysis. A comprehensive numerical scan consistently led to the same conclusion: the lowest-energy bound states occur at energy
\begin{equation}
    \Delta E \,\sim\, \mathcal{O}\left(\frac{1}{c^\frac{5}{3}} \right),
\end{equation}
and these always involve a large number of antimomenta, $\overline{n}_p = \cO(N_k N_p)$. \medskip

How many bound states lie between the lowest possible energy and the mass gap, i.e., between $c^{-5/3}$ and $c^{-1}$? Estimating the typical energy spacing is challenging, as it can involve large jumps in $(\overline{n}_1,\overline{n}_5,\overline{n}_p)$. Equation \eqref{eq:MinEnergyGap} shows that the minimal gap is at least of order $c^{-5/3}$. Our numerical scans reveal smaller gaps between successive excited states, indicating that the low-energy bound states densely populate the range between $c^{-5/3}$ and $c^{-1}$.

\subsubsection{Geometry}

While the geometry of the generic solutions given by \eqref{eq:MetBS} is quite intricate, the fields significantly simplify in most of the spacetime for the impossible states. The main simplification arises from the fact that for these geometries  $\ell^2$ is a very small parameter in string units,
\begin{equation}
    \ell^2 \,\ll\, \frac{g_s^2 l_s^8}{R_y^2 V_4}\,.
\end{equation}
This implies that taking $r^2 \gg \ell^2$ corresponds to considering $r^2$ not too close to zero, which is the Taub-bolt locus. By expanding the fields \eqref{eq:FieldsBS} for $r^2 \gg \ell^2 \sim 0$,\footnote{We also have $\sqrt{\ell^4-M_p^2+Q_p^2} \ll \frac{g_s^2 l_s^8}{R_y^2 V_4}$.} the bound states become indistinguishable from
\begin{align}
ds_{10}^2 \= &\frac{1}{2\sqrt{Q_1 Q_5}} \left[-\frac{r^2}{1+\frac{Q_p}{r^2}}\,dt^2 +\left(r^2+Q_p\right) \,\left(dy+ \frac{Q_p}{r^2+Q_p} dt\right)^2 \right] \nn\\
&+ N_k \sqrt{Q_1 Q_5} \left[\frac{dr^2}{r^2} + d\Omega_3^2 \right]+ \sqrt{\frac{Q_1}{Q_5}}\,ds(T^4)^2  \,,\\
C^{(2)} \= & N_k Q_5 \cos^2 \theta \,d\varphi_1 \wedge d\varphi_2 +\frac{r^2}{2Q_1} \,dt\wedge dy \,,\qquad  e^\Phi \= \sqrt{\frac{Q_1}{Q_5}}\,, \nn
\end{align}
with corrections of order $\ell^2/r^2$. This is the four-charge BPS black hole in AdS$_3$, reviewed in Section \ref{sec:BTZ}.
Thus, the bound states are indistinguishable from this black hole up to very close to its horizon at $r^2=0$. More precisely, the solutions start to differ at $r^2\sim \ell^2$, where instead of the BPS black hole horizon we find the structure of our bound state: a smooth Taub-bolt bubble with the two black holes at its poles.

\subsubsection{Bound state deep down the throat} Of particular interest is the fact that the bound-state structure is entirely confined to the near-horizon region, despite being a large non-perturbative deformation of the BPS black hole. One might expect the many momenta and anti-momenta to produce a noticeable dipole at the boundary—but they do not. From the boundary perspective, these details are effectively invisible, ``entrapped'' deep within the high-redshift region, as also observed in asymptotically flat spacetimes \cite{Heidmann:2023thn}.

\begin{figure}[t]
    \centering
    \includegraphics[width=0.6\textwidth]{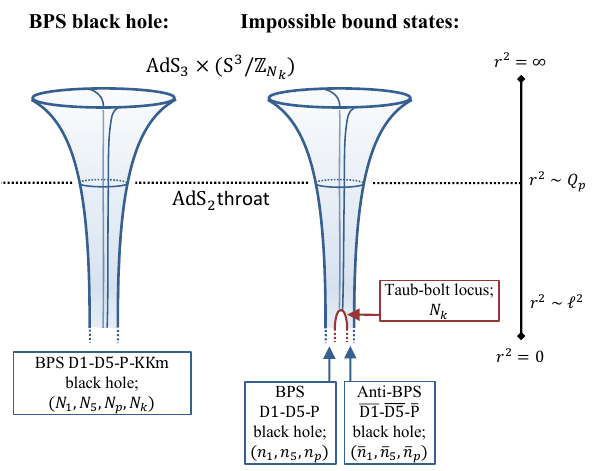}
    \caption{Geometry of the impossible bound states of D1-D5 branes and $\overline{\text{D1}}$-$\overline{\text{D5}}$ antibranes with P momenta and $\overline{P}$ antimomenta, compared to the BPS four-charge black hole.}
    \label{fig:LowEnStates}
\end{figure}

These impossible states therefore reside deep within the near-horizon region of the BPS black hole. They asymptote to AdS$_3$, flowing into an AdS$_2\times$S$^1$ extremal throat as $r^2$ decreases to $r^2 \ll Q_p$. Unlike the BPS black hole, where the horizon lies at $r^2 = 0$, the spacetime here differs at $0 \le r^2 \lesssim \ell^2$, capping off at a finite distance with the novel bound-state structure. Figure \ref{fig:LowEnStates} illustrates these geometries compared to the BPS black hole in AdS$_3$.

This explains how the states can be both highly non-perturbative and of very low energy despite the large number of antimomenta. From the cap perspective, the bound-state energy is large and the deformations are significant. From the boundary perspective, the long AdS$_2$ throat strongly redshifts these effects, resulting in states with tiny energy gaps.

This behavior resembles “scaling” BPS multicenter solutions, which can be made arbitrarily close to the supersymmetric four-charge black hole with a long AdS$_2$ throat by fine-tuning the centers’ positions \cite{Bena:2007qc,Heidmann:2017cxt,Bena:2017fvm}. In those solutions, at least three centers are required, and the quantization of spacetime sets a maximum throat length, ensuring compatibility with CFT expectations \cite{deBoer:2008zn}. In contrast, our impossible bound states are quantized by construction, with charges fine-tuned to produce the longest throat, yielding a mass gap far smaller than expected from the CFT.

\subsubsection{Throat length} The length of the throat can be estimated and compared to that of a near-extremal four-charge black hole with the same energy, given by the metric \eqref{eq:metBTZ}. To do so, we introduce
\begin{equation}
    L_\text{BTZ} \equi \int_{r=0}^{\sqrt{Q_p}} \sqrt{g_{rr}}\,dr\,, \qquad L_\text{BS} \equi \int_{r=0}^{\sqrt{Q_p}} \sqrt{g_{rr}} |_{\theta=\frac{\pi}{4}}\,dr\,,
\end{equation}
where $g_{rr}$ is the radial component of the metric, \eqref{eq:metBTZ} for $L_\text{BTZ}$ and \eqref{eq:MetBS} for $L_\text{BS}$, and we have considered that the AdS$_2$ throat starts at $r^2\approx Q_p$ for near-extremal solutions which is when $g_{yy}$ starts to be constant as a function of $r$. For the BTZ black hole, $r=0$ corresponds to the location of the non-extremal horizon, whereas for the bound state, it corresponds to the Taub-bolt. Note that for the bound state, we evaluate the integral at $\theta = \pi/4$, the equator of the bolt, to avoid the poles where two infinite AdS$_2$ throats are present.

The throat length for the near-extremal BTZ black hole with $g_{rr}=4 R_{\text{AdS}_2}^2 (r^2+a^2)^{-1}$ is analytically derivable and yields the known result in terms of energy and momentum charge,\footnote{We have expressed $a^2$ in terms of $N_p$ and $\Delta E$ using \eqref{eq:a&lasNp&DeltaE}.}
\begin{equation}
    L_\text{BTZ}\=\frac{R_{\text{AdS}_2}}{2} \,\log \frac{N_p}{\Delta E}.
\end{equation}
where $R_{\text{AdS}_2} = \frac{1}{2} (N_k^2 Q_1 Q_5)^{1/4}$ is the AdS$_2$ radius.\footnote{The AdS$_2$ radius can be derived by taking the near-horizon geometry of \eqref{eq:metBTZ} with $\bar{r}^2 = \frac{1}{4}\,Q_1 Q_5 N_k M_p r^4$ which yields $ds_{tr}^2 \sim R_{\text{AdS}_2}\left(\frac{d\bar{r}^2}{\bar{r}^2}-\bar{r}^2 dt^2 \right).$}

To evaluate the throat length of the low-energy bound states, we consider the length difference to that of a BTZ with the same energy,
\begin{equation}
    \Delta L \equi L_\text{BS} - L_\text{BTZ}\,.
\end{equation}
We use approximations that hold in the low-energy regime to bound the difference as
\begin{equation}
    \log \left(\frac{a^2}{\ell^2}\right) - \frac{M_p^2 - Q_p^2}{2\ell^4}<\frac{\Delta L}{R_{\text{AdS}_2}} < \log \left(\frac{a^2}{\ell^2}\right)\,.
\end{equation}
 We now express $\ell^2$ and $a^2$ in terms of the energy and momentum of the solutions using \eqref{eq:TempEntropEnBTZ} and \eqref{eq:EnergySpectrum}, and the fact that $\Delta E \ll N_p$. This yields
\begin{equation}
    a^2 \= \sqrt{2N_p\Delta E} \,\,\frac{4 g_s^2 l_s^8}{R_y^2 V_4}\,,\qquad \ell^2 \sim \sqrt{2N_p \,\Delta E}\,\, \frac{2 g_s^2 l_s^8}{R_y^2 V_4}.\label{eq:a&lasNp&DeltaE}
\end{equation}
This demonstrates that the two lengths are of the same order, differing only slightly in the large-charge limit. More precisely, we find,
\begin{equation}
    -\left(\frac{1}{2}+\log2 \right) < \frac{\Delta L}{R_{\text{AdS}_2}} < -\log 2\,.
\end{equation}
The throat of the bound state is slightly longer than that of the black hole. 

Thus, the low-energy bound states exhibit a very long AdS$_2$ throat, comparable in length to that of a near-extremal black hole with the same energy and charges, scaling as 
\begin{equation}
    L_\text{BS}\approx \frac{R_{\text{AdS}_2}}{2} \,\log \frac{N_p}{\Delta E}.
\end{equation}
However, this is significantly deeper than what the CFT gap would suggest since $\Delta E \ll \Delta E_\text{CFT}$. 

In \cite{Lin:2022rzw}, it was argued, via a Euclidean calculation in $\mathcal{N}=2$ JT gravity, that quantum effects become significant in the AdS$_2$ throat once a certain maximal depth is reached. This maximal depth is given by $L_\text{max}= R_{\text{AdS}_2} \log S_\text{ext}$,\footnote{In \cite{Lin:2022rzw}, the depth is expressed in units of AdS$_2$. Taking the near-horizon geometry of \eqref{eq:metBTZ} with $\bar{r}^2 = \frac{r^4}{4}$ yields $R_{\text{AdS}_2} = R_\text{AdS}/2$, where $R_\text{AdS}$ is the AdS$_3$ radius defined in section~\ref{sec:BTZ}.} which implies 
\begin{equation}
    L_\text{max}=\frac{R_{\text{AdS}_2}}{2} \log \frac{N_p}{\Delta E_\text{CFT}}\,.
\end{equation}
Our low-energy states therefore develop an AdS$_2$ throat whose length exceeds $L_\text{max}$, as expected, since it corresponds to that of a near-extremal BTZ black hole with energy much smaller than the CFT gap. According to \cite{Lin:2022rzw}, this suggests that quantum effects are likely to become important deep within the throat of these geometries. However, the analysis in \cite{Lin:2022rzw} is restricted to $\mathcal{N}=2$ JT gravity and does not capture the full dynamics of our solutions. A proper description requires the $\mathcal{N}=4$ version of the theory, and JT gravity neglects the degrees of freedom associated with the internal space in type IIB supergravity---primarily the S$^3$---which play a central role in our constructions and are therefore absent from the two-dimensional description.

In any case, our analysis makes it clear that whether the impossible states remain within the gap after quantum Schwarzian corrections depends crucially on how these effects affect the throat length.

\subsection{The zero-energy states}
\label{sec:ZeroEnBS}

Now we analyze in detail the zero-energy bound states without antimomenta, which, despite being non-BPS, share the same energy as the BPS four-charge black hole.

\subsubsection{No antibranes: BPS black hole and a NUT center}  To build intuition, we first examine the bound states in the absence of antibranes, $\overline{n}_1=\overline{n}_5=\overline{n}_p=0$, for which the regularity condition \eqref{eq:RegCond} implies $N_k=2$. In Appendix~\ref{App:SUSYLim}, we demonstrate—using the solution in pole-centered coordinates \eqref{eq:FieldsPoleCoor}—that the configuration consists of a BPS D1-D5-P black hole located at the South pole and a NUT center at the North pole. The solution takes the standard form of a BPS multicenter configuration \cite{Bates:2003vx}, where all nontrivial deformations induced by the warp factors $K_I$ vanish, yielding fields expressed in terms of harmonic functions on a three-dimensional flat base \cite{Gutowski:2003rg}.

This shows that zero-energy bound states without antibranes correspond to a rewriting of a BPS two-center solution. Therefore, the existence of these configurations---with the same energy as the BPS four-charge black hole, and unbounded length $\ell^2$---is not surprising.

\subsubsection{Non-supersymmetric zero-energy states} The situation is, however, much more interesting for solutions with antibranes. These are described by the fields \eqref{eq:FieldsPoleCoor} and \eqref{eq:BasePoleCoor} with $M_p=Q_p$. Such a configuration would be BPS if the four-dimensional base were hyper-Kähler and if $Z_1$, $Z_5$, and $Z_p$ were harmonic functions \cite{Gutowski:2003rg}---but they are not.

Although the absence of antimomenta ($M_p = Q_p$) makes the $P$ fields $(Z_p, A_p)$ take the standard BPS single-center form,
\begin{equation}
Z_p = 1 + \frac{Q_p}{4 r_S},, \qquad A_p = \frac{1}{Z_p} - 1,
\end{equation}
and the functions $Z_{1,5}$ for the D1–D5 fields retain a remnant of harmonic structure through the $\frac{1}{r_{S,N}}$ dependence in \eqref{eq:FieldsPoleCoor}, an additional coupling factor between the two centers spoils the purely harmonic form of these functions.
Furthermore, the four-dimensional base \eqref{eq:BasePoleCoor} is not Ricci-flat, owing to the conformal factor $K_1 K_5$ that encodes the brane–antibrane interaction.

These bound states are thus non-BPS excitations of the four-charge BPS black hole. Still, it is striking that they retain a flat direction, with the separation between centers remaining unconstrained---at least at the leading supergravity level.

\subsubsection{Flat direction and scaling} The metric and fields \eqref{eq:BasePoleCoor} and \eqref{eq:FieldsPoleCoor} provide insight into the nature of the flat direction. While the factor $K_1 K_5$ breaks the hyper-Kähler property of the base, it preserves the scale invariance associated with the BPS two-center solution. The base retains a scale invariance set solely by $\ell^2$, which can be rescaled without modifying the angle periodicities or asymptotic data. Specifically, the transformation $r^2 \to \lambda r^2$ modifies the base metric as $ds_4^2(\ell^2) \to \lambda\, ds_4^2(\ell^2/\lambda)$, allowing us to set $\ell^2=1$ without affecting the properties of the base, including regularity at $r=0$. Thus, from the point of view of the base, all values of $\ell^2$ corresponds to the same solution, which explains why the regularity condition does not constrain $\ell^2$.

However, this scale invariance does not extend to the ten-dimensional solution, suggesting that different values of $\ell^2$ correspond to fundamentally different solutions. The ten-dimensional metric remains invariant under the rescaling $r^2 \to \lambda r^2$ only if we also transform $(t,y) \to \lambda^{-1} (t,y)$. Since $y$ is $2\pi R_y$-periodic, with $R_y$ being the AdS$_3$ circle radius at infinity, this transformation alters $R_y \to R_y/\lambda$, indicating that $\ell^2$ cannot be reabsorbed freely. Consequently, solutions with different $\ell^2$ relative to $R_y$ should yield distinct physical configurations.

The existence of a continuum of bound-state solutions along this flat direction remains unexplained. It resembles the moduli space of multicenter BPS solutions, where inter-center distances can be dialed arbitrarily, but now appearing in a non-BPS context and with two centers only. 

\subsubsection{Constraints on antibranes} 
Next, we examine the regularity constraints on the number of antibranes for given conserved charges $N_k, N_1, N_5,$ and $N_p$. While $\ell^2$ is free, these numbers must satisfy the restrictive arithmetic equation \eqref{eq:RegCond2}, which typically admits only a few solutions.

For illustration, consider equal D1–D5 charges, $N_1 = N_5 = N$, so that the antibrane numbers $(\overline{n}_1, \overline{n}_5)$ satisfy
\begin{equation}
N_k N^2 = 2 (N + 2 \overline{n}_1)(N + 2 \overline{n}_5),.
\end{equation}
The existence and number of solutions depend on the parity of $N$ and $N_k$. If both are odd, no solutions exist since 2 does not divide $N_k N^2$. Otherwise, the solutions are discrete and roughly correspond to the number of divisors of $N_k$, with explicit factorizations determining $\overline{n}_I$. For example, if both $N_k$ and $N$ are even, write $N_k = 2k$ and $N = 2p$; each divisor $k_I$ of $k$ gives a solution $\overline{n}_I = k_I(p-1)$. A similar counting applies when $N_k$ is odd and $N$ is even, yielding the number of solutions as the number of divisors of $N_k$ minus one.

Thus, the existence of solutions depends sensitively on the arithmetic properties of the KKm charge and defies simple intuition or resemblance to previously known constructions. First, because the number of divisors is not a smooth or predictable function, the number of solutions can vary abruptly: there may be many for a given $N_k$, none for $N_k+1$, and then many again for $N_k+2$. Second, one might expect the number of antibranes to vary continuously up to a threshold where equilibrium fails due to competing brane/antibrane attraction and bubble pressure. Instead, the allowed numbers of antibranes are sharply quantized and highly discontinuous.

This second point echoes the discussion in Section~\ref{sec:Stability}. There, we saw that both the brane/antibrane forces and the KKm bubble pressure scale nonlinearly with distance, following a Hookean rather than a Coulomb behavior. In typical systems with Coulomb and pressure forces, equilibrium arises below a charge threshold, with the system size adjusting accordingly. Here, by contrast, equilibrium exists only when the Hookean “spring constant’’ vanishes---requiring precise fine-tuning of the charges. If the KKm charge is too large, the pressure drives unlimited expansion; if too small, the brane/antibrane attraction leads to total structural collapse.\medskip

In conclusion, zero-energy bound states of branes and antibranes correspond to peculiar states in AdS$_3\times$S$^3/\mathbb{Z}_{N_k}$. These are non-supersymmetric states with the same energy as the supersymmetric black hole states---however, their non-BPS nature leaves them unprotected, allowing them to be lifted as the string coupling increases. 
The number of allowed antibranes is strongly constrained by the KKm charge, but the size of the bound state remains entirely unrestricted.

\section{Warming up the bound states}
\label{sec:BSNonExt}

Recent studies have revealed that strictly extremal black hole horizons are problematic: they fail to provide reliable semiclassical saddles of supergravity \cite{Iliesiu:2020qvm,Heydeman:2020hhw,Lin:2022rzw}, and even at the classical level, they can develop non-smooth features \cite{Horowitz:2024kcx}. These issues can be regulated by introducing a small but finite temperature above extremality.

In our setting, just as the BTZ black hole admits thermal excitations away from extremality, we wish to investigate whether our bound states can also display thermal behavior. To this end, we construct bound states of near-extremal, rather than strictly extremal, black holes by introducing a small temperature into its constituents.

We will show that the zero-energy bound states of extremal black holes do \emph{not} admit thermal fluctuations. These are isolated $T=0$ solutions which, for their given charges, possess no regular deformations to small but finite temperature. Even infinitesimal heating appears to destabilize them, casting doubt on their relevance as valid and physical states.

By contrast, positive-energy bound states carrying antimomentum---whose energies may nevertheless lie below the CFT mass gap---\emph{do} support thermodynamic perturbations. Small finite-temperature deformations smoothly drive these configurations slightly away from extremality, yielding near-extremal bound states that remain continuous perturbations of their extremal counterparts.

\subsection{The solution}

The solution is built using the same procedure as the previous bound states, via the generalized Ernst formalism in type IIB supergravity (Section~\ref{sec:ErnstTypeIIB}). Extremal centers are replaced by rods: in the D1–D5–P sector, two rods carry opposite charges with the correct asymptotics \eqref{eq:AsympBehav}, locating the non-extremal black holes; in the KKm sector, the rods form non-extremal four-charge black holes connected by a smooth bolt where the $\psi$ circle degenerates.

\begin{figure}[t]
\centering
    \begin{tikzpicture}
\def\deb{-10} 
\def\inter{0.7} 
\def\ha{2.8} 
\def\zaxisline{4.5} 
\def\rodsize{1.7} 
\def\numrod{1.7} 
\def\fin{\deb+1+2*\rodsize+\numrod*\rodsize} 


\draw (\deb+0.5+\rodsize+0.5*\numrod*\rodsize,\ha+2-\inter) node{{{\it Two Nonextremal Black Holes on a Taub-Bolt}}}; 

\draw[draw=black] (\deb+0.5+\rodsize-0.3+0.5,\ha+0.5*\inter) rectangle (\deb+0.5+\rodsize-0.3-0.5,\ha-\zaxisline*\inter-0.3) ;
\draw[gray] (\deb+0.5+\rodsize-0.3,\ha+0.5*\inter+0.2) node{{\tiny Near-Extremal BH}};

\draw[draw=black] (\deb+0.5+3*\rodsize-0.3+0.4,\ha+0.5*\inter) rectangle (\deb+0.5+3*\rodsize-0.3-0.4,\ha-\zaxisline*\inter-0.3) ;
\draw[gray] (\deb+0.5+3*\rodsize-0.3,\ha+0.5*\inter+0.2) node{{\tiny Near-Extremal BH}};

\draw [decorate, 
    decoration = {brace,
        raise=5pt,
        amplitude=5pt},line width=0.2mm,gray] (\deb-1,\ha-1.5*\inter+0.05) --  (\deb-1,\ha+0.5*\inter-0.05);
        \draw [decorate, 
    decoration = {brace,
        raise=5pt,
        amplitude=5pt},line width=0.2mm,gray] (\deb-1,\ha-2.5*\inter+0.05) --  (\deb-1,\ha-1.5*\inter-0.05);
                \draw [decorate, 
    decoration = {brace,
        raise=5pt,
        amplitude=5pt},line width=0.2mm,gray] (\deb-1,\ha-3.5*\inter+0.05) --  (\deb-1,\ha-2.5*\inter-0.05);
        
\draw[gray] (\deb-2,\ha-0.5*\inter) node{S$^3$};
\draw[gray] (\deb-2,\ha-2*\inter) node{{\scriptsize time}};
\draw[gray] (\deb-2.2,\ha-3*\inter) node{{\scriptsize S$^1$ in AdS$_3$}};


\draw[black,thin] (\deb+1,\ha) -- (\fin-1,\ha);
\draw[black,thin] (\deb,\ha-\inter) -- (\fin,\ha-\inter);
\draw[black,thin] (\deb,\ha-2*\inter) -- (\fin,\ha-2*\inter);
\draw[black,thin] (\deb,\ha-3*\inter) -- (\fin,\ha-3*\inter);

\draw[black,->,line width=0.3mm] (\deb-0.4,\ha-\zaxisline*\inter) -- (\fin+0.2,\ha-\zaxisline*\inter);

\draw (\deb-0.8,\ha) node{$\phi$};
\draw (\deb-0.8,\ha-\inter) node{$\psi$};
\draw (\deb-0.8,\ha-2*\inter) node{$t$};
\draw (\deb-0.8,\ha-3*\inter) node{$y$};

\draw (\fin+0.2,\ha-\zaxisline*\inter-0.3) node{$z$};


\draw[black, dotted, line width=1mm] (\deb,\ha) -- (\deb+0.5,\ha);
\draw[black,line width=1mm] (\deb+0.5,\ha) -- (\deb+0.5+\rodsize-0.6,\ha);
\draw[black,line width=1mm] (\deb+0.5+3*\rodsize-0.1,\ha) -- (\fin-0.58,\ha);
\draw[black, dotted,line width=1mm] (\fin-0.5,\ha) -- (\fin,\ha);


\draw[amazon,line width=1mm] (\deb+0.5+\rodsize,\ha-\inter) -- (\deb+0.5+3*\rodsize-0.5,\ha-\inter);

\draw[gray,line width=1mm] (\deb+0.5+\rodsize-0.6,\ha-2*\inter) -- (\deb+0.5+\rodsize,\ha-2*\inter);
\draw[gray,line width=1mm] (\deb+0.5+3*\rodsize-0.5,\ha-2*\inter) -- (\deb+0.5+3*\rodsize-0.1,\ha-2*\inter);


\draw[amazon,line width=1mm,opacity=0.25] (\deb+0.5+\rodsize,\ha-\zaxisline*\inter) -- (\deb+0.5+3*\rodsize-0.5,\ha-\zaxisline*\inter);

\draw[gray,line width=1mm,opacity=0.25] (\deb+0.5+\rodsize-0.6,\ha-\zaxisline*\inter) -- (\deb+0.5+\rodsize,\ha-\zaxisline*\inter);
\draw[gray,line width=1mm,opacity=0.25] (\deb+0.5+3*\rodsize-0.5,\ha-\zaxisline*\inter) -- (\deb+0.5+3*\rodsize-0.1,\ha-\zaxisline*\inter);

\draw[black,<->,line width=0.2mm] (\deb+0.5+\rodsize-0.57,\ha-\zaxisline*\inter+0.3) -- (\deb+0.5+\rodsize-0.03,\ha-\zaxisline*\inter+0.3) node[pos=0.5,above] {{\small $\frac{\sigma_1}{2}$}};

\draw[black,<->,line width=0.2mm] (\deb+0.5+3*\rodsize-0.47,\ha-\zaxisline*\inter+0.3) -- (\deb+0.5+3*\rodsize-0.13,\ha-\zaxisline*\inter+0.3) node[pos=0.5,above] {{\small $\frac{\sigma_2}{2}$}};


\draw[gray,dotted,line width=0.2mm] (\deb+0.5+\rodsize-0.6,\ha) -- (\deb+0.5+\rodsize-0.6,\ha-\zaxisline*\inter);
\draw[gray,dotted,line width=0.2mm] (\deb+0.5+\rodsize,\ha) -- (\deb+0.5+\rodsize,\ha-\zaxisline*\inter);

\draw[gray,dotted,line width=0.2mm] (\deb+0.5+3*\rodsize-0.1,\ha) -- (\deb+0.5+3*\rodsize-0.1,\ha-\zaxisline*\inter);
\draw[gray,dotted,line width=0.2mm] (\deb+0.5+3*\rodsize-0.5,\ha) -- (\deb+0.5+3*\rodsize-0.5,\ha-\zaxisline*\inter);

\draw[line width=0.3mm] (\deb+0.5+\rodsize-0.3,\ha-\zaxisline*\inter+0.1) -- (\deb+0.5+\rodsize-0.3,\ha-\zaxisline*\inter-0.1);
\draw[line width=0.3mm] (\deb+0.5+3*\rodsize-0.3,\ha-\zaxisline*\inter+0.1) -- (\deb+0.5+3*\rodsize-0.3,\ha-\zaxisline*\inter-0.1);

\draw (\deb+0.5+1*\rodsize-0.3,\ha-\zaxisline*\inter-0.6) node{{\small $-\frac{\ell^2}{8}$}};
\draw (\deb+0.5+3*\rodsize-0.3,\ha-\zaxisline*\inter-0.6) node{{\small $\frac{\ell^2}{8}$}};

\draw[gray] (\deb+0.5+2*\rodsize-0.3,\ha-\zaxisline*\inter-0.3) node{{\tiny Taub-bolt along $\psi$}};

\node[anchor=south,inner sep=-1cm] at (\fin+2.6,0) {\includegraphics[width=.2\textwidth]{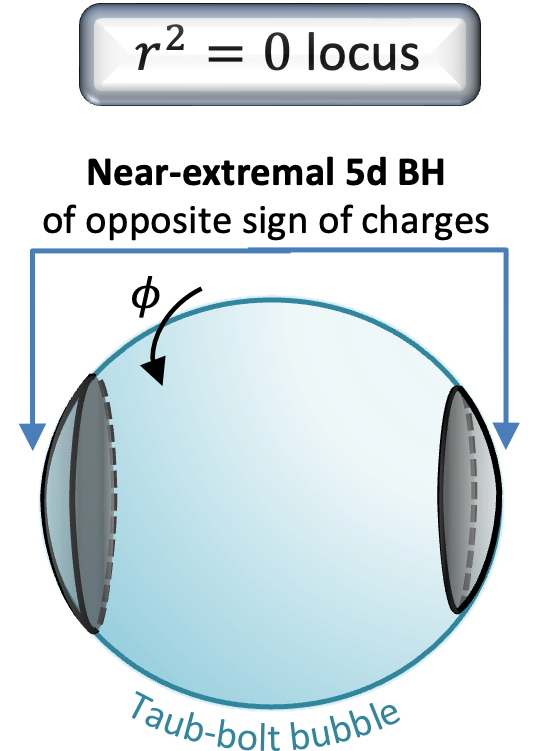}};

\draw (\fin+0.5,\ha-2.5*\inter) node{{\LARGE $\Rightarrow$}};

\end{tikzpicture}
\caption{Spacetime structure of the bound state of two non-extremal black holes in AdS$_3\times$S$^3/\mathbb{Z}_{N_k}\times$T$^4$. The black holes are separated by a Taub-bolt. The thick lines indicate which direction degenerates on the symmetry axis. The $r^2=0$ locus corresponds to the segment between $-(\ell^2+2\sigma_1)/8$ to $(\ell^2+2\sigma_2)/8$ on the symmetry axis.}
\label{fig:D1D5PBSNear-Ex}
\end{figure}
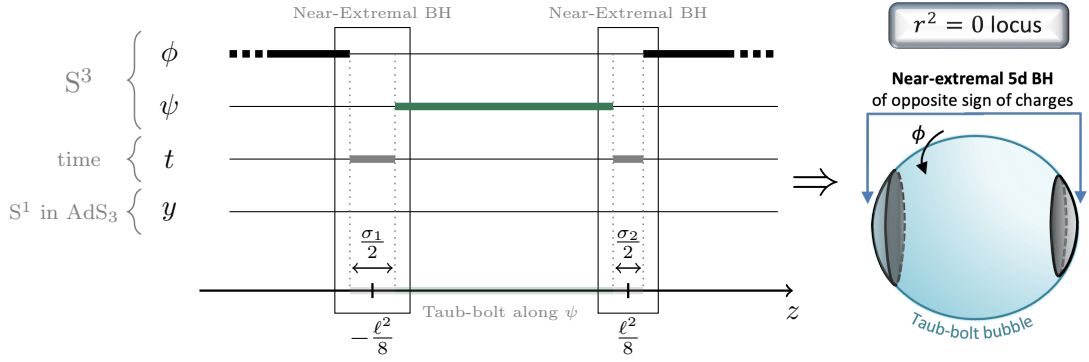

Because the solution is considerably more intricate than the extremal case, details are in Appendices~\ref{App:Ernst} and \ref{App:ErnstTypeIIB}; here we summarize the main results and regularity conditions. The spacetime structure is shown in Fig.~\ref{fig:D1D5PBSNear-Ex}. The new parameters $\sigma_1$ and $\sigma_2$ set the black-hole rod lengths and quantify the non-extremality of the black holes.

The solutions are parametrized by ten parameters: 
\begin{itemize}
    \item[•] \underline{Taub-bolt features:} The segment between the black holes corresponds to a bolt where the $\psi$ circle degenerates (see green segment in Fig.\ref{fig:D1D5PBSNear-Ex}). It is characterized by the bolt length, proportional to $\ell^2 - \sigma_1 - \sigma_2$, and the KKm charge $q^{(b)}_{kkm}$ that it carries. This KKm charge is related to the total quantized KKm charge by
    \begin{equation}
        q^{(b)}_{kkm} \= \frac{\ell^2-\sigma_1-\sigma_2}{\ell^2}\,N_k\,.
    \end{equation}
    \item[•] \underline{Black-hole features:} The D1-D5-P charges of the first and second black holes are $(q_1, q_5, q_p)$ and $(\overline{q}_1, \overline{q}_5, \overline{q}_p)$ respectively. We will assume later on that they have opposite sign: $(q_1, q_5, q_p)>0$ and $(\overline{q}_1, \overline{q}_5, \overline{q}_p)\leq 0$. In addition, each momentum charge contributes an energy $(m_p, \overline{m}_p) \geq 0$, and the energies and charges are constrained by the non-extremality parameters as
    \begin{equation}
        \sigma_1^2 \= m_p^2 - q_p^2 + \frac{2q_p(\overline{m}_p q_p - m_p \overline{q}_p)}{\ell^2+m_p+\overline{m}_p}\,,\qquad \sigma_2^2 \= \overline{m}_p^2 -\overline{q}_p^2 - \frac{2\overline{q}_p(\overline{m}_p q_p - m_p \overline{q}_p)}{\ell^2+m_p+\overline{m}_p}\,.
    \end{equation}
    One can verify the existence of two extremal limits, $\sigma_1 = \sigma_2 = 0$. One limit corresponds to both sources being BPS, $(q_p, \overline{q}_p) = (m_p, \overline{m}_p) \geq 0$, so that the energy equals the charge at each center. The other is the BPS/anti-BPS limit,
    \begin{equation}
        q_p \= m_p \sqrt{\frac{(\ell^2+\overline{m}_p)^2-m_p^2}{(\ell^2-\overline{m}_p)^2-m_p^2}}\,,\qquad \overline{q}_p \= -\overline{m}_p \sqrt{\frac{(\ell^2+m_p)^2-\overline{m}_p^2}{(\ell^2-m_p)^2-\overline{m}_p^2}},\label{eq:ExtremalLimit}
    \end{equation}
    which reproduces the previous bound state with \eqref{eq:PmomentaParam}
    \begin{equation}
        M_p\=m_p+\overline{m}_p\,,\qquad Q_p\=q_p+\overline{q}_p\,.
        \label{eq:MpQpdef}
    \end{equation}

    For clarity, we introduce three constants that capture the interaction between the black holes and appear in several expressions,
    \begin{equation}
        \gamma_I \equi \frac{\ell^2 q_I \overline{q}_I - \sqrt{\ell^4 q_I^2 \overline{q}_I^2+(q_I^2-\overline{q}_I^2)(q_I^2 \sigma_2^2-\overline{q}_I^2 \sigma_1^2)}}{q_I^2 -\overline{q}_I^2}, \qquad \gamma_p \equi \frac{\overline{m}_p q_p - m_p \overline{q}_p}{\ell^2+m_p+\overline{m}_p}
    \end{equation}
    where $I = 1,5$. Note that $\gamma_I = \gamma_p = 0$ when both black holes are mutually BPS D1-D5-P black holes, $\sigma_1 = \sigma_2 = 0$, $(q_I, \overline{q}_I) > 0$, and $(q_p, \overline{q}_p) = (m_p, \overline{m}_p) \geq 0$.
    
    Additionally, the black holes carry nonzero positive KKm charges along the $\psi$ direction given by
    \begin{equation}
        q_{kkm} \= \frac{\sigma_1}{\ell^2} \,N_k\,,\qquad \overline{q}_{kkm} \= \frac{\sigma_2}{\ell^2} \,N_k\,,
    \end{equation}
    and we retrieve that the net KKm charge is $N_k=q_{kkm}+\overline{q}_{kkm}+q^{(b)}_{kkm}$.
\end{itemize}

The solution is asymptotic to AdS$_3 \times$ S$^3/\mathbb{Z}_{N_k} \times$ T$^4$ and terminates in the IR as a bound state of two non-extremal black holes with opposite charges, as shown in Fig.~\ref{fig:D1D5PBSNear-Ex}. The energy and net D1-D5-P charges are
\begin{equation}
E= \frac{R_y^2 V_4\,(m_p+\overline{m}_p)}{2 g_s^2 l_s^8}\,,\quad N_1 \= \frac{V_4\,(q_1+\overline{q}_1)}{g_s l_s^6}\,,\quad N_5 \= \frac{(q_5+\bar{q}_5)}{g_s l_s^2}\,,\quad N_p \= \frac{R_y^2 V_4\,(q_p+\overline{q}_p)}{2 g_s^2 l_s^8}\,.
\end{equation}
As in the previous bound state, the physics is set by regularity along the symmetry axis. Here, however, the two non-extremal black holes have nonzero temperatures, which fix the non-extremality parameters $\sigma_1$ and $\sigma_2$. We construct bound states in thermal equilibrium, $T = T_1 = T_2$, at sufficiently low temperature so that the near-extremal solutions are small deformations of the extremal bound state,
\begin{equation}
\sigma_1, \sigma_2 ,\ll, \ell^2,.
\end{equation}
The geometry is then indistinguishable from the extremal bound state except very close to each black hole, where non-extremality is manifest.

\subsection{Regularity condtions}
\label{sec:RegCondNonEx}

The regularity constraints along the symmetry axis follow from expanding the fields near each segment: a degenerating spacelike circle must locally approach $\mathbb{R}^2$, while a degenerating timelike direction must produce a smooth black-hole horizon at temperature $T$.

\begin{itemize}
    \item[•] \underline{Above and below the sources:}
\end{itemize}

The regularity conditions above and below the sources (the black rods in Fig.~\ref{fig:D1D5PBSNear-Ex}) are the same as for the extremal bound state (Section~\ref{sec:RegAxisOutside}), requiring $N_k$ to be an integer and the $(\psi, \phi)$ periodicities to satisfy \eqref{eq:psi&phiPerio}.

\begin{itemize}
    \item[•] \underline{At the bolt:}
\end{itemize}

The bolt, where the $\psi$ circle degenerates (green rod in Fig.~\ref{fig:D1D5PBSNear-Ex}), yields an origin in $\IR^2$,
\begin{align}
    &ds^2_{\rho\psi} \= d\rho^2 + \rho^2 \delta_b^2 \left( d\psi + \frac{(\sigma_2-\sigma_1)N_k}{2\ell^2}\,d\phi- \frac{N_k(\ell^2-\sigma_1-\sigma_2)}{2\ell^2} \cos 2\theta \,d\phi \right)^2 \,, \label{eq:RegCondNonEx}\\
    &\delta_b^4 \= \frac{16\ell^8\,\,\frac{\ell^4 - (m_p-\overline{m}_p)^2+\left(q_p-\overline{q}_p- 2\gamma_p\right)^2}{\ell^4 - (m_p+\overline{m}_p)^2+(q_p+\overline{q}_p)^2} }{N_k^4 \left(\ell^4-(\sigma_1-\sigma_2)^2 \right)\left(\ell^4-(\sigma_1+\sigma_2)^2 \right)}\prod_{I=1,5} \frac{(q_I-\overline{q}_I)\left(\ell^4 (q_I+\overline{q}_I)-(q_I-\overline{q}_I)(\sigma_1^2-\sigma_2^2+2\ell^2 \gamma_I)\right)}{(q_I+\overline{q}_I)\left(\ell^4 (q_I-\overline{q}_I)-(q_I+\overline{q}_I)(\sigma_1^2-\sigma_2^2-2\ell^2 \gamma_I)\right)}. \nn
\end{align}
This geometry is smooth when $\delta_b=1$, in which case the $\psi$ circle shrinks smoothly at fixed $\phi$ with the periodicity lattice \eqref{eq:psi&phiPerio}. One can show that when the black holes are extremal $\sigma_1=\sigma_2=0$, we find $\delta_b=2/N_k$ when they are both BPS (same sign of charges), and we retrieve the previous regularity condition \eqref{eq:RegCond} when they are BPS/anti-BPS (opposite sign of charges).
\begin{itemize}
    \item[•] \underline{At the black holes:}
\end{itemize}
The black holes lie at the two segments of length $\sigma_1/2$ and $\sigma_2/2$ at either end of the bolt. Their near-horizon regions are similar to that of the non-extremal BTZ black hole in type IIB, reviewed in Section \ref{sec:BTZ}, though the $S^3$ fibration is less trivial.

The induced metric on the $(\rho,t)$ space near each black hole takes the form:
\begin{equation}
    ds^2_{\rho t} \= d\rho^2 - (2\pi T_i)^2 \, \rho^2 \,\frac{dt^2}{R_y^2}\,,\qquad i=1,2,
\end{equation}
where $R_y$ is the asymptotic radius of the $y$-circle in AdS$_3$, and $T_1$, $T_2$ are the temperatures of each black hole,
\begin{equation}
\begin{split}
    T_1 &\= \frac{R_y \left((\ell^2-\sigma_1)^2-\sigma_2^2 \right) \sqrt{N_k}\, \sigma_1}{2\pi \ell \sqrt{\left((\sigma_1+m_p)(\ell^2-\sigma_1+\overline{m}_p)- q_p \overline{q}_p \right) \prod_{I=1,5} \left(\ell^2 q_I - (q_I-\overline{q}_I)(\sigma_1+\gamma_I)\right)}}, \\
    T_2 & \= \frac{R_y \left((\ell^2-\sigma_2)^2-\sigma_1^2 \right) \sqrt{N_k}\, \sigma_2}{2\pi \ell \sqrt{\left((\sigma_2+\overline{m}_p)(\ell^2-\sigma_2+m_p)- q_p \overline{q}_p \right) \prod_{I=1,5} \left(\ell^2 \overline{q}_I - (\overline{q}_I-q_I)(\sigma_2-\gamma_I)\right)}}. 
    \label{eq:TempNonEx} 
\end{split}
\end{equation}
The constraints fix $\sigma_1$ and $\sigma_2$ in terms of the black hole temperatures, which depend on the charges of the entire bound state, reflecting the interacting nature of the solution. Thermal equilibrium requires $T_1 = T_2 = T$, recovering $T_1 = T_2 = 0$ in the extremal limit $\sigma_1 = \sigma_2 = 0$.

Although \eqref{eq:RegCondNonEx} and \eqref{eq:TempNonEx} are not generally solvable analytically, for sufficiently small temperatures any extremal bound state admits a unique smooth non-extremal deformation. Bound states are expected to cease at large $T$, when stronger mutual attraction causes the black holes to merge. Unequal-temperature black holes are also possible. They are not in thermal equilibrium and do not maximize the entropy, so they are less typical than other bound states with the same total energy and charges.

The horizon areas $S_1$ and $S_2$ coincide with those of two noninteracting non-extremal BTZ black holes \eqref{eq:TempEntropEnBTZ}, as in the extremal case, but nontrivially so, since this holds far from extremality and at finite temperature. The bound-state entropy is then
\begin{equation}
     S \= S_1+S_2 \=\left( \frac{\sigma_1}{T_1} + \frac{\sigma_2}{T_2} \right) \frac{R_y^2 V_4}{g_s^2 l_s^8}.
\end{equation}

\subsection{Bound states at small temperature}

At low temperatures, we consider near-extremal configurations where the nonextremality parameters,
\begin{equation}
\sigma_1, \sigma_2 \ll \ell^2\,,
\label{eq:CondSmallDefo}
\end{equation}
are the smallest scales, such that solutions exist for infinitesimal $T$. This demonstrates that extremal bound states admit small thermodynamic perturbations that preserve the overall geometry, affecting only the IR structure, and yield configurations effectively indistinguishable from the extremal case.

First, the regularity condition at the bolt, \eqref{eq:RegCondNonEx}, reduces to the extremal regularity constraint \eqref{eq:RegCond}, up to small corrections,
\begin{equation}
\frac{N_k (q_1 + \bar{q}_1) (q_5 + \bar{q}_5)}{2(q_1 -\bar{q}_1) (q_5 - \bar{q}_5)}  \,\sqrt{1-\frac{M_p^2-Q_p^2}{\ell^4}}\= 1 + \cO\left( \frac{\sigma_i^2}{\ell^4}\right)\,,
\label{eq:RegCondApprox}
\end{equation}
where $M_p$ and $Q_p$ are the momentum energy and charge, as defined in \eqref{eq:MpQpdef}. This confirms that the near-extremal solutions are small, controlled perturbations of the extremal bound states.

Then the expansion of the temperatures \eqref{eq:TempNonEx} depends on the momentum charges at the black holes, $\overline{q}_p,q_p$.

\newpage
\begin{itemize}
    \item \underline{Black holes with $\overline{q}_p,q_p\neq 0$:}

We first consider bound states of near-extremal D1-D5-P and $\overline{\text{D1}}$-$\overline{\text{D5}}$-$\overline{\text{P}}$ black holes. Those can be considered as the non-extremal generalization of the positive-energy bound states of extremal black holes introduced in Section \ref{sec:Spectrum}. Expanding the temperatures in the limit \eqref{eq:CondSmallDefo} yields
\begin{equation}
    \begin{split}
        T_1^2 &\= \frac{N_k (q_1 + \bar{q}_1) (q_5 + \bar{q}_5) \,\sqrt{1-\frac{M_p^2-Q_p^2}{\ell^4}}}{2(q_1 -\bar{q}_1) (q_5 - \bar{q}_5)} \,\frac{R_y^2\,\sigma_1^2}{2\pi^2 q_1 q_5 q_p}\left(1\+ \cO\left( \frac{\sigma_i}{\ell^2}\right)\right)\,,\\
        T_2^2 &\= \frac{N_k (q_1 + \bar{q}_1) (q_5 + \bar{q}_5) \,\sqrt{1-\frac{M_p^2-Q_p^2}{\ell^4}}}{2(q_1 -\bar{q}_1) (q_5 - \bar{q}_5)} \frac{R_y^2\,\sigma_2^2}{2\pi^2 \overline{q}_1 \overline{q}_5 |\overline{q}_p|}\left(1\+ \cO\left( \frac{\sigma_i}{\ell^2}\right)\right)\,,
        \label{eq:TemperatureExpansion}
    \end{split}
\end{equation}
where the momentum and anti-momentum charges, $q_p$ and $\overline{q}_p$ are nonzero and given in terms of $M_p$ and $Q_p$ as \eqref{eq:QuantizedChargesLocal}
\begin{equation}
    q_p \= \frac{1}{2}\left(Q_p+\frac{M_p(\ell^2+M_p)-Q_p^2}{\sqrt{\ell^4-M_p^2+Q_p^2}} \right)\,,\qquad \overline{q}_p \= \frac{1}{2}\left(Q_p-\frac{M_p(\ell^2+M_p)-Q_p^2}{\sqrt{\ell^4-M_p^2+Q_p^2}} \right).
\end{equation}

It is highly nontrivial that all interaction terms in the temperatures \eqref{eq:TempNonEx} factor out in the near-extremal limit, and the overall prefactor becomes exactly 1 when the bolt regularity condition is satisfied. Therefore, the near-extremal temperatures reduce to
\begin{equation}
    T_1 \,\sim\, \frac{R_y \sigma_1}{\pi \sqrt{2 q_1 q_5 q_p}}\,,\qquad T_2 \,\sim\, \frac{R_y \sigma_2}{\pi \sqrt{2 \overline{q}_1 \overline{q}_5 |\overline{q}_p|}},
    \label{eq:TempNearExtr}
\end{equation}
These precisely match the temperatures of two isolated near-extremal BTZ black holes, as in \eqref{eq:TempEntropEnBTZ}. Hence, the small temperature deformations decouple and localize on each black hole, as if the two black holes were non-interacting. This shows that the low-temperature deformations of the bound state of extremal black holes are entirely determined by the low-temperature deformations of each of their black holes within the geometry. 

Then, once the $\sigma_i$ are fixed in terms of the temperature, $\ell^2$ will slightly adapt so that \eqref{eq:RegCondApprox} is satisfied. Since the left-hand side is already $1$ when the black holes are extremal $(\sigma_i=0)$, this will require a small change in $\ell^2$ to satisfy the regularity condition. This demonstrates that the bound states can generically admit low-temperature deformations without significantly modifying the geometries, exactly like the single BTZ black hole.

One can rewrite the expressions for the temperatures \eqref{eq:TempNearExtr} in terms of the quantized charges \eqref{eq:QuantizedChargesLocal}, and we obtain
\begin{equation}
    \sigma_1 \,\sim\, 4\pi \sqrt{n_1 n_5 n_p} \,T_1 \, \frac{g_s^2 l_s^8}{V_4 R_y^2}\,,\qquad \sigma_2 \,\sim\, 4\pi \sqrt{\overline{n}_1 \overline{n}_5 \overline{n}_p} \,T_2 \, \frac{g_s^2 l_s^8}{V_4 R_y^2}.
\end{equation}
Thus, at thermal equilibrium $T_1=T_2=T$, one has
\begin{equation}
    \frac{\sigma_1}{\sigma_2} \sim \sqrt{\frac{n_1 n_5 n_p}{\overline{n}_1 \overline{n}_5 \overline{n}_p}}.
\end{equation}
Since the net charges are positive, one has $n_1 n_5 n_p > \overline{n}_1 \overline{n}_5 \overline{n}_p$, implying that, at thermal equilibrium, the second black hole is smaller than the first one, $\sigma_2 < \sigma_1$. \medskip

Moreover, the regularity condition \eqref{eq:RegCondApprox} requires $\ell^2$ to be given by \eqref{eq:EnergySpectrum} when $\bar{n}_p\neq 0$
\begin{equation}
    \ell^2 \= \sqrt{\frac{E^2-N_p^2}{1-X^2}}\,\,\frac{2g_s^2 l_s^8}{R_y^2 V_4}\,,
\end{equation}
where $X$ is given in \eqref{eq:XDef}. Thus, the assumption that $\sigma_1$ and $\sigma_2$ corresponds to small deformations \eqref{eq:CondSmallDefo} requires the small temperatures to satisfy:
\begin{equation}
    n_1 n_5 n_p\, T^2 \,\ll\, \frac{E^2-N_p^2}{1-X^2}\,.
\end{equation}
Moreover, one has the bound $\frac{E^2-N_p^2}{1-X^2}> 2N_p \Delta E$ which is saturated for the low-energy states $E\sim N_p$ and $X\sim 0$. Thus, the condition is satisfied when the temperature satisfies
\begin{equation}
    T^2 \,\ll\, \frac{2N_p \Delta E}{n_1 n_5 n_p}.
    \label{eq:TemperatureRange}
\end{equation}
This is easily satisfied for generic bound states. However, for the impossible states where $\Delta E< \Delta E_\text{CFT}$, this requires the temperature to be smaller than the minimal temperature allowed by the CFT, which could be expected as those states break CFT expectations. 

For instance, when the quantized charges are equal $N=N_1=N_5=N_p=N_k$, one has $n_1,n_5 \sim N$, $n_p \sim N^2/4$ and $\Delta E \sim N^{-5}$ \eqref{eq:MinEnergyGap}. Thus, the required temperature so that the bound states of near-extremal black holes are small deformations is
\begin{equation}
    T \ll \frac{1}{N^4},
\end{equation}
which is much smaller than the minimal temperature allowed by the CFT mass gap at the boundary $T_\text{min} \sim \frac{1}{c} \sim \frac{1}{N^3}$. If the temperature is larger than this value, then we are guaranteed that $\sigma_1$ and $\sigma_2$ will not be negligible compared to $\ell^2$ so that the thermodynamic backreaction will induce large deformation on the bound state structure and the solution cannot be considered as small perturbation above the bound state of extremal black holes.

\item \underline{Black holes with $\overline{q}_p= 0$:}

We now consider bound states of near-extremal D1-D5-P and $\overline{\text{D1}}$-$\overline{\text{D5}}$ black holes. These can be viewed as non-extremal generalizations of the zero-energy bound states of extremal black holes introduced in Section~\ref{sec:Spectrum}. 

The temperature of the first black hole remains given by \eqref{eq:TemperatureExpansion} with $M_p = Q_p$, while for the second black hole, we find
\begin{equation}
    T_2^2 \= \frac{N_k (q_1 + \bar{q}_1) (q_5 + \bar{q}_5)}{2(q_1 -\bar{q}_1) (q_5 - \bar{q}_5)} \frac{R_y^2\,\sigma_2}{4\pi^2 \overline{q}_1 \overline{q}_5}\left(1\+ \cO\left( \frac{\sigma_i}{\ell^2}\right)\right)\,,
\end{equation}
which yields, assuming the regularity condition is satisfied:
\begin{equation}
    T_2 \sim \frac{R_y \sqrt{\sigma_2}}{2\pi \sqrt{\overline{q}_1\overline{q}_5}}.
\end{equation}
As for the previous bound state, this corresponds to the temperature of an isolated massless BTZ black hole, and the non-extremality, $\sigma_2$, is fixed as if the black holes were isolated.

However, a major difference arises compared to the previous case when considering the regularity condition at the bolt \eqref{eq:RegCondApprox}. The left-hand side of this condition is independent of $\ell^2$, so $\ell^2$ cannot be simply tuned to satisfy it. If we start with values of the charges for which a bound state of extremal black holes exists, we obtain the condition
\begin{equation}
    1 \= 1 + \cO\left( \frac{\sigma_i}{\ell^2}\right),
\end{equation}
which cannot be solved by a small adjustment of $\ell^2$. Therefore, there are no solutions with $\sigma_i \ll \ell^2$ for these charge values, not even at infinitesimal temperature. More significantly, we have verified that there are strictly no solutions even when allowing $\sigma_i \sim \ell^2$.

This demonstrates that the zero-energy bound states of extremal black holes without antimomenta—those with exotic properties such as having the same energy as the BPS states and a flat direction in $\ell^2$—do not admit thermodynamic perturbations. They are isolated at $T=0$ and are destabilized by both thermodynamic and quantum fluctuations.
\end{itemize}
\medskip

In conclusion, regular bound states of extremal black holes admit thermodynamic perturbations as long as they carry antimomenta and have energy strictly above the BPS limit. These perturbations backreact only slightly on the geometry, provided the temperature is below the bound in \eqref{eq:TemperatureRange}. The resulting solution is nearly indistinguishable from the extremal bound state, except in the immediate vicinity of each black hole. This suggests that the low-energy bound states below the CFT mass gap are stable under thermodynamic and possibly quantum fluctuations. In contrast, the exotic extremal bound states with energy equal to the BPS ground state do not survive such perturbations and are isolated $T=0$ solutions.

\section{Closing comments}
\label{sec:Conclusion}

We have uncovered a discrete family of physical bound states of branes and antibranes in AdS$_3\times$S$^3/\mathbb{Z}_{N_k}\times$T$^4$ realized within type IIB supergravity. These non-BPS configurations form a quantized spectrum determined solely by the number of anti-BPS constituents at fixed charges. Most strikingly, many of these states lie within the CFT mass gap---“impossible states” that defy the usual expectation of a strict energy gap $\sim 1/c$ between supersymmetric ground states and the first excitations. 

The ultimate significance of these states hinges on whether they remain within the gap once quantum effects---particularly those associated with the Schwarzian mode of the UV throat---are properly taken into account. This remains the most pressing question raised. But even if quantum corrections ultimately clear the gap, the states we have identified would still be of considerable interest. They show that the CFT spectrum harbors intricate, highly nonperturbative excited states near the lowest possible energies which (i) carry momentum in a qualitatively very different way from typical thermal states, (ii) are extremely numerous, with large entropies comparable to those of the four-charge black hole, and (iii) admit smooth supergravity descriptions as black hole/anti-black hole bound states. At the very least, this is a striking demonstration of the remarkable richness of the spectra of holographic CFTs, and of the power of holography to reveal it.

The interpretations proposed in Section~\ref{sec:Interpretation} represent only a first attempt to make sense of these findings, but their discovery opens a number of questions that call for further detailed investigation, beginning with the following:
\begin{itemize}
    \item \textbf{Entanglement entropy and quantum scars.} Entanglement entropy is a natural probe of the microscopic structure of black hole geometries. It can be derived via the Ryu-Takayanagi prescription \cite{Ryu:2006bv,Ryu:2006ef} and its extensions to geometries with nontrivial dependence on internal directions \cite{Bena:2024gmp}. Previous work \cite{Hayden:2020vyo,Wei:2022cpj} has shown that sufficiently many disentangled black hole microstates, with entanglement entropies parametrically smaller than those of typical microstates, exist and account for the leading contribution to the Bekenstein-Hawking entropy. These are called \emph{quantum scars} \cite{Bernien:2017ubn,Serbyn:2020wys}. Their existence has motivated the study of localized black hole geometries in AdS$_3$ \cite{Bena:2024gmp,Dias:2025csz} and the comparison of their entanglement entropy with that of BTZ black holes. However, the geometries in \cite{Bena:2024gmp} exist only at low energy, so their entanglement entropy differs only slightly from that of BTZ. Our brane/antibrane bound states also realize localized black hole geometries, but they persist at high energies and account for a fraction of the black hole entropy. It would be of great interest to compute their entanglement entropy and determine whether they provide AdS$_3$ realizations of quantum scars.
    \item  \textbf{Thermodynamics.} In the last section, we showed that the bound states admit finite-temperature generalizations, allowing one to study them in both the microcanonical ensemble (fixed energy and charges) and the canonical ensemble (fixed temperature and charges). Although these solutions remain subdominant compared to the nonextremal black hole, it would be worthwhile to derive their thermodynamic laws by applying the variational principle to the Euclidean action, as done for related multi-black hole geometries \cite{Gregory:2020mmi,Krtous:2019fpo,Garcia-Compean:2020gii}, and compare the results with those of the corresponding black hole.
    \item \textbf{Characterizing the instability.} Section~\ref{sec:Stability} indicated that the bound states possess an instability with a rather small decay rate. Deriving explicit perturbations would be challenging because of the lack of spherical symmetry, but one can explore this dynamics through simpler probe analyses. For instance, one could study the potential felt by various brane probes and extract tunneling rates using the Born-Infeld action, as in \cite{Bena:2015dpt}. Another possibility is to analyze Schwinger pair production---e.g., of momentum-antimomentum pairs---in regions of strong electromagnetic fields, such as the segment connecting the two extremal black holes following \cite{Pioline:2005pf}.
\end{itemize}

To conclude, although we focused on brane/antibrane bound states realized as two-center geometries, one can envision configurations where the brane/momentum and antibrane/antimomentum components localize at multiple centers. As their number increases, the geometries are expected to become increasingly ``typical,''\footnote{Ref.~\cite{Dulac:2024cso} showed that adding extremal centers makes the bound states progressively harder to distinguish from the nonextremal black hole. An analogy may help: consider a gas of two species in equal proportion. Configurations where each species occupies one side of the box capture half the entropy but are highly atypical. As the number of regions occupied by each species increases---up to scaling with the number of particles---the subset still carries a finite entropy fraction while becoming indistinguishable from random configurations in the phase space.} approaching the structure of the nonextremal four-charge black hole, though at the cost of greater complexity in both construction and analysis. A key question is whether the minimal energy gap found for two centers persists once additional centers are introduced. It would be interesting to explore three-center bound states and test whether they can reduce this gap, adapting the construction of \cite{Dulac:2024cso} to AdS$_3$ asymptotics.

\section*{Acknowledgments}

We are grateful to Iosif Bena, Guillaume Bossard, Soumangsu Chakraborty, Rapha\"el Dulac, Samir Mathur, Gela Patashuri, Nick Warner, and Zixia Wei for comments and stimulating discussions. We especially thank Joaquín Turiaci for sharing the unitarity argument discussed in Sec.~\ref{sec:Interpretation}.
R.E. is supported by MICINN grant PID2022-136224NB-C22, AGAUR grant 2021 SGR 00872, and grant CEX2024-001451-M funded by MICIU/AEI/10.13039/501100011033. 
P.H. is supported by the Department of Physics at The Ohio State University.

\appendix
\section{Ernst solution in AdS}
\label{App:Ernst}

In this appendix, we detail the Ernst technique, which enables the construction of novel non-BPS asymptotically AdS solutions sourced by multiple axially symmetric centers and rods.

\subsection{From flat to AdS asymptotics}

Many asymptotically flat solutions to the electrostatic Ernst equations \eqref{eq:EOMErnst} have been constructed within four-dimensional Einstein-Maxwell theory, including the most general configuration describing $N$ non-extremal sources with arbitrary mass and charge \cite{Ruiz:1995uh,NoraBreton1998}.

These solutions are expressed in terms of two Ernst potentials, $(\cE,\Phi)$:
\begin{equation}
    \cE \equi Z^{-2} - A^2\,,\qquad \Phi \equi A\,,
\end{equation}
along with the base conformal factor $\nu$, where $Z$ and $A$ are the gravitational and electric potentials defined in \eqref{eq:EOMErnst}. In four-dimensional Einstein-Maxwell theory, the potentials enter the metric and gauge field as
\begin{equation}
\begin{split}
ds_4^2 &\= - \frac{dt^2}{Z^2}+ Z^2\left[e^{8\nu}\left( d\rho^2+dz^2\right) +\rho^2 d\phi^2 \right],\quad F \=  -2 \, dA\wedge dt\,,
\end{split}
\end{equation}
whereas in type IIB supergravity, four decoupled Ernst sectors appear, as detailed in Section \ref{sec:ErnstTypeIIB}. For asymptotically flat solutions, we have $Z \to 1$ and $A \to 0$ at large distances, whereas for asymptotically AdS solutions, $Z$ must vanish and $A$ must diverge radially, as in \eqref{eq:AsympBehav}. \medskip

A systematic way to adapt any known asymptotically flat solution to AdS asymptotics is to apply an Ernst transformation that preserves the source structure and conformal factor $\nu$ but modifies the asymptotics of the Ernst potentials. A suitable transformation is the Harrison transformation \cite{harrison_new_1968,Harrison:1980fr}, which acts as
\begin{equation}
\mathcal{E} \to \frac{\mathcal{E}}{1-2c \Phi - c^2 \mathcal{E}}\,,\qquad \Phi \to \frac{c \mathcal{E}+\Phi}{1-2c \Phi - c^2 \mathcal{E}}\,,
\label{eq:Harrison}
\end{equation}
where $c$ is an arbitrary constant and $\nu$ remains unchanged. In terms of the gravitational and electric potentials, this transformation becomes
\begin{equation}
Z \to \pm \,\frac{\left(Z(1-c A)-c\right)\left(Z(1-c A)+c\right)}{Z}\,,\qquad A \to \frac{c+(1-cA)AZ^2}{\left(Z(1-c A)-c\right)\left(Z(1-c A)+c\right)}\,.
\label{eq:HarrisonTransformation}
\end{equation}
Starting from an asymptotically flat solution with $(Z,A)\to (1,0)$, the transformed asymptotics are
\begin{equation}
Z \to \pm (1-c^2)\,,\qquad A \to \frac{1}{1-c^2}\,.
\end{equation}
Thus, by setting $c=\pm 1$, the Harrison transformation maps any asymptotically flat solution to an asymptotically AdS one. Importantly, the internal structure is preserved: wherever $Z$ diverges (i.e., at sources), the transformed $Z$ also diverges. However, local masses and charges at the sources may change.

The only remaining field to determine is the magnetic dual potential $H$, which is obtained by integrating the electromagnetic duality equation \eqref{eq:EMDualEq}.

\subsection{Bound states of two non-extremal sources}
\label{app:2nonExSources}

The asymptotically flat solution of the electrostatic Ernst equations corresponding to two non-extremal sources was derived in \cite{Alekseev:2007re,Alekseev:2007gt,Manko:2007hi} and further simplified in \cite{Bah:2022yji}.\footnote{Solutions with two non-extremal four-charge black holes were constructed with another method in \cite{Emparan:2001bb}.}

The two sources are represented by rod segments of lengths $\sigma_1/2$ and $\sigma_2/2$ aligned along the $z$-axis in Weyl-Papapetrou coordinates. The centers of these segments are separated by a distance $\ell^2/4 > (\sigma_1 + \sigma_2)/4$, and we place the origin of the $z$-axis at the location of the first source. The sources generate masses $(m,\overline{m})$ and carry charges $(q,\overline{q})$, related to the rod lengths by
\begin{equation}
\sigma_1 \equi \sqrt{m^2-q^2+2 q \gamma }\,,\qquad  \sigma_2 \equi \sqrt{\overline{m}^2-\overline{q}^2-2 \overline{q} \gamma }, \qquad
\gamma \equi \frac{\overline{m} q \- m \overline{q}}{\ell^2+m+\overline{m}} .
\label{eq:ArbitraryChargeParam}
\end{equation}
We introduce spherical coordinates centered at each rod, $(r_1,\theta_1)$ and $(r_2,\theta_2)$, defined in terms of the Weyl-Papapetrou coordinates as:\footnote{We define the spherical coordinates as five-dimensional radial and angular coordinates. The relation between four-dimensional and five-dimensional coordinates are $r^{(4d)} = \frac{{r^{(5d)}}^2}{4}$ and $\theta^{(4d)}=2\theta^{(5d)}$.}
\begin{equation}
\begin{split}
    \frac{r_i^2}{2} &\equi \sqrt{\rho^2+\left(z-a_i-\frac{\sigma_i}{4} \right)^2}+\sqrt{\rho^2+\left(z-a_i+\frac{\sigma_i}{4} \right)^2}-\frac{\sigma_i}{2},\\
    \cos 2\theta_i &\equi \frac{2\left(\sqrt{\rho^2+\left(z-a_i+\frac{\sigma_i}{4} \right)^2}-\sqrt{\rho^2+\left(z-a_i-\frac{\sigma_i}{4} \right)^2}\right)}{\sigma_i}, \label{eq:DistanceDef}
\end{split}
\end{equation}
with $a_1 = 0$ and $a_2 = \ell^2/4$.

The asymptotically flat solutions to the electrostatic Ernst equations for these sources are given by:
\begin{align}\label{eq:ArbitraryChargeSol}
Z & \=  \frac{(r_1^2+\sigma_1+m)(r_2^2+\sigma_2+\overline{m})-\left(q-2\gamma \cos^2 \theta_2\right)\left(\overline{q}+2\gamma\sin^2 \theta_1\right)}{\sqrt{\left(r_1^2(r_1^2+2\sigma_1)+\gamma^2\sin^2 2\theta_2 \right)\left(r_2^2(r_2^2+2\sigma_2)+\gamma^2\sin^2 2\theta_1 \right)}}\,, \nonumber\\
A & \= \frac{(q-\gamma) (r_2^2+\sigma_2) +(\overline{q}+\gamma) (r_1^2+\sigma_1) +\gamma \left( m \cos 2\theta_1  +\overline{m} \cos 2\theta_2  \right)}{(r_1^2+\sigma_1+m)(r_2^2+\sigma_2+\overline{m})-\left(q-2\gamma \cos^2 \theta_2\right)\left(\overline{q}+2\gamma\sin^2 \theta_1\right)} \,,\\
e^{8\nu} & \=  \frac{\left(r_1^2(r_1^2+2\sigma_1)+\gamma^2\sin^2 2\theta_2 \right)\left(r_2^2(r_2^2+2\sigma_2)+\gamma^2\sin^2 2\theta_1 \right)}{(1+2\delta)^2\,\left(r_1^2(r_1^2+2\sigma_1)+\sigma_1^2\sin^2 2\theta_2 \right)\left(r_2^2(r_2^2+2\sigma_2)+\sigma_2^2\sin^2 2\theta_1 \right)}  \nonumber\\
& \hspace{-0.4cm} \text{ \scalebox{0.9}{$\times \left(1+2\delta \,\frac{(q-\gamma) \left( r_1^2+\sigma_1- \ell^2 \cos 2\theta_1\right)+(\overline{q}+\gamma) \left( r_2^2+\sigma_2+ \ell^2 \cos 2\theta_3\right)+\gamma \left( \overline{m} \cos 2\theta_1+m \cos 2\theta_2 \right)}{(q-\gamma) (r_2^2+\sigma_2) +(\overline{q}+\gamma) (r_1^2+\sigma_1) +\gamma \left( m \cos 2\theta_1  +\overline{m} \cos 2\theta_3  \right)} \right)^2$}} ,\nonumber
\end{align}
where we introduced an auxiliary parameter
\begin{equation}
\delta \equi \frac{m \overline{m} -(q-\gamma)(\overline{q}+\gamma)}{\ell^4-m^2-\overline{m}^2+(q-\gamma)^2+(\overline{q}+\gamma)^2}\,.
 \label{eq:deltaParam}
\end{equation}
The magnetic dual potential $H$ has a complicated expression that can be found in Appendix B of \cite{Bah:2022yji}. \medskip

Applying the Harrison transformation \eqref{eq:HarrisonTransformation} with $c = \pm 1$ yields a new solution $(Z', A', \nu)$ with AdS asymptotics:
\begin{equation}
    Z' \= \frac{Z^2(1\mp A)^2-1}{Z},\qquad A'\= \frac{\pm 1+(1\mp A)A Z^2}{Z^2(1\mp A)^2-1}. \label{eq:HarrisonTranformation2}
\end{equation}
Although we were unable to significantly simplify these new fields or find a closed-form expression for the transformed magnetic potential $H'$ from \eqref{eq:EMDualEq}, the transformation clearly shifts the asymptotics of the solution to AdS while leaving the rod structure intact.

Upon examining the behavior of the transformed fields near the rods, we found that the original parameters $(m, \overline{m}, q, \overline{q})$ no longer correspond to physical quantities. Instead, it is more natural to parametrize the new solution in terms of transformed charges $(q', \overline{q}')$ and the non-extremality parameters $(\sigma_1, \sigma_2)$:
\begin{align}
    m &\= \frac{q'^2+\sigma_1^2-\gamma^2}{2 q'},\quad \overline{m} \= \frac{\overline{q}'^2+\sigma_2^2-\gamma^2}{2 \overline{q}'},\quad q \= \frac{-q'^2+\left(1+\frac{\overline{q}'}{\ell^2} \right)(\sigma_1^2-\gamma^2)-\frac{q'^2}{\ell^2 \overline{q}'}(\sigma_2^2-\gamma^2)}{2q'}, \label{eq:ReparamNonExAdS}\\
    \overline{q} &\= \frac{-\overline{q}'^2+\left(1+\frac{q'}{\ell^2} \right)(\sigma_2^2-\gamma^2)-\frac{\overline{q}'^2}{\ell^2 q'}(\sigma_1^2-\gamma^2)}{2\overline{q}'},\quad \gamma \= \frac{\ell^2 q' \overline{q}' - \sqrt{\ell^4 q'^2 \overline{q}'^2+(q'^2-\overline{q}'^2)(q'^2 \sigma_2^2-\overline{q}'^2 \sigma_1^2)}}{q'^2 -\overline{q}'^2}.\nn
\end{align}

When interpreted in four-dimensional Einstein-Maxwell theory, this solution describes two non-extremal black holes in AdS$_2 \times$S$^2$ of charges $(q', \overline{q}')$ and nonextremality parameters $(\sigma_1, \sigma_2)$. The conical defect (or strut) between the sources can be eliminated by considering an orbifold S$^2/\mathbb{Z}_{N_k}$. In this paper, however, we are more interested in embedding these solutions into type IIB supergravity to construct regular geometries in AdS$_3\times$S$^3/\mathbb{Z}_{N_k} \times$T$^4$.

\subsection{Bound states of two extremal sources}
\label{app:BS2Ex}

The solutions above simplify significantly in the extremal limit, where the sources become BPS or anti-BPS:
\begin{equation}
    \sigma_1 \= \sigma_2 \= 0\,.
\end{equation}
In the asymptotically flat case, this condition fixes the charges $(q, \overline{q})$ in terms of the masses $(m, \overline{m})$. There are two distinct extremal configurations depending on the relative signs of the charges.

\begin{itemize}
    \item \underline{Mutually BPS or anti-BPS sources:}
\end{itemize}
When both sources carry charges of the same sign,
\begin{equation}
    (q,\overline{q}) \= \pm (m,\overline{m})\,,\qquad \gamma=0.
\end{equation}
the solution reduces to the well-known linear Majumdar-Papapetrou configuration:
\begin{equation}
    Z  \= \frac{1}{\mp A+1}\= 1 +\frac{m}{r_1^2}+ \frac{\overline{m}}{r_2^2},\quad H \= \pm \left(m \cos 2\theta_1+\overline{m} \cos 2\theta_2 \right),\quad \nu \= 0\,,
\end{equation}
where $(r_1, \theta_1)$ and $(r_2, \theta_2)$ are spherical coordinates centered at two locations along the $z$-axis at $z = 0$ and $z = \ell^2/4$, respectively, as defined in \eqref{eq:DistanceDef}. \medskip

The asymptotically AdS solution obtained by the Harrison transformation \eqref{eq:HarrisonTranformation2} corresponds to the common decoupling limit of the BPS (resp. anti-BPS) solution, consisting of ``removing the $1$'' in the harmonic function $Z$.

\begin{itemize}
    \item \underline{BPS/anti-BPS pair:}
\end{itemize}
When the sources have opposite charges, the extremal configuration becomes more intricate:
    \begin{equation}
        q = m \sqrt{\frac{(\ell^2+\overline{m})^2-m^2}{(\ell^2-\overline{m})^2-m^2}}\,,\quad \overline{q} = -\overline{m} \sqrt{\frac{(\ell^2+m)^2-\overline{m}^2}{(\ell^2-m)^2-\overline{m}^2}}, \quad \gamma = \frac{2\ell^2 m \overline{m}}{\sqrt{(\ell^4-(m+\overline{m})^2)(\ell^4-(m-\overline{m})^2)}}.\nn
    \end{equation}
The resulting solution is
    \begin{align}
        Z &\= 1+2\frac{M (2r^2+\ell^2+M )-Q\left(Q+\sqrt{\ell^4-M^2+Q^2}\cos2\theta \right)}{4r^2(r^2+\ell^2)+(\ell^4-M^2+Q^2)\sin^22\theta}, \nn \\
 A &\= -\frac{2Q(2r^2+\ell^2)-2M \sqrt{\ell^4-M^2+Q^2}\,\cos 2\theta}{(2r^2+\ell^2+M)^2-\left( Q+\sqrt{\ell^4-M^2+Q^2}\cos2\theta \right)^2}, \label{eq:ErnstSol2ExFlat}\\
 e^{2\nu} & \=  1- \frac{(M^2-Q)^2 \,\sin^2 2\theta}{4(r^2+\ell^2 \cos^2\theta)(r^2+\ell^2 \sin^2\theta)},\quad H = - \frac{Q}{4} \cos 2\theta + \frac{(Z-1)\sqrt{\ell^4-M^2+Q^2}}{2Q} \sin^2 2\theta, \nn
    \end{align}
where we redefined the total mass and charge as $M=m+\overline{m}$ and $Q=q+\overline{q}$, and used spherical coordinates $(r,\theta)$ centered around the segment in between both extremal centers:
\begin{equation}
\begin{split}
    \frac{r^2}{2} \equi \sqrt{\rho^2+\left(z-\frac{\ell^2}{4} \right)^2}+\sqrt{\rho^2+z^2}-\frac{\ell^2}{4},\quad \cos 2\theta \equi \frac{4\left(\sqrt{\rho^2+z^2}-\sqrt{\rho^2+\left(z-\frac{\ell^2}{4} \right)^2}\right)}{\ell^2}. \label{eq:DistanceDef2}
\end{split}
\end{equation}
One can verify that $M \geq Q$ and that the solution is well-defined as long as $\ell^4 - M^2 + Q^2 \geq 0$. This imposes a lower bound on the separation $\ell$, ensuring it exceeds the effective ``Reissner-Nordström radius'' of the bound state. \medskip

We now apply the Harrison transformation \eqref{eq:HarrisonTransformation} to the solutions with $c=1$. We can simplify the transformed fields and derive an explicit expression for the magnetic dual potential $H'$, using \eqref{eq:HarrisonTranformation2} and \eqref{eq:ReparamNonExAdS} with $\sigma_1 = \sigma_2 = 0$:
\begin{align}
Z' &\= \frac{2(q+\overline{q}) \left(r^2+\frac{\ell^2}{2} \left(1- \frac{q+\overline{q}}{q-\overline{q}}\cos 2\theta \right) \right)}{r^2(r^2+\ell^2)+\frac{\ell^4 (q+\overline{q})^2}{4(q-\overline{q})^2} \sin^2 2\theta}\,,\quad H' =  \frac{q+\overline{q}}{2}\left( \cos 2\theta  + \frac{\ell^2 \,Z}{4(q-\overline{q})}\,\sin^2 2 \theta \right)\,, \nn\\
 A' &\= -\frac{r^2}{2(q+\overline{q})}-\frac{\ell^2}{4(q-\overline{q})} \left( \cos 2\theta - \frac{2\ell^2 q \overline{q}}{(q^2 -\overline{q}^2)\left( r^2+\frac{\ell^2}{2} \left(1- \frac{q+\overline{q}}{q-\overline{q}}\cos 2\theta \right)\right)} \right),  \label{eq:ErnstSol2ExAdS}\\ 
e^{2\nu}&=1+ \frac{\ell^4q \overline{q} \,\sin^2 2\theta}{(q-\overline{q})^2(r^2+\ell^2 \cos^2\theta)(r^2+\ell^2 \sin^2\theta)}. \nn 
\end{align}

A particularly interesting limit arises when $(q, -\overline{q}) \to \infty$ with fixed total charge $Q = q + \overline{q}$. This corresponds to local sources with large individual charges and small net charge. In this regime, the solution approaches that of a single rod of length $\ell^2/4$ and total charge $Q$:
\begin{align}
Z' &\= \frac{2Q \left(r^2+\frac{\ell^2}{2} \right)}{r^2(r^2+\ell^2)}\,,\qquad H' =  \frac{Q}{2}\,\cos 2\theta  \,, \nn \\ 
A' &\= -\frac{r^2}{2Q}- \frac{\ell^4}{8Q\left(r^2+\frac{\ell^2}{2}\right)},  \qquad e^{2\nu}=\frac{r^2(r^2+\ell^2)}{(r^2+\ell^2 \cos^2 \theta)(r^2+\ell^2 \sin^2 \theta)}.  \label{eq:SingleRodAdS}
\end{align}

In Section \ref{sec:BTZ}, we have seen that the asymptotically flat nonextremal four-charge black hole can be modified into an asymptotically AdS black hole by taking the near-horizon and near-extremal decoupling limit \eqref{eq:NearHorizonLim} and \eqref{eq:NearExtremalLim}. The Harrison transformation used to go from \eqref{eq:ErnstSol2ExFlat} to \eqref{eq:ErnstSol2ExAdS} is also related to a ``near-bound-state" and near-extremal limit of the asymptotically flat solution. Indeed, one can go from \eqref{eq:ErnstSol2ExFlat} to \eqref{eq:ErnstSol2ExAdS} by considering the scaling of coordinates:
\begin{equation}
    r^2 \to \epsilon^2\, r^2\,,\qquad t \to \frac{1}{\epsilon} \,t\,,
\end{equation}
and the near-extremal limit and parameter redefinition:
\begin{equation}
    \ell^2 \to \epsilon^2 \ell^2,\qquad M \to 2(q+\overline{q}),\qquad Q\to 2(q+\overline{q}) \left(1-\frac{\ell^4}{8(q+\overline{q})^2}\left(1- \left(\frac{q+\overline{q}}{q-\overline{q}}\right)^2\right) \, \epsilon^4 \right),
\end{equation}
with the limit $\epsilon \to 0$.

However, while the decoupling limit for the nonextremal four-charge black hole has forced the fields to be extremal and BPS, the present decoupling limit forces extremality ($Q\to M$), but the solution remains non-BPS with a pair of BPS and anti-BPS centers.

\section{Regular bound states of two black holes in AdS$_3$}
\label{App:ErnstTypeIIB}

In this appendix we present the construction of regular bound states of two black holes in AdS$_3\times$S$^3/\mathbb{Z}_{N_k}\times$T$^4$ within type IIB supergravity. As reviewed in Section \ref{sec:ErnstTypeIIB}, static cohomogeneity-two solutions in type IIB supergravity with D1-D5-P-KKm flux decompose into four decoupled sectors governed by electrostatic Ernst equations:
\begin{itemize}
    \item The D1 sector $(Z_1,A_1,\nu_1)$.
    \item The D5 sector $(Z_5,H_5,\nu_5)$.
    \item The P sector $(Z_p,A_p,\nu_p)$.
    \item The KKm sector $(Z_0,H_0,\nu_0)$.
\end{itemize}
Note that in the D1 and P sectors, the primary gauge potential is the electric potential $A$, while in the D5 and KKm sectors, it is the magnetic potential $H$.

Each sector must be solved consistently to construct solutions describing two black holes with opposite D1-D5-P charges, separated by a Taub-bolt along the $\psi$ direction that carries $N_k$ units of KKm charge. Moreover, to ensure the solution is asymptotic to AdS$_3\times$S$^3/\mathbb{Z}_{N_k}\times$T$^4$ as in \eqref{eq:AsympBehav}, the P sector $(Z_p, A_p, \nu_p)$ must satisfy flat boundary conditions, while the D1, D5, and KKm sectors must exhibit AdS asymptotics.

\subsection{Bound states of extremal black holes}

To generate two Strominger-Vafa black holes with opposite D1-D5-P charges, we solve the D1, D5, and P Ernst sectors using the extremal solutions reviewed in Section \ref{app:BS2Ex}. Since the P sector must exhibit flat asymptotics, we use \eqref{eq:ErnstSol2ExFlat} for $(Z_p, A_p, \nu_p)$, and \eqref{eq:ErnstSol2ExAdS} for both the D1 and D5 sectors: $(Z_1, A_1, \nu_1)$ and $(Z_5, H_5, \nu_5)$.

To avoid a conical singularity (strut) between the black holes, the KKm sector $(Z_0, H_0, \nu_0)$ is solved using the asymptotically AdS solution \eqref{eq:SingleRodAdS}, corresponding to a single rod source carrying $Q = N_k$ units of KKm charge.

Since all sectors are written in terms of the spherical coordinates $(r, \theta)$ defined in \eqref{eq:DistanceDef2}, we change from Weyl-Papapetrou to spherical coordinates using:
\begin{equation}
\begin{split}
     &\rho \= \frac{r\sqrt{r^2+\ell^2}}{4} \sin 2\theta,\quad z \= \frac{2r^2+\ell^2}{8} \cos 2\theta+\frac{\ell^2}{8},\\
     &d\rho^2+dz^2 \= \frac{(r^2+\ell^2 \cos^2 \theta)(r^2+\ell^2 \sin^2 \theta)}{4} \left(\frac{dr^2}{r^2+\ell^2} +d\theta^2\right).
\end{split}
\end{equation}
Combining all components yields the full solution introduced and analyzed in Section \ref{sec:3}, whose source structure is illustrated in Fig.~\ref{fig:D1D5PBS}.

\subsection{Bound states of non-extremal black holes}

In Section \ref{sec:BSNonExt}, we extended the extremal black hole bound states to the non-extremal case, describing black holes at finite temperature. The construction closely follows the extremal procedure but involves additional subtleties.

The two non-extremal D1-D5-P black holes are obtained by solving the D1, D5, and P Ernst sectors using the general solutions presented in Section \ref{app:2nonExSources}. The P sector $(Z_p, A_p, \nu_p)$ is solved using the asymptotically flat solution \eqref{eq:ArbitraryChargeSol}, while the D1 and D5 sectors, $(Z_1, A_1, \nu_1)$ and $(Z_5, H_5, \nu_5)$, are solved using the asymptotically AdS solution \eqref{eq:HarrisonTranformation2}. Note that we could not obtain a closed-form expression for $H_5$; however, it is fully determined by the duality condition \eqref{eq:EMDualEq}, which suffices for analyzing the solution.

Solving the KKm sector is more involved than in the extremal case. First, the solution $(Z_0, H_0, \nu_0)$ must contain the same rod source along the segment between the black holes. Second, because $Z_1$ and $Z_5$ diverge at the black holes, $Z_0$ must also be sourced along the black hole rods to ensure the $\psi$-circle remains finite there, as required by \eqref{eq:TypeIIBAnsatz}. Thus, the KKm sector must be a composite of three rod sources with appropriate divergences. Using the method of \cite{Bah:2022pdn}, we find:
\begin{align}
    Z_0 &\= \frac{2N_k}{\ell^2} \sinh\left[\log \left(\left(1+\frac{\ell^2-\sigma_1-\sigma_2}{r_0^2}\right)\prod_{i=1,2} \left(1+\frac{2\sigma_i}{r_i^2} \right)^\frac{1}{2} \right)  \right], \nn\\
    H_0 & \= \frac{N_k}{2\ell^2}\left(\left(\ell^2-\sigma_1-\sigma_2 \right) \cos 2\theta_0+\sigma_1 \cos 2\theta_1 +\sigma_2 \cos 2\theta_2 \right), \\
    e^{2\nu_0} &\=   \prod_{i, j=0,1,2} \left( \frac{\left( \left(r_i^2+\ell_i^2 \right) \cos^2\theta_i +  \left(r_j^2+\ell_j^2\right) \sin^2\theta_j \right)\left(r_i^2 \cos^2\theta_i +  r_j^2 \sin^2\theta_j \right)}{\left( \left(r_i^2+\ell_i^2 \right) \cos^2\theta_i + r_j^2 \sin^2\theta_j \right)\left(r_i^2 \cos^2\theta_i +  \left(r_j^2+\ell_j^2\right) \sin^2\theta_j \right)}\right)^{\alpha_{ij}}\,, \nn
\end{align}
where $\ell_0^2 = \ell^2 - \sigma_1 - \sigma_2$, $\ell_1^2 = 2\sigma_1$, $\ell_2^2 = 2\sigma_2$, and the symmetric matrix $\alpha$ has components: $\alpha_{00} = 1$, $\alpha_{01} = \alpha_{02} = 1/2$, and $\alpha_{11} = \alpha_{22} = \alpha_{12} = 1/4$. The spherical coordinates $(r_0, \theta_0)$ centered around the inter-black-hole segment are defined as:
\begin{equation}
\begin{split}
    \frac{r_0^2}{2} &\equi \sqrt{\rho^2+\left(z-\frac{\ell^2-\sigma_2}{4} \right)^2}+\sqrt{\rho^2+\left(z-\frac{\sigma_1}{4} \right)^2}-\frac{\ell^2-\sigma_1-\sigma_2}{4},\\
    \cos 2\theta_0 &\equi \frac{4\left(\sqrt{\rho^2+\left(z-\frac{\sigma_1}{4} \right)^2}-\sqrt{\rho^2+\left(z-\frac{\ell^2-\sigma_2}{4}  \right)^2}\right)}{\ell^2-\sigma_1-\sigma_2},
\end{split}
\end{equation}
The full type IIB solution is obtained by implementing the four sectors into the general metric and field ansatz \eqref{eq:TypeIIBAnsatz}. \medskip

Regularity and thermodynamic properties of the solution, including the black hole temperatures, were analyzed in Section \ref{sec:RegCondNonEx} by expanding the fields near the rods along the $z$-axis ($\rho = 0$):
\begin{itemize}
    \item \underline{Regularity at the bolt:}
    
    The bolt lies along the segment $\rho = 0$ and $\frac{\sigma_1}{4} \leq z \leq \frac{\ell^2 - \sigma_2}{4}$, where the $\psi$-circle shrinks. The induced $(\rho,\psi)$ metric is:
    \begin{equation}
        d\rho^2 \+ \left(\frac{e^{-2\nu}}{Z_0^2} \right)_{\rho\to 0,z\in [\frac{\sigma_1}{4},\frac{\ell^2-\sigma_2}{4}]}\, (d\psi+H_0 d\phi)^2,
    \end{equation}
    and regularity requires
    \begin{equation}
       \left(\frac{e^{-2\nu}}{Z_0^2} \right)_{\rho\to 0,z\in [\frac{\sigma_1}{4},\frac{\ell^2-\sigma_2}{4}]} \= \rho^2\,. 
    \end{equation}
    \item \underline{Conditions at the black holes:}
    The black holes lie along the segments $\rho = 0$, $-\frac{\sigma_1}{4} \leq z \leq \frac{\sigma_1}{4}$ and $\frac{\ell^2 - \sigma_2}{4} \leq z \leq \frac{\ell^2 + \sigma_2}{4}$. Focusing on the first, the induced $(\rho, t)$ metric reads:
    \begin{equation}
        d\rho^2 - \left(\frac{e^{-2\nu}}{Z_0Z_1 Z_5 Z_p} \right)_{\rho\to 0,z\in [-\frac{\sigma_1}{4},\frac{\sigma_1}{4}]}\, dt^2
    \end{equation}
    and the temperature condition requires
    \begin{equation}
         \left(\frac{e^{-2\nu}}{Z_0Z_1 Z_5 Z_p} \right)_{\rho\to 0,z\in [-\frac{\sigma_1}{4},\frac{\sigma_1}{4}]} \= \left(2\pi\frac{T}{R_y} \right)^2\, \rho^2\,.
    \end{equation}
    The horizon area density can then be directly derived from the temperature since
    \begin{equation}
        \det g_{(y,z,\psi,\phi,T^4)}\= \rho^2\,e^{2\nu} Z_0 Z_1 Z_5 Z_p.
    \end{equation}
    which remains constant along the rod, equal to $\left(2\pi\frac{T}{R_y} \right)^{-1}$, and yields a Bekenstein-Hawking entropy of
    \begin{equation}
        S_1 \= \frac{A_1}{4G_{10}} \= \frac{R_y^2 V_4\,\sigma_1}{g_s^2 l_s^8\,T},
    \end{equation}
    using $G_{10} = 8\pi^6 g_s^2 l_s^8$. The analysis on the second black hole is identical by replacing $\sigma_1$ with $\sigma_2$.
\end{itemize}

\subsection{Energy and quantized charges, and central charge in AdS$_3$}
\label{app:EnergyAdS}

The procedure to derive the energy and quantized momentum charge in AdS$_3$ consists of expanding the asymptotic three-dimensional $(r,t,y)$ metric \eqref{eq:AsympCoord} as
\begin{equation}
ds^2  \underset{u\to 0}{\sim} N_k\sqrt{Q_1 Q_5} \left[ \frac{du^2+\eta_{\mu\nu} dw^\mu dw^\nu}{u^2}  + g_{\mu \nu}^{(2)} dw^\mu dw^\nu \right] + \ldots \,,\qquad w_\mu \equiv (\tau,\sigma) \equiv \frac{(t,y)}{R_y}\,,
\end{equation}
and $u\equiv \frac{\sqrt{2N_k Q_1 Q_5}}{R_y\,r}$.  For static solutions,  the conformal dimensions are derived from the subleading-order expansion:
\begin{equation}
h - \frac{c}{24} \= \frac{c}{24} \left[ g_{\tau\tau}^{(2)}+g_{\sigma\sigma}^{(2)} +2 g_{\tau\sigma}^{(2)} \right],\qquad \bar{h} - \frac{c}{24} \= \frac{c}{24} \left[ g_{\tau\tau}^{(2)}+g_{\sigma\sigma}^{(2)} -2 g_{\tau\sigma}^{(2)} \right],
\end{equation}
where the shifts $c/24$ represent half the energy of the NS-NS ground state (global AdS$_3$).  Then the energy and momentum are given by
\begin{equation}
N_p \= h - \bar{h}\,,\qquad E\= h+\bar{h}-\frac{c}{12}\,.
\end{equation}

The net quantized D1 and D5 brane charges are derived from the integral of the three-form RR field strength:
\begin{equation}
    N_1 \= \frac{1}{g_s (2\pi l_s)^6} \int_{S^3\times T^4} \star F_3\,,\qquad N_5 \= \frac{1}{g_s (2\pi l_s)^2} \int_{S^3} F_3.
\end{equation}

Finally, the central charge using the standard formula in the string frame:
\begin{equation}
ds^2 \sim e^{2\varphi} \left[ds(AdS_3)^2+ds(S^3)^2 \right] + ds(T^4)^2\,, \qquad c= \frac{3 \int_{S^3\times T^4} e^{4\varphi-2\Phi} \sqrt{g_{S^3}\,g_{T^4}}}{2 G_{10}}\,,
\end{equation}
where $G_{10} = 8\pi^6 g_s^2 l_s^8$.

\section{Supersymmetric limit: configurations without antibranes and antimomenta}
\label{App:SUSYLim}

In this section, we investigate the supersymmetric limit of the solutions. We show that this limit corresponds to configurations without antibranes and antimomenta, $\overline{n}_1=\overline{n}_5=\overline{n}_p=0$, while still featuring a nonvanishing bolt, $\ell^2\neq 0$. Consequently, the supersymmetric limit does not yield the expected single-center four-charge black hole, but instead a BPS two-center configuration: one center corresponding to the BPS Strominger-Vafa black hole, and the other to a Taub-NUT center.

We will use the solution \eqref{eq:MetBS} with the fields and base given in the pole-centered coordinates \eqref{eq:FieldsPoleCoor} and \eqref{eq:BasePoleCoor}. In this frame, the solution consists of BPS D1-D5-P black hole located at the South pole with a NUT center at the North pole. Note that the regularity condition \eqref{eq:RegCond} requires $N_k=2$ for smooth solutions, but we will keep it arbitrary for now.

For $\overline{q}_I=0$, the fields and the four-dimensional hyper-Kähler base become
\begin{align}
    Z_I &=  \frac{q_I}{2r_S},\qquad Z_p = 1+\frac{Q_p}{4 r_S},\qquad A_1 = \frac{1}{Z_1},\qquad A_p = \frac{1}{Z_p}-1 ,  \nn \\
    H_5 &= -\frac{q_5}{2} \cos \theta_S, \qquad V = \frac{1}{N_k}\left(\frac{1}{r_S}+\frac{1}{r_N}\right)\,,\qquad H = -\frac{1}{N_k} \left(\cos \theta_S + \cos \theta_N \right),
\end{align}
and
\begin{equation}
    ds_4^2 \= \frac{1}{V} \left( d\phi + H \,d\psi \right)^2 + V \left[\frac{N_k^2}{4} \left(dr_S^2+{r_S}^2 d\theta_S^2 \right) + {r_S}^2 \sin^2 \theta_S\,d\psi^2\right]\,,
\end{equation}
According to \cite{Gutowski:2003rg}, the solutions satisfy the BPS conditions for type IIB solutions on T$^4$ with a null Killing vector. The base is hyper-Kähler in its Gibbons-Hawking form \cite{Gibbons:1987sp}, the $Z$'s and $V$ are harmonic functions with the dual relation between their associated magnetic gauge fields
\begin{equation}
    dV = - \star_3 d H\,,\qquad dZ_5 = - \star_3 dH_5.
\end{equation}
This shows that zero-energy bound states without antibranes are supersymmetric solutions of type IIB supergravity. However, there is one subtlety compared to the traditional BPS two-center solution and forces the regularity condition $N_k=2$: the conformal factor $\frac{N_k^2}{4}$ in the four-dimensional base. This factor can be absorbed by rescaling the radial distances to the centers and the $\psi$ angle as
\begin{equation}
    \bar{r}_{S/N} \equi \frac{N_k}{2} \,r_{S/N}\,,\qquad \bar{\psi} \equi \frac{2}{N_k} \psi\quad \Rightarrow\quad \bar{V} = \frac{1}{2} \left(\frac{1}{\bar{r}_S}+\frac{1}{\bar{r}_N} \right), \quad \bar{H} \= - \frac{1}{2} \left(\cos \theta_S + \cos \theta_N \right) ,\nn
\end{equation}
and we retrieve the usual Gibbons-Hawking form $ds_4^2 = \bar{V}^{-1}\left( d\phi + H \,d\bar{\psi} \right)^2 + \bar{V} ds(\IR^3)^2$. 

However, in this picture, the azimuthal angle of the $\IR^3$, $\bar{\psi}$ is not $2\pi$ periodic but $4\pi/N_k$. If $N_k \neq 2$, one has a conical singularity on the $z$-axis in between the centers. We therefore retrieve the regularity condition \eqref{eq:RegCond} without antibranes and antimomenta: $N_k=2$.

These results demonstrate that zero-energy and regular bound states without antibranes correspond to a subtle reformulation of a BPS two-center solution, where one center represents a BPS Taub-NUT locus and the other a BPS D1-D5-P black hole. The fact that this solution can be rewritten as a smooth bolt requires each center to carry a unit of KKm charge, leading to an asymptotic structure of AdS$_3\times$S$^3/\mathbb{Z}_{2}\times$T$^4$. Therefore, the existence of these solutions, the fact that they have the same energy as the BPS four-charge black hole of KKm charge $N_k=2$, or that the length $\ell^2$ is unbounded, is now unsurprising.


\bibliographystyle{utphys}      

\bibliography{microstates}       

\end{document}